\documentclass[conference,compsoc]{IEEEtran}

\def\BibTeX{{\rm B\kern-.05em{\sc i\kern-.025em b}\kern-.08em
    T\kern-.1667em\lower.7ex\hbox{E}\kern-.125emX}}

\usepackage{caption}
\usepackage{subcaption}

\usepackage[framemethod=tikz]{mdframed}
\usepackage[most]{tcolorbox}
\usepackage[normalem]{ulem}
\usepackage{algorithm}
\usepackage{algpseudocode}
\usepackage[utf8]{inputenc}
\usepackage{amsfonts}
\usepackage{amsmath}
\usepackage{arydshln}
\usepackage{balance}
\usepackage{blindtext}
\usepackage{booktabs} 
\usepackage{cancel}
\usepackage{caption}[font={footnotesize}]
\usepackage{color,soul}
\usepackage{comment}
\usepackage{enumitem}
\usepackage{fontawesome}
\usepackage{graphicx}
\usepackage{hyperref,xcolor}% http://ctan.org/pkg/{hyperref,xcolor}
\usepackage{hyperref}
\hypersetup{hidelinks}
\usepackage{listings}
\usepackage{mathtools}
\usepackage{multirow}
\usepackage{pifont}
\usepackage{tabularx}
\usepackage{threeparttable}
\usepackage{tikz}
\usepackage{xcolor,colortbl}
\usepackage{xspace}

\usepackage{makecell}
\usepackage{svg}

\usepackage{etoolbox}
\usepackage{varwidth}

\definecolor{winered}{rgb}{0.7,0,0}
\definecolor{gray}{gray}{0.7}
\definecolor{darkpastelgreen}{rgb}{0.01, 0.75, 0.24}
\definecolor{cadmiumgreen}{rgb}{0.0, 0.42, 0.24}
\definecolor{brickred}{rgb}{0.8, 0.25, 0.33}
\definecolor{cornellred}{rgb}{0.7, 0.11, 0.11}
\definecolor{burgundy}{rgb}{0.5, 0.0, 0.13}
\definecolor{frenchblue}{rgb}{0.0, 0.45, 0.73}
\definecolor{light-gray}{gray}{0.92}
\definecolor{lightlight-gray}{gray}{0.97}
\definecolor{codegray}{gray}{0.90}
\definecolor{inputgray}{gray}{0.90}
\definecolor{darkgreen}{RGB}{40,125,40}

\hypersetup
{
colorlinks=true,
linkcolor=winered,
urlcolor={winered},
filecolor={winered},
citecolor={winered},
allcolors={winered}
}

\newcommand{\cmt}[1]{}

\newcommand{\homedir}{\raise.17ex\hbox{$\scriptstyle\sim$}}

\global\mdfdefinestyle{rtboxstyle}{%
linecolor=black,%
leftmargin=0cm,rightmargin=0cm,linewidth=0.5pt,
roundcorner=3,
skipbelow=0pt,backgroundcolor=lightlight-gray
}

\newcommand{\code}[1]{\texttt{#1}}

\mdfdefinestyle{MyFrame}{%
     roundcorner=1pt,
     innertopmargin=6pt,
     innerbottommargin=6pt,
     innerrightmargin=6pt,
     innerleftmargin=6pt,
    }

\graphicspath{{figure/}{figures/}}
\DeclareGraphicsExtensions{.pdf,.eps,.png,.jpg,.jpeg}

\long\def\comment#1{}

\renewcommand{\paragraph}[1]{\smallskip\noindent\emph{#1}\quad}
\def\Snospace~{\S{}}

\newcommand{\heading}[1]{{\vspace{2pt}\noindent\bf{#1}}} % inside section
\newcommand{\eg}{{\it e.g.,~}}
\newcommand{\ie}{{\it i.e.,~}}

\newcommand{\nm}[1]{{\it (#1)\xspace}} % at beginning of section -- no extra space

\newcommand{\optional}[1]{}

\definecolor{paneTitle}{HTML}{0B3C5D}
\definecolor{paneBack}{HTML}{F7F9FC}
\definecolor{logBack}{HTML}{F3F7FF}
\definecolor{edgeBack}{HTML}{FFF7F0}
\definecolor{okGreen}{HTML}{2E7D32}
\definecolor{warnAmber}{HTML}{EF6C00}
\definecolor{badRed}{HTML}{C62828}

\tcbset{
  pane/.style={
    enhanced,
    colback=paneBack,
    colframe=paneTitle,
    coltitle=white,
    fonttitle=\bfseries\small,
    boxrule=0.6pt,
    arc=2pt,
    left=4pt,right=4pt,top=4pt,bottom=4pt,
    title filled,
  },
  logpane/.style={
    pane,
    colback=logBack,
    fonttitle=\bfseries\small\ttfamily,
  },
  edgepane/.style={
    pane,
    colback=edgeBack,
    fonttitle=\bfseries\small\ttfamily,
  }
}

\newcommand{\yesflag}{\textcolor{okGreen}{\normalsize\ttfamily YES}}
\newcommand{\notflag}{\textcolor{okGreen}{\normalsize\ttfamily SAFE}}
\newcommand{\flagged}{\textcolor{badRed}{\normalsize\ttfamily FLAGGED (MISSING EVIDENCE)}}
\newcommand{\noflag}{\textcolor{badRed}{\normalsize\ttfamily NO}}
\newcommand{\flaged}{\textcolor{badRed}{\normalsize\ttfamily FLAGGED}}

\def\pname{{\textsc{ProvSEEK}}\xspace}
\usepackage{fp}

\newif\ifbiblatex
\biblatextrue

\newif\ifcomments
\commentsfalse

\ifcomments
\usepackage{todonotes}
	\newcommand{\murat}[1]{\textcolor{blue}{[MK: #1]}}
	\newcommand{\wc}[1]{}
	\newcommand{\kunal}[1]{\textcolor{orange}{[{\bf KUNAL:} #1]}}
	\newcommand{\fix}[1]{\textcolor{red}{\hl{#1}}}
	\newcommand{\corr}[2]{\sout{#1} \hl{#2}}
\else
	\newcommand{\murat}[1]{}
	\newcommand{\wc}[1]{}
	\newcommand{\kunal}[1]{}
	\newcommand{\fix}[1]{}
	\newcommand{\corr}[2]{}
\fi

\def\darpa{DARPA\xspace}%
\def\kairos{Kairos\xspace}%
\def\magic{MAGIC\xspace}%
\def\flash{Flash\xspace}%
\def\orthrus{Orthrus\xspace}%

\RequirePackage[normalem]{ulem} %DIF PREAMBLE
\RequirePackage{color}\definecolor{RED}{rgb}{1,0,0}\definecolor{BLUE}{rgb}{0,0,1} %DIF PREAMBLE
\providecommand{\DIFdel}[1]{{\protect\color{red}\sout{#1}}}                      %DIF PREAMBLE
\providecommand{\DIFdel}[1]{}                      %DIF PREAMBLE
\usepackage[acronym]{glossaries}

\newacronym{hdl}{HDL}{High-level Dynamic Language}

\newacronym{ml}{ML}{Machine Learning}
\newcommand{\ml}{\gls*{ml}\xspace}

\newacronym{ai}{AI}{Artificial Intelligence}

\newacronym{nn}{NN}{Neural Network}

\newacronym{gnn}{GNN}{Graph Neural Network}

\newacronym{dnn}{DNN}{Deep Neural Network}

\newacronym{rnn}{RNN}{Recurrent Neural Network}

\newacronym{llm}{LLM}{Large Language Model}
\newcommand{\llm}{\gls*{llm}\xspace}
\newcommand{\llms}{\glspl*{llm}\xspace}

\newacronym{rag}{RAG}{Retrieval-Augmented Generation}
\newcommand{\rag}{\gls*{rag}\xspace}

\newacronym{mlm}{MLM}{Masked Language Model}

\newacronym{gan}{GAN}{Generative Adversarial Network}

\newacronym{nlp}{NLP}{Natural Language Processing}

\newacronym{pl}{PL}{Programming Language}

\newacronym{vm}{VM}{Virtual Machine}

\newacronym{bert}{BERT}{Bidirectional Encoder Representations from Transformers}

\newacronym{csn}{CSN}{Code Search Net}

\newacronym{sota}{SOTA}{State-Of-The-Art}
\newcommand{\sota}{\gls*{sota}\xspace}

\newacronym{pypi}{PyPI}{Python Package Index}

\newacronym{cdm}{CDM}{Common Data Model}

\newacronym{tc}{TC}{Transparent Computing}
\newcommand{\tc}{\gls*{tc}\xspace}

\newacronym{dt}{DT}{Decision Tree}

\newacronym{apt}{APT}{Advanced Persistent Threat}
\newcommand{\apt}{\gls*{apt}\xspace}
\newcommand{\apts}{\glspl*{apt}\xspace}

\newacronym{cve}{CVE}{Common Vulnerabilities and Exposures}

\newacronym{ids}{IDS}{Intrusion Detection System}
\newcommand{\ids}{\gls*{ids}\xspace}

\newacronym{pids}{PIDS}{\emph{Provenance-based Intrusion Detection System}}
\newcommand{\pids}{\gls*{pids}\xspace}

\newacronym{av}{AV}{Anti-Virus}

\newacronym{edr}{EDR}{Endpoint Detection and Response}

\newacronym{poi}{POI}{Point-Of-Interest}

\newacronym{vlan}{VLAN}{Virtual Local Area Network}

\newacronym{ctwo}{C2}{Command and Control}

\newacronym{csg}{CSG}{Computer Security Group}

\newacronym{wicys}{WiCyS}{Women In Cyber Security}

\newacronym{nsa}{NSA}{National Security Agency}

\newacronym{dod}{DoD}{Department of Defense}

\newacronym{lsm}{LSM}{Log Structured Merge}

\newacronym{me}{ME}{Microelectronics}

\newacronym{ci}{CI}{CyberInfrastructure}

\newacronym{soc}{SoC}{System-on-Chip}

\newacronym{tle}{TLE}{Two-line element set}

\newacronym{wal}{WAL}{Write Ahead Log}

\newacronym{obc}{OBC}{On-Board Computer}

\newacronym{dos}{DoS}{Denial-of-Services}

\newacronym{ttp}{TTP}{Tactics, techniques, and Procedures}

\newcommand{\ttps}{\glspl*{ttp}\xspace}

\newacronym{cot}{CoT}{chain-of-
thought}
\newcommand{\cott}{\gls*{cot}\xspace}

\newacronym{cti}{CTI}{Cyber Threat Intelligence}
\newcommand{\cti}{\gls*{cti}\xspace}

\newacronym{ioc}{IOC}{Indicators Of Compromise}
\newcommand{\ioc}{\gls*{ioc}\xspace}
\newcommand{\iocs}{\glspl*{ioc}\xspace}

\newacronym{jspoc}{JSpOC}{The Joint Space Operations Center}

\newacronym{cfg}{CFG}{Control Flow Graph}

\newacronym{cdg}{CDG}{Control Dependency Graph}

\newacronym{dgl}{DGL}{Deep Graph Library}

\usepackage{listings}
\usepackage[utf8]{inputenc}
\usepackage[english]{babel}
\usepackage{multicol}
\usepackage[
  style=ieee,
  citestyle=numeric-comp,
  mincrossrefs=99
]{biblatex}

\graphicspath{{figs/}}

\begin{document}

% \title{\pname{}: LLM-Powered Threat Intelligence Extraction and Detection Framework for System Provenance}
\title{LLM-driven Provenance Forensics for Threat Intelligence and Detection}

\author{
{\rm Kunal Mukherjee}\\
Virginia Tech
\and
{\rm Murat Kantarcioglu}\\
Virginia Tech
} % end author

\maketitle

%! root=./main.tex
\begin{abstract}

System provenance provides a rich \textit{forensic trail} for analyzing stealthy cyberattacks (\eg \apt campaigns). However, traditional detection pipelines rely heavily on recognized patterns and lack end-to-end automated mechanisms to integrate external intelligence or reason iteratively about complex attack traces. Therefore, analysts are required to craft ad-hoc queries, correlate disparate evidence, and iteratively reconstruct attack narratives. These approaches suffer from scalability bottlenecks, limited integration of external threat intelligence, and a lack of automated reasoning support for complex, multi-stage attack campaigns.

We introduce \pname, an LLM-powered agentic framework for automated provenance-driven forensic analysis and threat intelligence extraction. \pname employs specialized toolchains to dynamically retrieve relevant context by generating precise, context-aware queries that fuse knowledge from threat reports with evidence from system provenance data. The framework resolves provenance queries, orchestrates multiple role-specific agents, and synthesizes structured, ground-truth verifiable forensic summaries. 
%MK-Oak: Do we want to add the ML model we use to the next sentence? -- added
By combining agent orchestration with \rag and \cott reasoning, data-guided filtration using a behavioral model, \pname enables adaptive multi-step analysis that iteratively refines hypotheses, verifies supporting evidence, and produces scalable, interpretable forensic explanations of attack behaviors. 
% NEW novelty sentence:
\pname is designed for automated threat investigation without task-specific training data, enabling forensic-style investigation even when no prior knowledge of the environment.
% The framework's novelty is that it unifies provenance-aware querying, CTI-grounded retrieval, and multi-agent LLM reasoning into a single closed-loop pipeline that both scales to enterprise-scale traces and yields verifiable, analyst-ready explanations.

% By combining provenance data with agentic reasoning, \pname establishes a new paradigm for \textit{grounded agentic forecics} to investigate \apts. 

We conduct a comprehensive evaluation on publicly available \darpa datasets, demonstrating that \pname outperforms retrieval-based methods for the intelligence extraction task, achieving a 34\% improvement in contextual precision/recall; and for threat detection task, \pname achieves 22\%/29\% higher precision/recall compared to both a strawman agent and \sota \pids. In our scalability study, we show \pname increases its token usage  by 1.42x and time required by 1.63x when the database size increases 50x making it optimal for large-scale deployment. 
%MK-Oak: I wonder whether we can add a sentence in the abstract above why our combination is novel. -- added above
\end{abstract}
%! root = ./main.tex
\section{Introduction}\label{sec:intro}

Advanced Persistent Threats (\apts) remain one of the most significant challenges facing enterprises and government agencies today. Characterized by their stealthy, multi-step campaigns, \apts achieve persistence by blending into normal system activity and maintaining long-term access to critical infrastructures~\cite{apt1,apt2}. They not only disrupt operations but also inflict severe socio-economic consequences.

Investigating these sophisticated campaigns requires fine-grained visibility into a system's complete execution history. System provenance has emerged as a powerful foundation for \sota forensic investigation and intrusion detection~\cite{inam2023sok, bilot12399sometimes}. Provenance data encodes rich causal semantics of processes, files, and network interactions, enabling analysts to reconstruct attack chains~\cite{killchain}, reason about dependencies, and uncover hidden adversarial behaviors. 
%System provenance has become indispensable for forensic reconstruction and incident response in real-world settings.

Despite this promise, current provenance-based analysis workflows~\cite{sigl2021sec, wang2020ndss, mukherjee2023sec, rehman2024flash, jia2024magic, jian2025, wang2024incorporating} face fundamental challenges as analysts are required to craft ad-hoc queries, manually correlate evidence, and iteratively reason about attack progression across billions of events per day~\cite{jian2025}, which slows investigations. Moreover, forensic tasks demand context-rich, explainable narratives, yet existing pipelines provide little support for automated reasoning or integration of external threat intelligence, which was highlighted by a recent study~\cite{bilot12399sometimes} as well.
%has also highlighted scalability and adaptability as bottlenecks for effective provenance-based forensics.

A key challenge in provenance-driven security is scalability. Existing \pids solutions are not capable of handling large datasets, where billions of logs accumulate daily. Several recent attempts have sought to alleviate scalability shortcomings, such as using federated learning (\eg \cite{mukherjee2023proviot}) or optimizing inference through vector recycling (\eg \cite{rehman2024flash}).
While these methods improve efficiency, they remain limited for forensics: \emph{(i)} distributed heuristics are prone to false positives, and \emph{(ii)} centralized ML pipelines still fail to integrate external threat intelligence or provide fine-grained, analyst-friendly forensic explanations.

This gap motivates a new research question: \emph{Can we design a provenance forencis framework that combines scalability, adaptability, and interpretability by orchestrating multiple reasoning steps and role-specific planning through an intelligent agentic architecture?}

To address this question, we propose \pname, an \llm-powered agentic framework for provenance forensic analysis. \pname employs three \llm-powered agents (\eg Investigation Agent, Follow-Up Agent, and Safety Agent) as evidence collection and correlation, and reasoning hub that dynamically plans, coordinates, and evaluates investigations. This architecture allows \pname to ingest unstructured \cti reports, extract \iocs, use them to formulate database queries, and return contextualized attack summaries with evidence.

\pname's agents uses chain-of-thought (\cott) reasoning and routes tasks among a set of specialized tools (\eg retrieval-augmented generation (\rag) over \cti reports, vector-similarity search for semantic retrieval, provenance database queries, query filtration using behavioral model, and correlating evidence from database queries with retrieved CTI artifacts). \cott reasoning helps the agents to iteratively refines hypotheses, validates evidence against provenance data, and synthesizes results into human-readable forensic narratives. \pname addresses the scalability issue by rethinking scalability as an \emph{agentic orchestration problem}. Instead of considering every event, \pname leverages an LLM to translate analyst intent or IoCs into \emph{precise database queries}. By retrieving only the relevant events and further filtering them using a behavioral model to identify only the relevant events, \pname avoids the scalability bottleneck.
%and enables adaptive workflow.
 
Attack campaigns evolve continuously, reusing known \ttps while adopting new methods to evade detection. \pname is designed for automated threat investigation without task-specific training data, enabling forensic-style investigation even when no prior knowledge of the environment. \pname adapts by employing a dynamic query-planning loop: the Investigation Agent understands the security analyst's intent (investigation, exploration, or identification), the agent issues an initial SQL or vector search, inspects the intermediate results, and the Follow-up Agent iteratively refines the investigation plan. For example, if intermediate evidence suggests lateral movement, \pname can automatically expand the query to include different system artifacts (\eg process, files, IPs) used for lateral movement. \pname's iterative refinement allows it to adapts its investigation path in real-time, mirroring how a human analyst adapts their line of questioning based on partial evidence.

In security domain, a system's output is useful if the analysts can verify it with actual system evidence. Therefore, \pname is built around \emph{verification-first design}. Each agentic step produces not only an answer but also supporting evidence from the system database. \pname uses the Safety Agent to identify if there is missing or unclear evidence, and it uses the Follow-up Agent to investigate and gather the missing evidence. By design, the framework ensures that every LLM-generated claim is tied to verifiable ground truth in the underlying database, thereby minimizing the risk of hallucination and \cott reasoning errors.

%MK-Oak: sql query planning mentioned before. Please double to check to make sure that we do not have too much overlap
While \rag usage has shown promise in retrieving relevant context, but simply applying \rag to the provenance pipeline introduces new risks. Provenance queries can easily return millions of candidate events, and dumping this unfiltered evidence into a prompt quickly exceeds token limits, forcing truncation and loss of critical context. 
% This will lead to the omission of subtle but crucial dependencies. 
\pname addresses this by using a provenance domain-aware SQL query that returns targeted results, and further filters these results using a behavioral model. This yields a compact, high-signal subset of events that \pname can precisely correlate with the extracted \cti artifacts. 
%This design prevents uncontrolled context expansion, ensuring that outputs remain grounded in verifiable system events.
% the LLM plans bounded SQL queries, retrieves only the relevant evidence slices, and incrementally fuses them into verified data before summarization. 

% The combination of scalability, adaptability, and interpretability culminates in a new paradigm: \emph{grounded agentic forencis}. Unlike traditional systems that either optimize for efficiency at the cost of detail or focus on semantics at the cost of scalability, \pname unifies the two by using an agentic query planner that continuously updates its reasoning path based on retrieved evidence, while its verification module ensures that outputs remain faithful to the ground-truth provenance data. 

%MK-Oak: Please double the check claims mentioned in the next paragraph, and make sure it is consistent with the final version.  
We conduct a comprehensive evaluation using six publicly available \darpa \tc and OpTC datasets. 
For \emph{intelligence extraction}, \pname outperforms retrieval-based methods (vanilla and type-filtered \rag) by 34\%/34\%/30\% in contextual precision, contextual recall, and relevance, demonstrating its ability to produce more accurate and context-aware forensic insights. For \emph{threat detection}, \pname achieves 22\%/29\% higher precision and recall compared to baselines, including a naive strawman agent and \sota \pids~\cite{cheng2024kairos,jia2024magic,rehman2024flash,jian2025}, highlighting its effectiveness in identifying stealthy attacks. For comprehensive evaluation, we also conduct an ablation study to show how the different components of \pname affect detection, an overhead study, and an error analysis.

This paper makes the following contributions:
\begin{itemize}[noitemsep,topsep=0pt]
    \item \emph{Agentic Provenance Forensic Framework.} We introduce \pname, the first agentic framework for forensic provenance analysis that uses LLM agents to automatically plan investigations by unifing provenance-aware querying, CTI-grounded retrieval, and multi-agent LLM reasoning into a single, verifiable, closed-loop pipeline providing analyst-focused explanations.
    \item \emph{Comprehensive Evaluation.} Using seven publicly available provenance datasets, we demonstrate that in intelligence extraction task \pname outperforms retrieval-based methods (vanilla and type-filtered \rag) by 34\%/34\%/30\% in contextual precision, contextual recall, and relevance. For threat detection, \pname achieves 22\%/29\% higher precision and recall compared to the naive strawman agent and \sota \pids~\cite{rehman2024flash,jian2025}. Finally, we conduct an ablation and error analysis study to show how different components of \pname affect the detection performances. 
    \item \emph{Scalability and Error Analysis.}  We conduct a thorough investigation to show the scalability of \pname as the token usage increases by 1.42x and time required by 1.63x even when the database size increases 50x. In our error analysis study, we show the majority of errors originate from \llm's limitation of extracting relevant artifacts from threat reports.

\end{itemize}
%! root=../main.tex
\section{Background}

% We provide an overview of system provenance and provenance-based \ml research.
%, and public provenance datasets that form the foundation of our work. We highlight how \llm and \rag introduce new opportunities and challenges in provenance-driven security analytics. Importantly, \pname demonstrates that LLMs can finally be applied in security without sacrificing the requirements of verifiability, interpretability, and analyst trust.   

\heading{System Provenance.}
System provenance records fine-grained audit data (\eg \code{syscall}, Windows ETW~\cite{winETW} and Linux audit logs~\cite{audit}) from real-world enterprise environments. It records information flow and control dependencies across system resources (\eg process, file, socket)~\cite{king:2003sosp, inam2023sok}. The resulting provenance log offers rich semantics making it invaluable for forensic analysis. Formally, a provenance log $\mathcal{L} = { \ell_1, \ell_2, \dots }$ is a sequence of timestamped events (\ie interactions between system resources).
Each resources is associated with attributes such as file paths, IP addresses, or executable names.
Each event $\ell = (u, v, r, t)$ captures a relation $r$ occurring at time $t$ between resources $u$ and $v$ (e.g.,``process $u$ writes file $v$''). It preserves causal structure while maintaining the chronological order of interactions, enabling analysts to trace attack progression, discover entry points, and assess the scope of compromise. Provenance logs efficiently capture attackers' intents and \ttps.
%, since adversaries find difficult to obfuscate. 

\heading{Provenance-based \ml Research.}
Provenance logs were leveraged for manual forensic reconstruction and has become the base for \ml-based \pids~\cite{sigl2021sec, jia2024magic, cheng2024kairos, rehman2024flash, wang2020ndss, zengy2022shadewatcher, watson2021ndss}, adversarial analysis~\cite{goyalsometimes, mukherjee2023sec} and interpretability~\cite{mukherjee2023interpreting}.
% Early approaches relied on frequency counts, handcrafted rules, or sequence features, yielding detectors with high false positive rates.
%MK-Oak: I added the next sentence, please check whether it is correct. 
As shown by recent work~\cite{bilot12399sometimes}, there is large performance degradation of \pids against stealthy \apt-style attacks, and 
\ml-based \pids are limited by scalability and interpretability: billions of events per day cannot be efficiently embedded, and black-box predictions without explanations are unsuitable for real-world deployment. 
%\sota methods incorporate graph representation learning and anomaly scoring, modeling the full structural context of provenance graphs. 

We surveyed representative provenance-based research published in top-tier security venues over the past seven years. These works span detection~\cite{wang2020ndss, sigl2021sec, yang2023prographer, zengy2022shadewatcher, cheng2024kairos, rehman2024flash, goyal2024rcaid, bilot12399sometimes, jian2025, wang2024incorporating}, evasion and robustness~\cite{goyalsometimes, mukherjee2023sec}, differential privacy~\cite{mukherjee2025provdp}, synthetic graph generation~\cite{wang2025provcreator} and explanation~\cite{mukherjee2023interpreting}. All rely heavily on the \darpa \tc and OpTC datasets for reproducibility. Therefore, in this work, to enable easy reproducibility, we chose the seven challenging publicly available \darpa datasets where previous \sota \pids struggled (\eg E3 and E5 \code{CADETS}, \code{THEIA}, \code{CLEARSCOPE}, and OpTC datasets).

\heading{LLMs and Provenance Analytics.}
%MK: please make sure we do not miss any new work especially in this area. I think this is very they can attack us.
%MK-Oak: My previous comment is still valid. Ie. let us make sure we do not miss anything. Also, once we submit to the Oakland, let us put an updated version of our work to arXiv. 
Large Language Models (LLMs) have recently transformed domains such as software engineering and social networks~\cite{gpt4, rag-survey}, largely through techniques such as prompt engineering~\cite{liu2022pretrainpromptpredict}, \cott reasoning~\cite{wei2022chainofthought}, tool usage and \rag~\cite{lewis2020rag}. In threat intelligence, LLMs have been utilized to extract knowledge graphs from CTI reports~\cite{cheng2024ctinexus} and anomaly detection. 
%To our knowledge, we are the first to develop this novel provenance domain aware agent orchestration for forensics. 
%MK-Oak: Please check my rewrite of the previous sentence.
To the best of our knowledge, we are the first to develop a provenance-domain-aware agentic framework specifically designed for digital forensics.
%MK-Oak: I think the next sentence is not super clear. It sounds like generic Agentic AI. Please clarify it further.
\pname differentiates itself by employing an \emph{agentic framework} rather than a sole \llm: the LLM is not the sole decision-maker but a reasoning hub that issues bounded queries, retrieves verifiable provenance evidence, and synthesizes narratives.

% A prompt provides task-specific context that guides the model toward useful outputs. While \rag enriches prompts with external knowledge, naïvely applying it to provenance analytics is problematic: provenance queries can yield millions of results, overwhelming token budgets and introducing risks of hallucination if unverified or inconclusive.
%In security~\cite{deng2024pentestgpt, huang2023penheal, qi2023loggpt, li2024iris,ullah2024llms}, where each claim must be auditable, this limitation is particularly acute.  
% This motivates agentic frameworks such as \pname, where the LLM acts not as a monolithic oracle but as a coordinator of specialized tools.
% Therefore, rather than embedding entire data, \pname generates bounded database queries and aggressively filters them to extract the relevant partion of the result and send it to the Agent for verification of the results. \pname's agentic orchestration ensures that outputs remain grounded in verifiable events while still benefiting from the LLM’s reasoning capabilities.
% deploys different agents with different tools, and iteratively refines its reasoning chain.
%! root = ./main.tex

\section{Problem Statement and Threat Model}\label{sec:threat-model}

\heading{Problem Statement.} 
Provenance-based forensic analysis plays a crucial role in investigating stealthy attacks (\eg \apt campaigns). Currently, forensic analysis is an extremely labor-intensive process that requires specialized domain knowledge. As a result, defenders face a critical gap: timely and thorough investigations require scalable mechanisms to retrieve and correlate evidence, while effective response demands interpretability and verifiable explanations.

To illustrate the challenge, consider an enterprise under attack by an APT actor. The provenance logs would contain colossal amounts of irrelevant noise from background benign processes, and stealthy attackers would craft their attacks so that each step, when viewed in isolation, would resemble benign system activity. Conventional forensic workflows require analysts to craft queries manually and iteratively shift through numerous events, which is both time-consuming and error-prone. This results in missing critical connections across attack stages or discovering them after extensive effort.

\pname's design goal is to provide: \nm{1} scalable query execution to retrieve relevant context, \nm{2} adaptive investigation planning that iteratively refines reasoning based on intermediate results, and \nm{3} fine-grained contextual evidence of adversarial activity. \pname employs agentic orchestration, where it automatically orchestrates provenance queries, evidence retrieval, and evidence validation to synthesize coherent forensic narratives. This approach strikes a balance between efficiency and analytical depth, enabling rapid and trustworthy forensic investigations.

\begin{figure}[t]
\centering
\small
\resizebox{\columnwidth}{!}{%
\begin{minipage}{\columnwidth}

\begin{tcolorbox}[
  colback=black!2, colframe=black!50, arc=2mm, boxrule=0.3pt,
  left=2.5mm, right=2.5mm, top=1.5mm, bottom=1.5mm
]
\footnotesize
\begin{itemize}[leftmargin=*,nosep]
  \item Based on Drakon dropper attack artifacts, was I attacked?
  % \item Based on privilege escalation attack artifacts, was I attacked?
  \item Can you give information about event $<$EDGE\_ID$>$?
  \item What is the source label of edge $<$EDGE\_ID$>$?
  \item What is the label of node $<$NODE\_ID$>$?
  \item What edges involve node $<$NODE\_ID$>$?
  \item What processes interact with file \code{/var/log/mail}?
\end{itemize}
\label{tbox:queries}
\end{tcolorbox}

\end{minipage}}
\caption{Example Security Analyst Queries}
\label{fig:sec-analyst}
\end{figure}

Let $\mathcal{L} = {\ell_1, \ell_2, \dots}$ denote a provenance log consisting of timestamped events, where each event $\ell = (u, v, r, t)$ describes a relation $r$ between system entities $u$ and $v$ at time $t$. 
%MK-Oak: queries are in plaintext language. Maybe clarify this? -- done
Let $\mathcal{Q}$ represent a set of analysts' natural language queries (as seen in \autoref{fig:sec-analyst}) containing generic or attack specific \ioc (\eg user got from \cti reports). \pname employs an \llm-based agent planner $\mathcal{M}$ that decomposes each query $q \in \mathcal{Q}$ into a sequence of tool calls $\pi = \langle a_1, a_2, \dots, a_T \rangle$, where each $a_t$ corresponds to an operation (\eg threat-intelligence extraction, investigation, data retrieval, artifact correlation, safety validation, or summary generation). Each intermediate result is verified through a grounding function $\mathcal{V}erify: (a_t, \mathcal{L}) \mapsto \mathcal{L}_t$, which maps tool outputs back to specific events in the provenance log. The overall objective is to generate an summary $\mathcal{X} = f({\mathcal{L}_1, \dots, \mathcal{L}_T})$ consisting of (i) an annotated subset of the log $\mathcal{L}^\prime \subseteq \mathcal{L}$ capturing the relevant evidence, and (ii) a natural language summary aligned with analyst intent. 
%MK: this discussion exactly says we are doing automated forensic analysis.

\heading{Threat Model.}  
Our threat model assumes the integrity of on-device provenance collection and secure audit pipelines, consistent with prior provenance-based security research~\cite{mukherjee2023sec, cheng2024kairos, rehman2024flash, jian2025}. Specifically, we trust the Trusted Computing Base (TCB), which includes the operating system, auditing framework, and provenance storage mechanisms.
% Attacks such as hardware trojans, side-channel leaks, or manipulations outside the scope of audit logging are not considered.
The integrity of the output provenance data is assumed to be preserved by existing secure logging and tamper-evident provenance systems.  

Our work focuses on automating scalable and interpretable provenance forensic analytics through an agentic framework. We do not consider adversarial attacks against the agents itself (\eg prompt injection or adversarial example generation) as part of our threat model. These remain important open research challenges orthogonal to this work. 
% Similarly, dataset poisoning and adversarial manipulation of provenance detectors~\cite{mukherjee2023sec} are out of scope. 
Instead, our primary focus is ensuring that \pname’s outputs remain tethered to verifiable ground-truth provenance data, enabling analysts to validate results even in the presence of stealthy APT activity.

%! root=./main.tex

\begin{figure*}[!htp]
\centering
    \includegraphics[width=1.\linewidth]{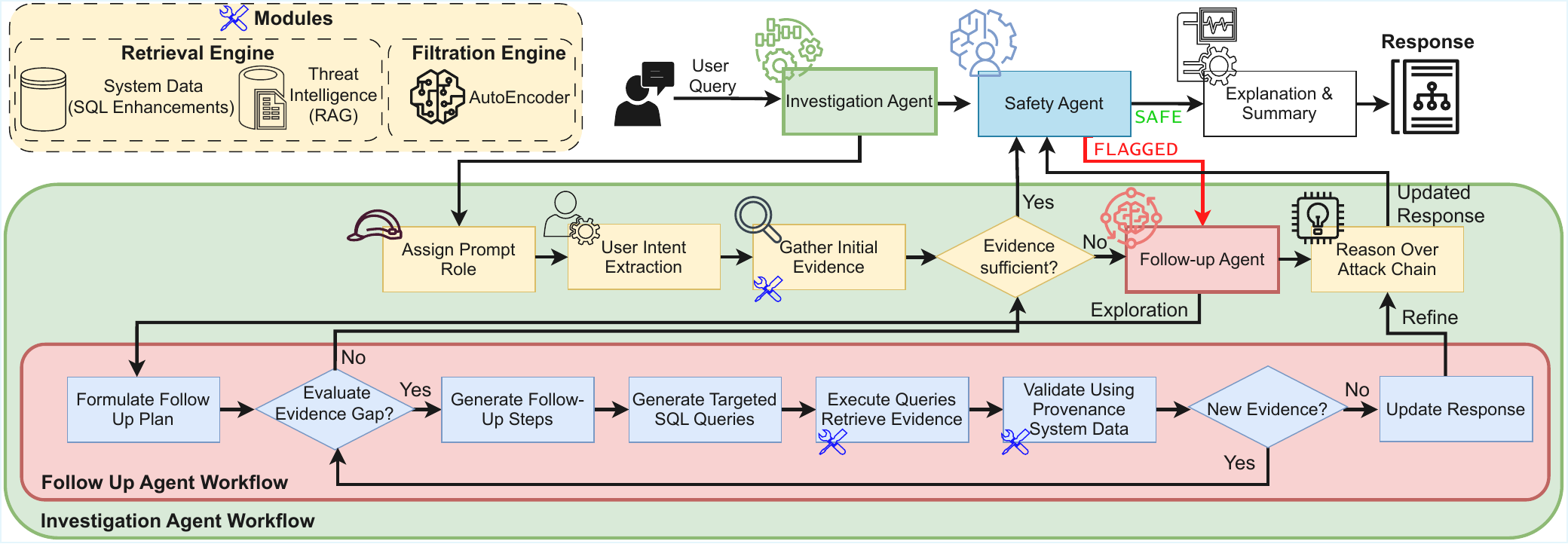}
    \caption{\pname Agentic Workflow.}
    \label{fig:agentwf}
\end{figure*}

\begin{figure}[!htb]
\centering
    \includegraphics[width=1.\linewidth]{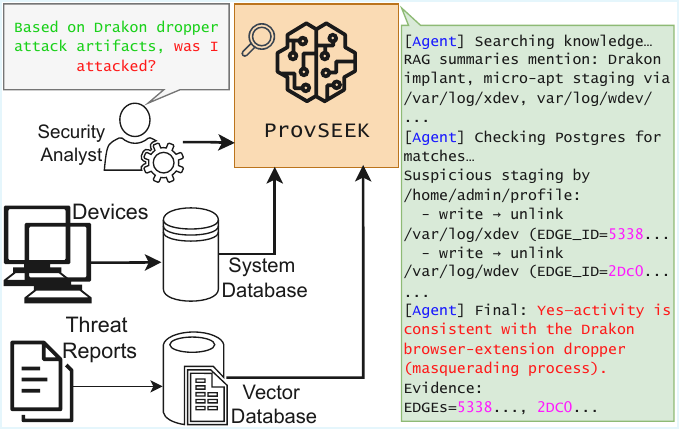}
    \caption{\pname Overview.}
    \label{fig:provseek}
\end{figure}

\section{\pname: Architecture}\label{sec:sysOverview}
%MK-Oak: Can we add forensic analysis to the first sentence ? -- done
\pname is a \llm-powered threat intelligence extraction and forensic analysis framework as seen in \autoref{fig:provseek}, where a user can use natural language to conduct forensic analysis. \pname expects that it has access to a vector database where \cti reports are parsed and stored, and system provenance databases where system provenance logs are stored. \pname will extract the intent of the user and conduct investigations until it has gotten concrete evidence from the system database to justify its answer.

% \pname has three main components: \nm{1} threat report parsing and vector DB creation, \nm{2} system provenance database exploration, and \nm{3} attack detection and explanation summary creation. The first component is related to parsing and storing the threat reports to extract actionable information (\eg artifacts (process, files, IP) and \ttp used) regarding the attacks. The second component queries the system provenance database to identify if the attack artifacts are present and how they are related to the attackers' \ttp. The third component uses the results of the first and second components to determine whether the system was attacked and generate a summary. 

%MK-Oak: the flow from the previous paragraph to this one seems missing. Please edit.  -- done
\pname utilizes three main agents: Investigation Agent, Follow-Up Agent, and Safety Agent; and three critical modules that are utilized by the agents: Threat Intelligence Retrieval Engine (using \rag), System Data Retrieval Engine (using SQL enhancements), and Filtration Engine (using AutoEncoder~\cite{autoencoder}), as seen in \autoref{fig:agentwf}.

The six main components of \pname are:
\begin{enumerate}[noitemsep, topsep=0pt]
    \item \textbf{Investigation Agent}: Interprets the user’s intent and gathers initial evidence to answer the query. For insufficient evidence, it delegates to the Follow-Up Agent.
    \item \textbf{Follow-Up Agent}: Collects additional evidence to close evidence gaps and reasons over the updated context. After new evidence is extracted and verified, control is handed over to the Safety Agent.
    \item \textbf{Safety Agent}: Ensures that the draft answer contains verifiable evidence. If the answer relies on inconclusive, missing, or unverifiable evidence, it invokes the Follow-Up Agent to gather the missing information. Once satisfied, it summarizes and returns it to the user.
    \item \textbf{Threat Intelligence Retrieval Engine}: Uses \rag to retrieve relevant context from \cti reports and build contextual background for the user’s query.
    \item \textbf{System Data Retrieval Engine}: Executes provenance-aware SQL queries to fetch system-level evidence, ensuring that \pname does not hallucinate artifacts that cannot be corroborated by system data.
    \item \textbf{Filtration Engine}: Applies a AutoEncoder-based behavioral model to filter out common benign SQL results so that only rare relevant context is passed to the agents when reasoning about the threat scenario.
\end{enumerate}

We will use a motivating security analyst query, \textit{``Based on Drakon dropper attack artifact, was I attacked?''} to understand \pname's components.
% (shown in \autoref{fig:mo-ex-user})
\begin{figure}[!ht]
    \centering
    \includegraphics[width=1.\columnwidth]{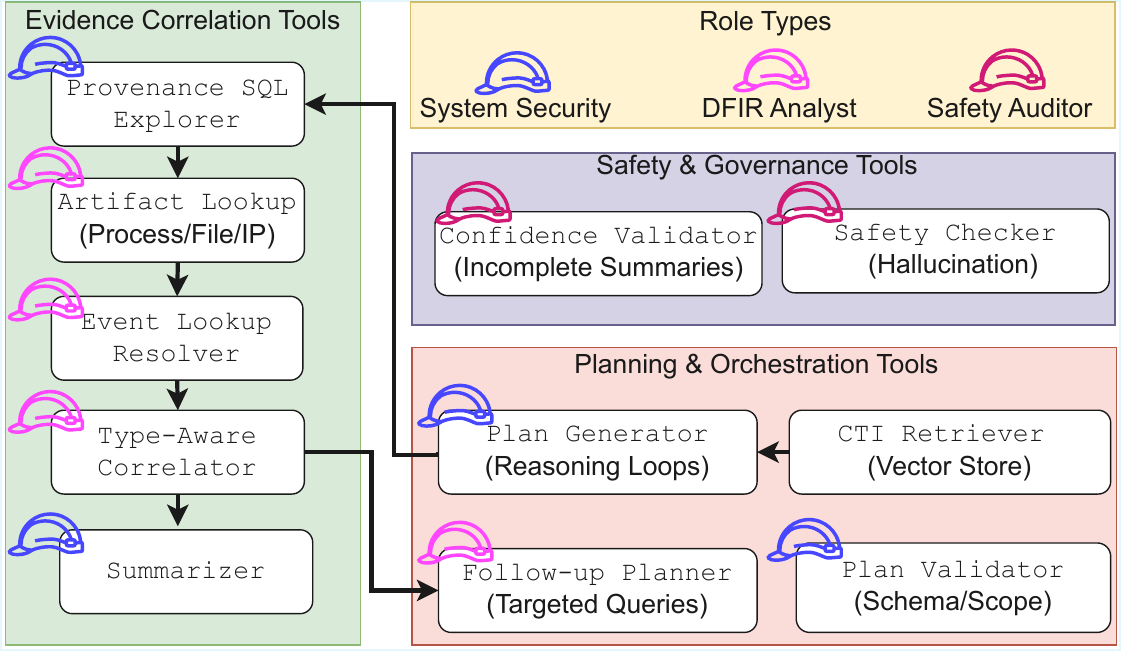}
    \caption{\pname tools.}
    \label{fig:tools}
\end{figure}

\subsection{Investigation Agent}

The Investigation Agent first extracts the intent of the analyst's query and, based on the intent it gathers initial evidence using different tools~\autoref{fig:tools}. These tools can help in \textit{Evidence Correlation} or \textit{Planning \& Orchestration}. \textit{Evidence Correlation} tool include: \code{Provenance SQL Explorer}, \code{Artifact Lookup}, \code{Event Lookup Resolver}), and \textit{Planning \& Orchestration} tool include: \code{CTI Retriver}, \code{Plan Generator}). By orchestrating these tools in bounded reasoning loops, the agent ensures investigations remain efficient, precise, and auditable. 
%\pname allows analysts to validate not just the evidence but also the reasoning process, thereby instilling greater trust in automation.

%MK-Oak: I missed where the figure 3 is discussed. Also in text, figure 4 and figure 5 are not mentioned. Only in the caption of figure 3. Please refer to these figures in the discussion as well.
\pname takes on different roles for tools, as shown in ~\autoref{fig:tools} and described in ~\autoref{sec:tools}. The Investigation Agent employs differentiated prompt roles that explicitly guide the behavior of underlying \llm. Rather than relying on a single monolithic role, the system frames prompts such as ``You are a Digital Forensics and Incident Response (DFIR) analyst'' when prioritizing artifact-level correlation, ``You are a system security expert'' when validating provenance patterns, and ``ou are a safety auditor for incident-response reports'' when reviewing outputs for compliance. This role specialization ensures that the tool is contextually aligned with its intended purpose, using concrete context and improving the interpretability of the generated analysis. The Investigation Agent aggregates the retrieved evidence and correlates them to see if it has enough evidence to answer the user's query, as seen in \autoref{fig:mo-inv-agent}. If the Investigation Agent could not answer the user's query because of lack of evidence, it invokes the Follow-up agent.

\subsection{Follow-Up Agent}
Follow-up agent generates structured evidence probing plans when initial evidence is insufficient or new leads emerge, as seen in \autoref{fig:mo-fol-agent} and uses the following prompt template \autoref{fig:followup-plan-prompt}. For example, some artifacts are recovered (\eg \code{/var/log/\{main/xdev\}}) but it is still missing evidence for artifact (\eg \code{/var/log/wdev}), it formulates follow-up steps to find those evidences. These plans are bounded in scope, ensuring that exploratory reasoning remains efficient while still progressively expand its investigation scope.  

Execution of follow-up plans is handled iteratively, with each step validated against provenance evidence before moving forward. This design ensures that new queries are not only semantically relevant but also grounded in prior results, enabling a logical flow of investigation. Analysts thus benefit from an adaptive system that can pivot to related evidence on demand, reconstruct longer attack chains, and refine hypotheses, all while maintaining accountability and minimizing redundant computation. Once, the Follow-up agent has either found enough evidence to answer the user's query or have exhausted all its \cott steps it hands over the answer to the Safety Agent.

% ============= FIGURE 2: Final Assessment & Analyst Actions ==================
\begin{figure*}[t]
\centering
\small
\resizebox{0.8\textwidth}{!}{%
\begin{minipage}{\textwidth}

\begin{tcbraster}[
  raster columns=1,
  raster valign=top,
  raster row skip=6pt
]

  % Final Assessment (prose)
  \begin{tcolorbox}[edgepane,title={\pname Response: Final Assessment \& Recommend Actions}]
  \textbf{Conclusion:} \textbf{Compromised (high confidence).} \\
  Observed Drakon-aligned TTP: \code{/home/admin/profile} writing then unlinking \code{/var/log/mail}, preceded by \code{firefox} writing to \code{/home/admin/profile}. Some auxiliary IoCs were not observed. \\

%    \textbf{Context:} Relevant Events are listed below: \\
%     - process(216927) \code{/home/admin/profile} $\rightarrow$ WRITE \code{/var/log/mail} (EDGE ID \code{C284C9F8...}) \\
%     - process(216981) \code{/home/admin/profile} $\rightarrow$ UNLINK \code{/var/log/mail} (EDGE ID \code{52B5094D...}) \\
%      - process(284573) \code{/home/admin/profile} $\rightarrow$ WRITE \code{/var/log/xdev} (EDGE ID \code{5338C80F...}) \\
%     - process(216394) \code{/home/admin/profile} $\rightarrow$ UNLINK \code{/var/log/xdev} (EDGE ID \code{63B9DC37...}) \\
%      - process(216368) \code{/home/admin/profile} $\rightarrow$ WRITE \code{/var/log/wdev} (EDGE ID \code{2DC0BB3B...}) \\
%     - process(216395) \code{/home/admin/profile} $\rightarrow$ UNLINK \code{/var/log/wdev} (EDGE ID \code{4850F9E4...}) \\
%     Artifact negatives:\code{/deploy/archive/*}, 104.228.117.212:80, 5.214.163.155.\\

\textbf{Assessment.} The host exhibits evidence of log tampering and artifact clean-up consistent with MITRE ATT\&CK: \texttt{T1070.004} (Indicator Removal on Host: file deletions of \code{/var/log/mail} and transient \code{/tmp/memtrace.so}), \texttt{T1070.002} (Clear Linux System Logs: writes to \code{/var/log/mail} immediately preceding deletion), \texttt{T1574.006} (Hijack Execution Flow: plausible \code{LD\_PRELOAD} via transient SO \code{memtrace.so} under \code{/tmp}), and \texttt{T1036} (Masquerading: benign-looking executables/paths such as \code{./gtcache} and \code{/home/admin/profile}). Overall, the sequence shows staged creation of plausible log/working files followed by removal, aligning with anti-forensic clean-up patterns. \\

\textbf{Grounded evidence (selected).}\\
process(216927) \code{/home/admin/profile} $\rightarrow$ \textsc{WRITE} \code{/var/log/mail} \;(\textsc{EDGE} \code{C284C9F8...})\\
process(216981) \code{/home/admin/profile} $\rightarrow$ \textsc{UNLINK} \code{/var/log/mail} \;(\textsc{EDGE} \code{52B5094D...})\\
process(284573) \code{/home/admin/profile} $\rightarrow$ \textsc{WRITE} \code{/var/log/xdev} \;(\textsc{EDGE} \code{5338C80F...})\\
process(216394) \code{/home/admin/profile} $\rightarrow$ \textsc{UNLINK} \code{/var/log/xdev} \;(\textsc{EDGE} \code{63B9DC37...})\\
process(216368) \code{/home/admin/profile} $\rightarrow$ \textsc{WRITE} \code{/var/log/wdev} \;(\textsc{EDGE} \code{2DC0BB3B...})\\
process(216395) \code{/home/admin/profile} $\rightarrow$ \textsc{UNLINK} \code{/var/log/wdev} \;(\textsc{EDGE} \code{4850F9E4...})\\[2pt]

\textbf{Negative indicators (evidence not found).}\\
Files under \code{/deploy/archive/*}, C2 endpoints \code{104.228.117.212:80} and \code{5.214.163.155}. \\

\textbf{Interpretation.} The WRITE\,$\rightarrow$\,UNLINK cadence on log-like paths (\code{/var/log/*}) by \code{/home/admin/profile} strongly supports anti-forensic behavior (\texttt{T1070.002}, \texttt{T1070.004}). The mention of transient \code{memtrace.so} in \code{/tmp} is compatible with an \code{LD\_PRELOAD}-style hook (\texttt{T1574.006}); the use of innocuous names (\code{./gtcache}, \code{/home/admin/profile}) aligns with \texttt{T1036}. Absence of known \code{/deploy/archive/*} payloads or the listed C2 IPs in this excerpt limits attribution strength to \emph{medium}. \\

  \textbf{Immediate actions:}
  Isolate host (EDR/VLAN) while preserving forensic access; acquire memory+disk images; preserve \code{/home/admin/profile} and \code{/var/log/mail}; monitor-first then block suspected egress (141.43.176.203:80, 149.52.198.23:80) via change control.
  \\

  \textbf{Further Notes:} Validate privilege context (writes under /var/log/*), confirm browser provenance, and sandbox process node IDs 216927/216981.

  \end{tcolorbox}

\end{tcbraster}

\end{minipage}}

\caption{\pname final assessment and recommended actions with concrete context.}
\label{fig:mo-ans}
\end{figure*}
\subsection{Safety Agent}
Autonomous reasoning introduces risks: such as the answer can have inconclusive evidence or an overly confident answer without having verifiable evidence. The Safety Agent enforces that all the answer contains verifiable evidence required and in case of inconclusive evidence it accurately relays that to the user, as seen in \autoref{fig:mo-saf-agent}. 

% operational constraints on all downstream agents.
% Built using \cite{smolagents2024}, it validates query syntax, filters anomalous behaviors, and prevents excessive token or resource usage. It also detects and neutralizes adversarial inputs designed to subvert the reasoning process.  

To further strengthen its operation, the Safety Agent is also assigned an explicit role persona that guides its reasoning as seen in the prompt template \autoref{fig:safety-agent-prompt}. Role definition \code{"You are a safety auditor for incident-response automation"} direct the agent to critically review investigative actions before execution and further validate the response to decrease the chance of hallucination. In this role, the agent validates query structures, ensures that requests fall within acceptable scopes, and flags potentially unsafe or ambiguous instructions for further scrutiny.

The safety agent can invoke the Follow-up agent when it identifies evidence that needs system justification or identifies an evidence that needs further collaboration as seen in \autoref{fig:mo-fol-safe-agent} where the Safety Agent flags answer and invokes the Follow-up agent since it is missing Event IDs that are required to verify the evidence. The Follow-up agent gathers the missing evidence and returns the new answer to the Safety Agent that marks the answer safe and returns the answer (as seen in \autoref{fig:mo-ans}) to the user since it already contains all the information to verify the evidence and the execution trace of \pname while investigating the query can be seen in \autoref{fig:mo-trace}. 

% By giving the Safety Agent a well-defined auditing persona, \pname constrains its autonomy to a watchdog function preventing unsafe tool usage, guarding against prompt injection attempts, and ensuring that investigations proceed within reproducible and trustworthy boundaries.

% By embedding these controls at the architectural level, \pname ensures that autonomy does not come at the cost of safety. Analysts can rely on the system to execute tasks responsibly, preserving reproducibility and mitigating the risks of automated misuse.  
%MK: are these guardrails are at the agent level only? we may do some rule based checks as well as future work.
% \begin{tcolorbox}[colback=green!5!white,colframe=green!60!black,title=Insight 5: Guardrails Enforced]
% Safety constraints ensure that autonomous investigation remains controlled, reproducible, and resistant to misuse.
% \end{tcolorbox}

\subsection{Threat Intelligence Retrieval Engine}
Cyber Threat Intelligence (CTI) reports are a rich source of adversarial knowledge, but are presented in unstructured and verbose formats. Analysts often spend significant time sifting through these documents to extract Indicators of Compromise (\ioc), leading to delays and inconsistencies in investigative workflows. To alleviate this bottleneck, \pname integrates a Threat Intelligence Retrieval Engine that applies retrieval-augmented generation (\rag) across an embedded corpus of CTI reports using the prompt template \autoref{fig:rag-prompts}. The capability of the engine is dependent on the underlying \llm. 

This Engine extracts relevant, complete, and actionable artifacts (\eg \iocs), separated by OS, process, files and IP as guided by the one-shot example in the prompt template. This translation from unstructured natural language to machine-actionable intelligence reduces cognitive burden on analysts and accelerates the transition from intelligence gathering to forensic validation.  

% By leveraging a vector database~\cite{chromadb2023} populated with domain-specific intelligence, the system ensures that IoCs are contextualized with semantically relevant evidence rather than generated in isolation. This minimizes hallucinations and preserves fidelity to the original CTI source material.

% Extracted IoCs and structured investigative plans must ultimately be validated against system-level provenance. However, provenance corpora such as the DARPA Transparent Computing (TC) datasets are large, heterogeneous, and stored in complex relational schemas. Without careful design, querying these databases risks inefficiency, duplication, and incomplete evidence recovery.  

\subsection{System Data Retrieval Engine}
The Data-Retrieval Engine interfaces directly with provenance databases, supporting artifact lookups across various entities and events. Optimized SQL templates handle schema complexity, normalize event records, and collapse duplicates across repeated system activities. This ensures that the evidence retrieved is both accurate and concise, preventing \llm from context explosion and from being overwhelmed by redundant records.  By abstracting away the intricacies of database schemas, this engine enables consistent evidence collection across diverse TC datasets. It ensures that each investigative step can reliably ground itself in factual provenance evidence, thereby anchoring the investigation in verifiable system behavior.  

A key design choice in the Data Retrieval Engine is the use of event type-specific SQL queries for processes, files, and IPs. Provenance data imposes constraints where the semantics of an entity are tightly coupled with its type: a process may be identified by its command or path, a file by its absolute or relative path, and a network flow by its IP and port tuple. To respect these constraints, the system issues distinct queries tailored to each type rather than relying on a single generalized lookup. This separation reduces ambiguity, ensures semantic correctness, and aligns with the structural guarantees inherent to provenance datasets.  

% Further efficiency is achieved by incorporating basename lookups for files and processes, as well as flexible IP and port matching strategies for network flows.
% For instance, when querying files, both the full path and the basename (with or without extensions) are considered, improving recall when attackers rename or move artifacts.
% Similarly, for network flows, queries differentiate between IP-only lookups and combined IP:port searches, ensuring precise recovery of connections even in noisy datasets. 
% These optimizations are critical in the provenance domain, where misclassification of a node type or incomplete event retrieval can break causal chains and undermine the fidelity of attack reconstruction.  

\subsection{Filtration Engine}
Even after the System Data Retrieval Engine issues targeted, provenance-aware SQL queries, the resulting candidate set of SQL events can still contain a large number of benign, high-frequency events (\eg benign background events). Passing all such results to the agents would both inflate the context window and increase the risk of false positives. To address this, \pname incorporates a Filtration Engine that performs rarity-aware post-processing over the retrieved SQL results using an AutoEncoder-based behavioral model~\cite{autoencoder}. The goal of this component is to down-select to only those events whose behavior is statistically uncommon with respect to benign system activity.

Concretely, each retrieved SQL result, which corresponds to a provenance event, is first embedded into a fixed-dimensional feature vector $x \in \mathbb{R}^d$ (capturing event type, participating node arguments). The AutoEncoder consists of an encoder $f_\theta(\cdot)$ and a decoder $g_\phi(\cdot)$ that are trained jointly on benign, publicly available provenance datasets so as to minimize reconstruction error on common events. During training, the model learns a compressed representation $z = f_\theta(x)$ and reconstructs $\hat{x} = g_\phi(z)$, and the reconstruction loss
\begin{equation}
    \mathcal{L}(x) = \bigl\lVert x - \hat{x} \bigr\rVert_2^2
    = \bigl\lVert x - g_\phi\bigl(f_\theta(x)\bigr) \bigr\rVert_2^2
    \label{eq:ae-loss}
\end{equation}
serves as a proxy for the rarity of the corresponding SQL event. Intuitively, provenance patterns that frequently occur in benign data are well reconstructed by the AutoEncoder (low $\mathcal{L}(x)$), whereas unusual or attack-like behaviors incur higher reconstruction error.

At inference time, the Filtration Engine computes $\mathcal{L}(x)$ for each candidate SQL result and assigns an anomaly score $s(x) = \mathcal{L}(x)$. A threshold $\tau$ is selected on a held-out benign validation set, and the filtration decision is given by
\begin{equation}
    x \text{ is anomalous } \Longleftrightarrow \mathcal{L}(x) > \tau.
    \label{eq:ae-threshold}
\end{equation}

Let $\mathcal{C}$ denote the set of all SQL results returned by the Data Retrieval Engine for a given query. The Filtration Engine passes forward only the subset $\mathcal{A} = \{\, x \in \mathcal{C} \mid \mathcal{L}(x) > \tau \,\}$, ensuring that the downstream agents reason primarily over rare and potentially attack-relevant events while filtering out common, benign provenance patterns.

\subsection{Explanation \& Summary}
\pname's Safety Agent uses the summarization tool with the prompt template as seen in \autoref{fig:mo-summ}, to translates structured evidence into human-interpretable narratives as seen in \autoref{fig:mo-ans}, where it not only provides the conclusion and interpretation but also grounded evidence, what evidence it could not find, and the possible immediate actions. 
%Using LLM reasoning, it organizes results into intuitive formats such as operating systems, processes, files, and IP addresses involved in the attack. These summaries present not just raw data but contextualized narratives that highlight causal relationships and adversary tactics. 
% To ensure accuracy in its outputs, the Explanation \& Summary Agent is guided through three-shot learning examples embedded in its prompt. These examples illustrate how to correctly identify and categorize processes, files, and IPs when parsing investigative context, and how to structure summaries into standardized fields. By showing the model explicit demonstrations of expected behavior, the agent learns to consistently extract the right artifact types while avoiding overgeneralization. 
This ensures the summaries are concise, actionable, and aligned with analyst needs, producing outputs that highlight causal relationships while maintaining semantic precision. This interpretive layer enhances analyst trust, reduces triage times, and ensures accountability in security workflows. 
\section{Provenance Tools and Considerations}

The architecture of \pname was designed not only as a modular pipeline but also as a resilient system capable of handling the nuanced challenges of provenance-driven security investigations. This section explains the tools and design considerations required for \textit{verifiable forecics}.

\begin{table}[htbp]
\centering
\caption{Summary of \pname tools.}
\label{tab:tools}
\resizebox{\columnwidth}{!}{%
\begin{tabular}{ll}
\toprule
\textbf{Name} & \textbf{Description} \\
\midrule
\multicolumn{2}{@{}c}{\textbf{Evidence Correlation Tools}} \\
\midrule 
\makecell[l]{\code{Provenance SQL} \\\code{Explorer}} & \makecell[l]{Executes provenance-aware SQL queries \\to extract system-level evidence.} \\
\addlinespace
\makecell[l]{\code{Artifact} \\ \code{Lookup}} & \makecell[l]{Searches for processes, files, and IPs in \\ provenance databases.} \\
\addlinespace
\makecell[l]{\code{Event Lookup}\\ \code{Resolver}} & \makecell[l]{Retrieves event for given process IDs \\to characterize system interactions.} \\
\addlinespace
\makecell[l]{\code{Type-Aware}\\ \code{Correlator}} & \makecell[l]{Correlates artifacts (\eg process$\rightarrow$file) \\to reconstruct causal chains.} \\
\addlinespace
\makecell[l]{\code{Summarizer}} & \makecell[l]{Summarizes the artifacts explored.} \\
\midrule
\multicolumn{2}{@{}c}{\textbf{Planning \& Orchestration Tools}} \\
\midrule 
\makecell[l]{\code{CTI} \\ \code{Retriever}} & \makecell[l]{Performs \rag over a CTI vectorDB \\to build context regarding IoCs.} \\
\addlinespace
\makecell[l]{\code{Plan} \\ \code{Generator}} & \makecell[l]{Decomposes analyst queries into structured \\and reproducible investigation steps.} \\
\addlinespace
\makecell[l]{\code{Follow-up} \\ \code{Planner}} & \makecell[l]{Formulates new targeted queries when evidence\\gaps are detected during investigation.} \\
\addlinespace
\makecell[l]{\code{Plan} \\ \code{Validator}} & \makecell[l]{Enforces schema and scope constraints to ensure\\ generated queries are correct and consistent.} \\
\midrule
\multicolumn{2}{@{}c}{\textbf{Safety \& Governance Tools}} \\
\midrule
\makecell[l]{\code{Safety} \\ \code{Checker}} & \makecell[l]{Validates agent reasoning and enforces \\ operational guardrails.} \\
\addlinespace
\makecell[l]{\code{Confidence}\\ \code{Validator}} & \makecell[l]{Detects incomplete or speculative summaries \\and enforces evidence-backed reporting.} \\
\bottomrule
\end{tabular}%
}
\end{table}

\subsection{Tools.}\label{sec:tools}
The tools in \pname span three dimensions: \textit{evidence correlation}, \textit{planning \& orchestration}, and \textit{safety \& governance} as described in \autoref{tab:tools}. Evidence correlation tools provide direct interfaces to the provenance databases, abstracting schema complexity and supporting artifact-centric lookups. Planning tools bridge the gap between unstructured CTI reports and structured forensic workflows, enabling agents to generate reproducible and auditable query sequences. Safety and governance tools act as guardrails, validating both inputs and outputs to prevent answers without evidence and mitigate hallucinations. Finally, \pname integrates \textit{role specialization}, where prompts are assigned professional personas (\eg DFIR analyst) to align reasoning context with investigative goals. 
%This layered tooling ensures that investigations remain grounded in verifiable provenance, systematically orchestrated, and bounded by security controls.

\subsection{Considerations.}

\heading{Ground Truth Verification.} A primary challenge in provenance analysis is ensuring that retrieved evidence corresponds to verifiable system-level events. \pname integrates ground truth verification by enforcing database-backed validation of artifacts before they are sent to analysts. Rather than relying on inferred associations, evidence must be corroborated against system provenance logs. This avoids speculative reasoning and anchors each investigative step in concrete, queryable system behavior.  

% \begin{tcolorbox}[colback=blue!5!white,colframe=blue!75!black,title= ]
% All evidence retrieved by \pname is validated against provenance databases to prevent reliance on inferred or unverified associations. This ensures that analysts can trust that surfaced artifacts reflect actual system events.
% \end{tcolorbox}

\heading{Follow-up Exploration.} Another design enhancement concerns follow-up exploration. Investigations rarely end with a single query; analysts often need to refine questions, pivot to related artifacts, or expand the scope of their search. \pname supports iterative exploration through bounded CoT reasoning loops, where outputs from one stage can seed subsequent queries. This enables the agent to progressively refine hypotheses, recover related processes or files, and map broader attack surfaces without requiring manual re-specification from the analyst.  

% Another crucial enhancement lies in handling multi-modal evidence correlation using different tools (\eg a CTI Retriever over a vector store to ground context, a Provenance SQL Explorer for process/file/IP lookups, an Event-Type Resolver to enforce edge semantics, and a Type-Aware Correlator that links processes$\rightarrow$files and processes$\rightarrow$IPs without cross-type leakage), as seen in \autoref{fig:tools}. 

Investigations often involve joining heterogeneous artifacts, such as linking processes with files and IP connections. Therefore, using an Event-Type Resolver to enforce edge semantics, and a Type-Aware Correlator that links processes$\rightarrow$files and processes$\rightarrow$IPs without cross-type leakage, \pname decomposes complex tasks into small manageable tasks. Also, instead of overloading a single agent with complex forensic task, \pname decomposes tasks across specialized Agents using different engines and tools.
%(\textit{Digital Forensics and Incident Response (DFIR) analyst} persona for artifact correlation, \textit{System Security expert} persona for provenance validation, and \textit{Safety Auditor} persona for compliance checks) and tool-specialized executors (\code{Artifact Lookup}, \code{Correlator}, \code{Plan Validator}, \code{Follow-up Planner}).  

% Finally, usability considerations influenced the system’s architecture. The web-based interface enables analysts to interact with agents naturally while still benefiting from underlying safety and validation layers. Analysts receive outputs that are enriched, contextualized, and filtered, ensuring that the complexity of provenance analysis is hidden behind a transparent and interpretable workflow. 

% \begin{tcolorbox}[colback=blue!5!white,colframe=blue!75!black,title= ]
% The system supports iterative reasoning loops, allowing results from one query to drive the next. This facilitates dynamic investigations that adapt to emerging evidence without restarting from scratch.
% \end{tcolorbox}

\heading{Hallucination Mitigation.} A recurring risk with LLM-driven agents is hallucination when generated outputs diverge from ground truth. To mitigate this, \pname anchors reasoning in authoritative CTI reports and provenance logs. 
%Retrieval-augmented generation ensures that analyst queries are supplemented with semantically relevant context from embedded reports, while SQL-based validation cross-checks outputs against system events. This hybrid design prevents the agent from introducing spurious processes, files, or IPs, thus curbing the risks associated with unconstrained generation.  
%MK: I think how the sql-based validation is happening is not clear. More details and examples could be useful.
\pname validates every hypothesized artifact against provenance DBs before sending it to the analyst. Concretely, the \code{CTI Retriever} proposes candidate artifacts from CTI (\eg process/file/IP), which are then verified via \code{Artifact Lookup}. This tool performs entity-specific searches with adaptive SQL templates, emitting concrete edge-level evidence (edge id, event type, source/destination labels) rather than free-text matches. Subsequently, \code{Event Lookup Resolver} cross-checks the \emph{operation/event\_type} by node id, and \code{Source/Destination Label Resolvers} extracts human-readable labels from edge ids. 
%Finally, \code{Type-Aware Correlator} assembles typed relations (process$\rightarrow$file) using schema-bound SQL.
Because artifact hypotheses must resolve to concrete node and edge identifiers,
the agent cannot “invent” a process or file without a corresponding row match, and typed validators further prevent cross-type leakage (\eg a file basename being misinterpreted as a process). This strong coupling forces summaries to be traceable to specific events. 

% \begin{tcolorbox}[colback=blue!5!white,colframe=blue!75!black,title= ]
% By coupling retrieval-augmented generation with provenance-backed SQL validation, the system minimizes fabricated outputs and ensures alignment with both CTI and observed system events.
% \end{tcolorbox}

%MK: I think our novelty in these query optimization strategies may need to be better explained.
\heading{Query \& Token Management.} Large provenance databases introduces a significant operational challenge of query explosion. Certain artifact lookups may return excessive results if not carefully constrained, which can overwhelm the \llm's context window. To address this, \pname applies query optimization strategies such as duplicate collapse, type-specific queries, and scoped matching, and further incorporates an AutoEncoder-based Filtration Engine to suppress high-frequency benign patterns before they reach the reasoning layer.

\pname issues \emph{type-specific} SQL queries where, for files and processes, it searches both \path{full paths} and \code{basenames}; for IPs, it attempts exact \code{IP:port} matches before falling back to IP-only lookups. 
% These templates enforce semantic correctness (path vs.\ command vs.\ \code{IP:port}) and enable the agent to collapse repeated system activities (\eg identical \code{<source,relation,destination>} records) and annotate the surviving record with an identifier count. 
Results are de-duplicated to prevent inflation of evidence counts and then passed through the Filtration Engine, to  discard the results whose whose reconstrcution loss falls below the rarity threshold, effectively filtering out common, routine system activity. These mechanisms keep queries bounded and prevent unnecessary token consumption during the reasoning process by limiting both the number and diversity of events that reach the agents.

In addition, reasoning depth is explicitly bounded: the \emph{Follow-up Agent} conditionally increases the step budget only when prior outputs \emph{explicitly} recommend deeper probing. Follow-up Agent use explicit tool calls for each extracted artifact, ensuring that expansions remain structured, semantically relevant, and focused on statistically rare, potentially attack-relevant events.

\heading{Evidence Verification.} \pname will provide answers that can be verified using system events. When an answer cannot be fully verified, \pname will explicitly state which artifact(s) it could not validate and outline potential follow-up steps the analyst should take. In addition, \pname clearly indicates when its results are inconclusive. Thus, beyond efficiency, verifiability remains a central focus of \pname.

Collectively, these design details demonstrate how \pname addresses real operational constraints that undermine many provenance-based systems. By embedding ground truth verification, iterative exploration thorugh bounded queries, and hallucination control, \pname provids an analyst-aligned investigative platform that is both powerful and trustworthy.

%MK-Oak: We did not mention the auto-encoder. Also, if we can again emphasize our novelty compared to basic implementations could be useful

% \input{sections/imple}
%! root=./main.tex
\section{Evaluation}

In this section, we evaluate \pname's effectiveness in automated intelligence extraction and forensic investigation. We aim to answer the following research questions:

\begin{itemize}[noitemsep, leftmargin=0.9cm]
    \item[{\bf RQ1:}] How effective is \pname in extracting actionable artifacts from threat reports (\autoref{sec:extract})? 
    \item[{\bf RQ2:}] How well does \pname identify threats and summarize post-detection context (\autoref{sec:detect})?
    \item[{\bf RQ3:}]How much does each component of \pname contribute to its overall performance (\autoref{sec:ablation})?
    \item[{\bf RQ4:}] How scalable is \pname (\autoref{sec:scale})?
    \item[{\bf RQ5:}] What is the impact of different LLM backends on \pname's performance (\autoref{sec:ablation_llm})?
    \item[{\bf RQ6:}] What errors are encountered by \pname (\autoref{sec:error})?
\end{itemize}

\subsection{Experiment Protocol}\label{sec:expr_protocol}

We first measure \pname performance in extracting actionable system artifacts (\eg process, files and IP addresses) from threat reports. We compare \pname to vanilla RAG pipeline and a type-filtered RAG variant. The vanilla RAG pipeline retrieves artifacts without provenance awareness, often introducing irrelevant or semantically inconsistent evidence. The type-filtered RAG pipeline improves upon this by constraining retrieval to artifacts of the correct provenance entity and event types, but it still lacks reasoning about multi-hop causal relationships. By comparing against these baselines, we \textit{highlight how \pname's provenance domain-aware query not only retrieves better artifacts but also captures richer attack context.}

Next, we measure how well \pname identifies threats and summarizes the post-detection context, so we compare \pname against multiple baselines spanning both provenance-based \ids and Strawman Agent. Specifically, for threat detection, we benchmark against \sota \pids Kairos~\cite{cheng2024kairos}, MAGIC~\cite{jia2024magic}, Flash~\cite{rehman2024flash} and Orthurus~\cite{jian2025}. Then, we conduct an ablation study by removing the components of \pname and measuring its detection accuracy to understand each component's significance. Next, we study the scalability of \pname across database sizes to understand how much token usage and latency is expected as database size grow.

Then, to understand the impact of different backend \llms, we perform an ablation study across different \llms, GPT-4o, GPT-4.1, and GPT-5, measuring detection accuracy, token usage, cost and time required. Finally, we conduct an error analysis study to understand what kind of errors are encountered by \pname. 
% Therefore, the comprehensive evaluation protocol allows us to demonstrate both the accuracy and practical trade-offs of deploying \pname in real-world security environments. 

\heading{Datasets.} 
Our evaluation leverages seven publicly available provenance datasets that capture diverse attack scenarios, including the largest publicly released provenance dataset to date~\cite{rehman2024flash} (\eg DAPRA OpTC dataset). Following prior work~\cite{bilot12399sometimes}, we use the most challenging publicly available provenance datasets: \darpa OpTC~\cite{darpaoptc}, \tc E3~\cite{darpae3} and E5~\cite{darpae5}. E3 and E5 datasets contain multi-day system traces from \code{CADETS}, \code{THEIA}, and \code{CLEARSCOPE}. DARPA OpTC dataset also contains multi-host device traces. We utilize the ground truth annotations for E3 and E5 datasets from~\cite{jian2025}. But for DAPRA OpTC dataset we manually created and verified the ground truth, since they are not publicly available. 
% These datasets contain traces from the Red Team simulated three attacker profiles: ``Nation State'', ``Common Threat,'' and ``Metasploit'' to model different levels of sophistication, while the Blue Team engaged in realistic enterprise defense. 
To evaluate retrieval and identification quality, we additionally employ synthetic Q\&A pairs generated using \cite{syntheticdatakit2024}, which produces security-specific question-answer pairs from \darpa threat reports tailored for RAG pipelines. 

To avoid information leakage, we explicitly exclude benign data from the dataset under evaluation when training the AutoEncoder and calibrating $\tau$. That is, for any target provenance dataset used in experiments, its benign events are never incorporated into the training or validation sets of the Filtration Engine. This setting accurately reflects deployment scenarios where the model must generalize to previously unseen systems and workloads.
%This separation guarantees that the rarity scores reflect generalizable behavioral deviations rather than memorization of dataset-specific distributions, and it more accurately reflects deployment scenarios where the model must generalize to previously unseen systems and workloads.

\heading{Metrics.}
To rigorously assess \pname, we employ a comprehensive set of evaluation metrics spanning intelligence extraction, threat detection, and ablation study tasks. For intelligence extraction, we use \emph{contextual precision} and \emph{recall} to measure whether retrieved evidence fragments are both relevant and complete relative to the analyst's query. We further measure \emph{relevance}, which captures semantic alignment between generated answers and threat reports, and \emph{faithfulness}, which evaluates whether responses remain grounded in retrieved provenance evidence rather than introducing unsupported hallucinations. These metrics are particularly important in the security domain, where spurious or misleading intelligence can significantly degrade analyst trust and decision-making. 

For detection and ablation studies, we measure \emph{precision}, \emph{recall}, \emph{F1-score}, \emph{Token Usage} (for LLM-based approaches), and \emph{Time} by comparing the post-detection summaries against the ground-truth \ioc. This allows us to evaluate not only whether threats are identified, but also the relative utilizing of the different components of \pname. For overhead study, token usage and time is measured against the database size. For the error analysis study, we measure the number of errors produced by \pname as it attempts to answer the queries. Together, these metrics provide a balanced view of both effectiveness and operational feasibility of \pname.

\begin{figure}[t]
  \centering
  \resizebox{0.9\columnwidth}{!}{%
    \begin{tabular}{c}
      \includegraphics{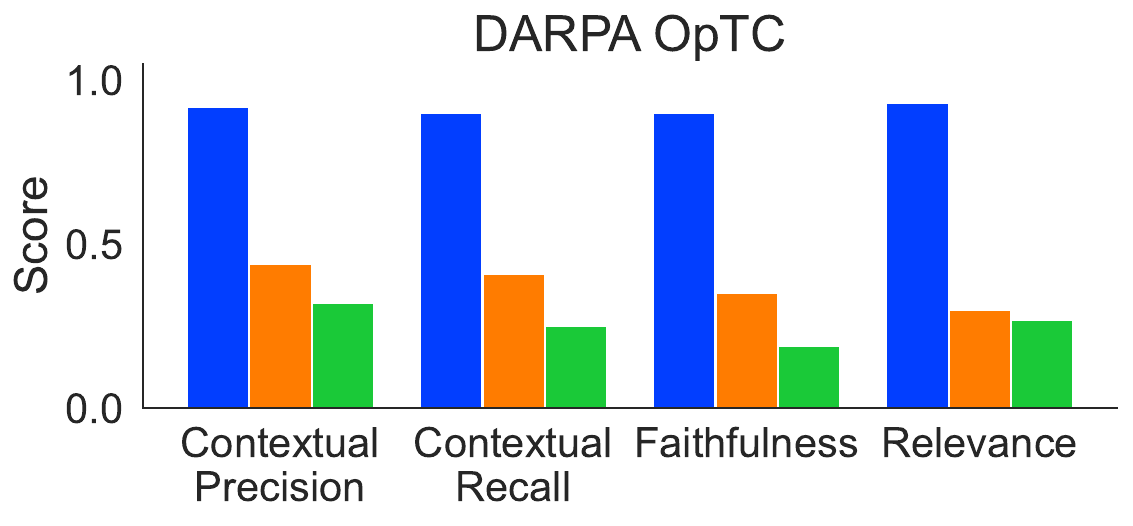} \\
      \includegraphics{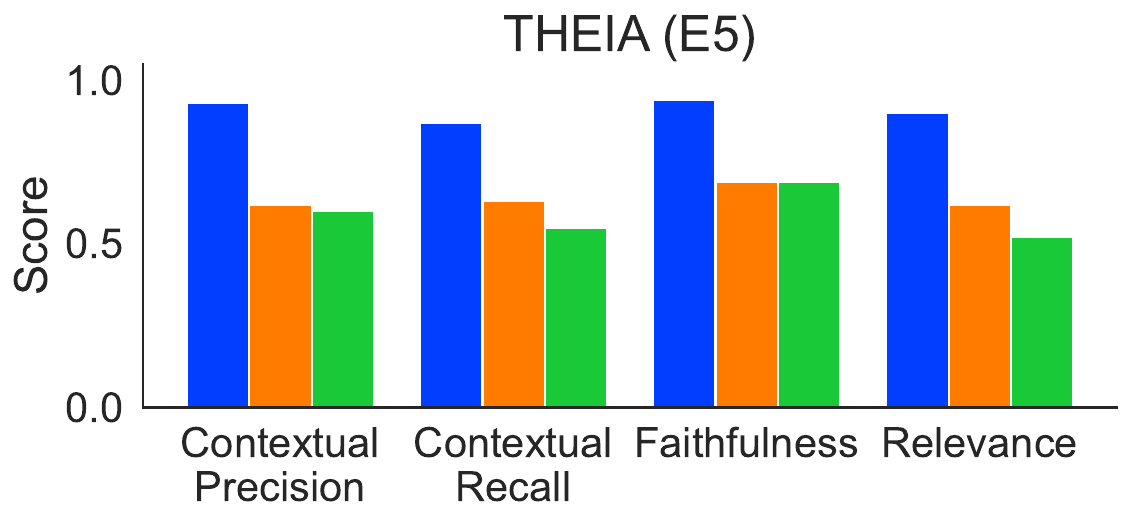} \\
      \includegraphics{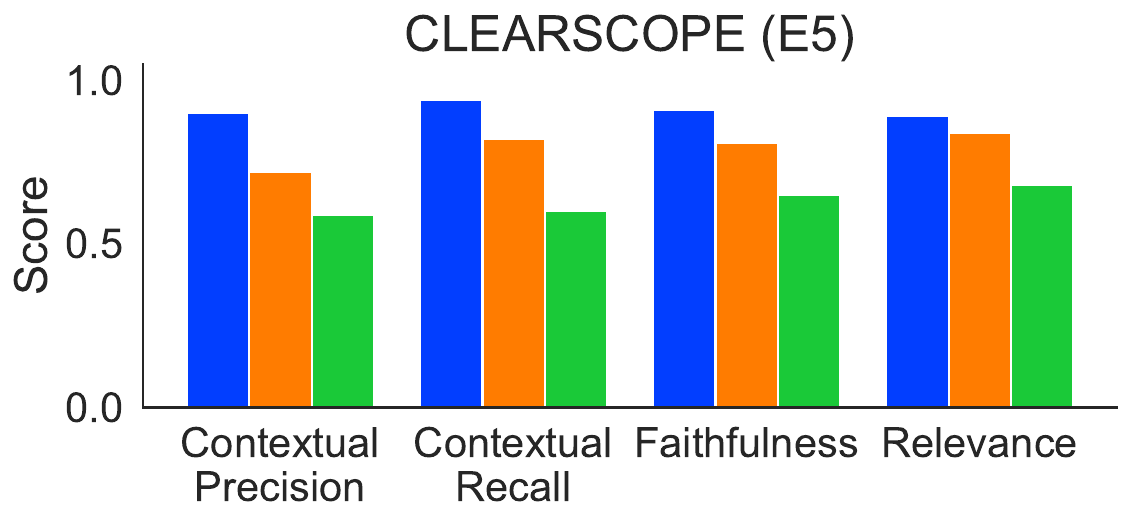} \\
      \includegraphics{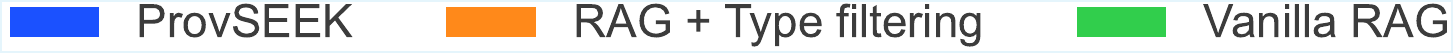}
    \end{tabular}%
  }
  \caption{Intelligence extraction performance on \darpa (E5) dataset. (complete results in \autoref{tab:intel_performance}).}
  \label{fig:intel_performance_e5}
\end{figure}
\subsection{Artifact Extraction Performance}\label{sec:extract}

\pname's threat artifact extraction performance snapshot is shown in \autoref{fig:intel_performance_e5} measured across DARPA OpTC and E5 datasets (\code{THEIA} and \code{CLEARSCOPE}); detailed results across all the datasets (\eg E3, E5, OpTC) are provided in \autoref{tab:intel_performance}. Across all the datasets, \pname achieves 34\% improvements in contextual precision and recall compared to vanilla RAG, with corresponding gains in relevance and faithfulness. Even when compared to type-filtered RAG, which is a stronger baseline, \pname improves by 18\%, showcasing the impact of context reasoning. These results show that provenance-guided intelligence extraction not only suppresses spurious associations but also enables richer contextual reasoning about attacks. 
    
Both vanilla and type-filtered RAG perform poorly in \code{THEIA} with contextual precision often dipping near 0.5-0.6 and worse for DARPA OpTC (going below 0.3). But, \pname consistently performs well (\ie precision and recall in the 0.87-0.94 range), demonstrating robustness to complex persistent attack traces. Type-filtered RAG narrows the recall gap on \code{CLEARSCOPE} (E5), but fails to maintain precision, highlighting the difficulty of reconciling relevance with correctness without provenance reasoning.
    
On all the datasets, \pname sustains both high contextual precision ($\geq$0.90) and recall ($\geq$0.91), while baselines fluctuate significantly. In particular, \code{DARPA OpTC} shows how naive retrieval pipelines cannot be deployed in the real-world where large \cti reports can mislead relevant intelligence extraction (recall dropping to 0.25 for vanilla RAG), whereas provenance-aware correlation stabilizes results.

\begin{figure}[t]
  \centering
  \resizebox{0.95\columnwidth}{!}{%
    \begin{tabular}{c}
        \includegraphics{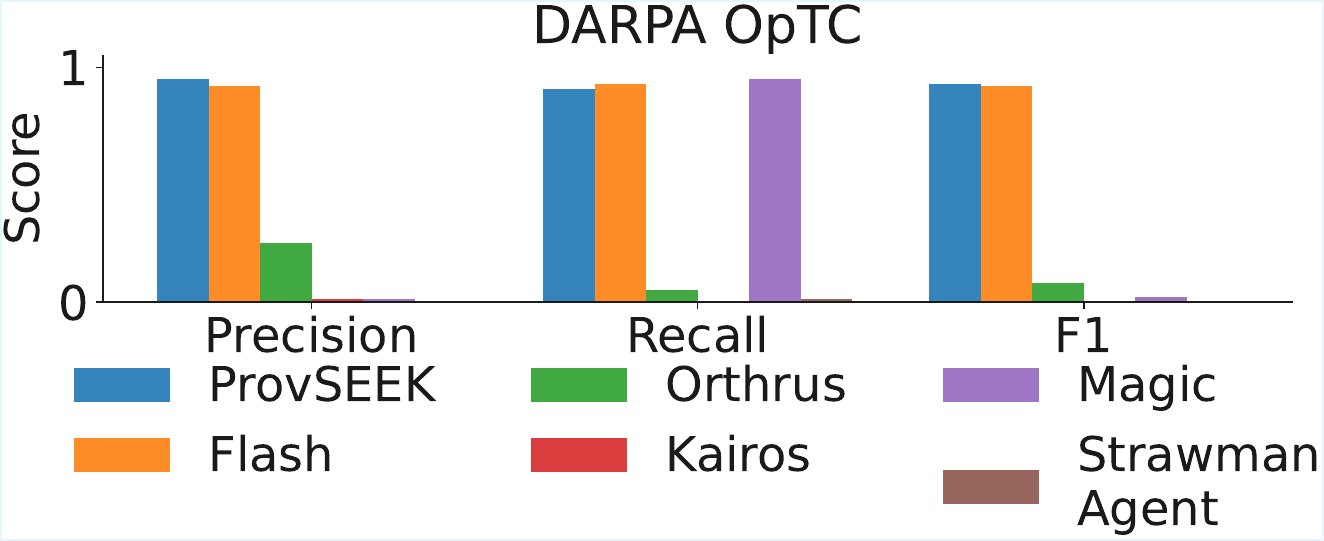} \\
        \includegraphics{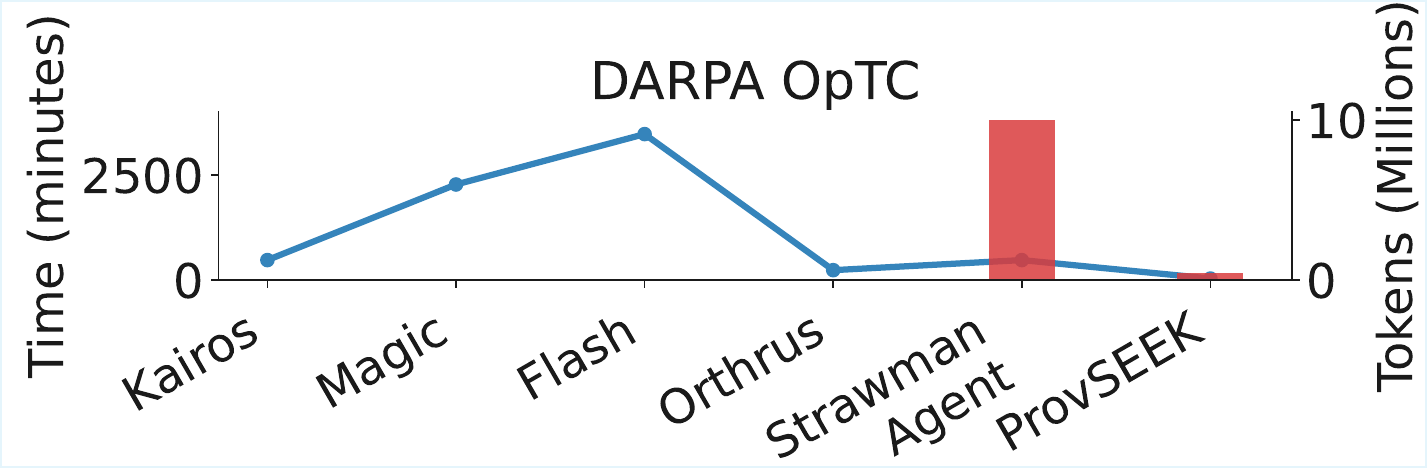} \\
        \includegraphics{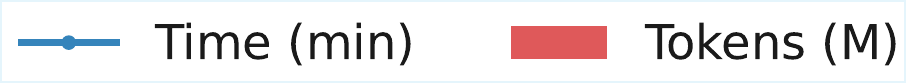} \\
        \includegraphics{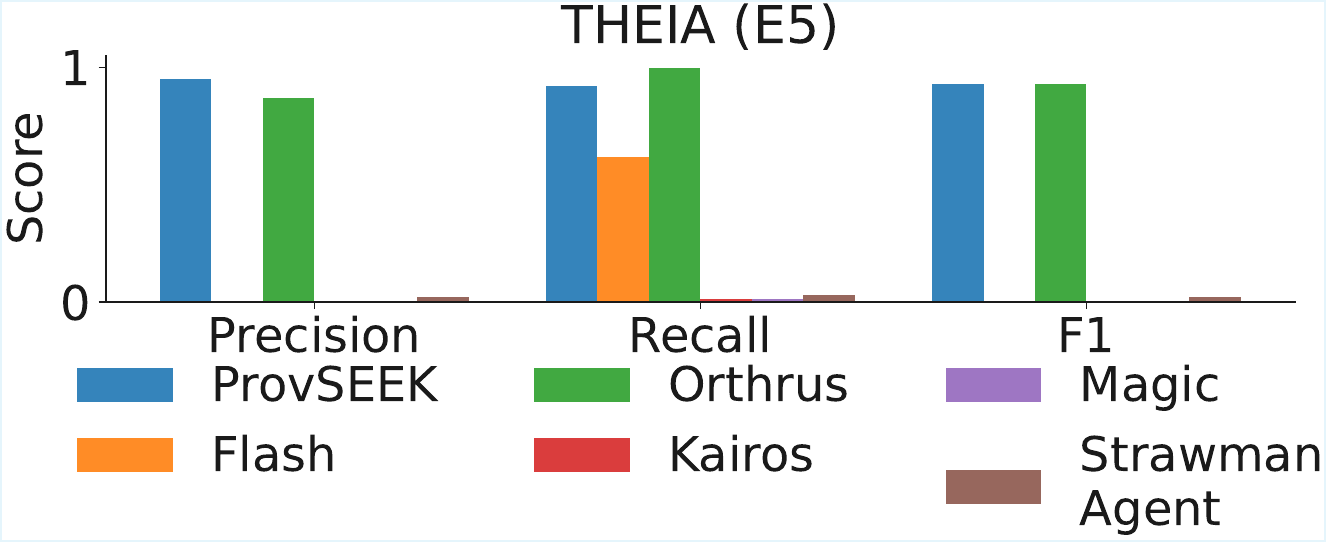} \\ 
        \includegraphics{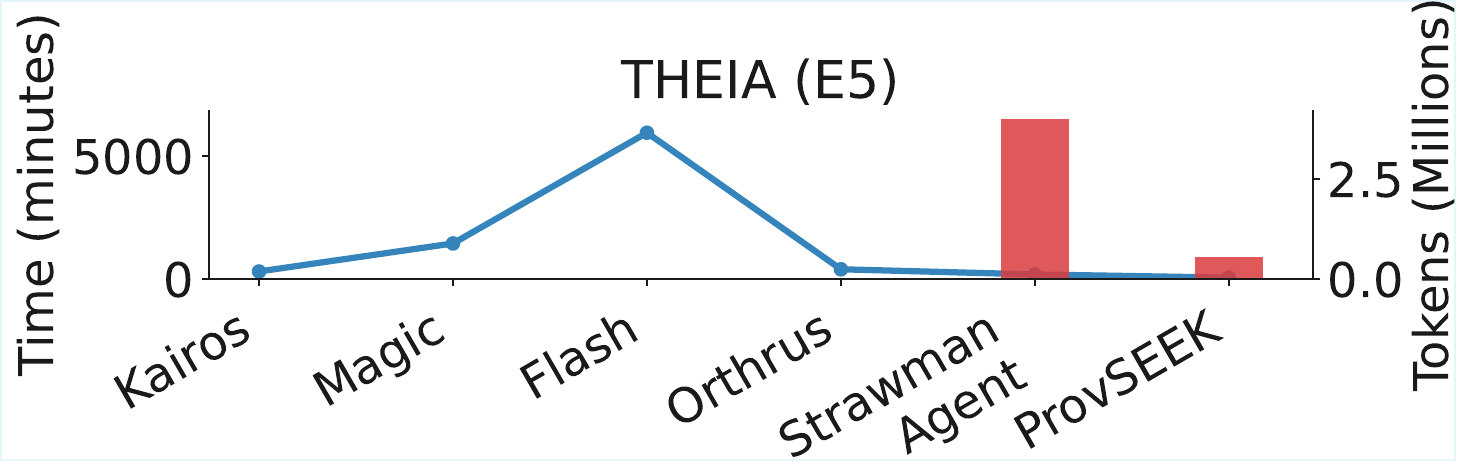} \\
        \includegraphics{figs/script/detection/perf_dual_legend.pdf} \\
    \end{tabular}%
  }
  \caption{Threat detection performance on \darpa (OpTC and E5-Theia) dataset, (complete results in \autoref{tab:detection_performance}).}
  \label{fig:det_performance_e5}
\end{figure}
\subsection{Threat Detection Performance}\label{sec:detect}
%MK-Oak: showing both on the same figure could be problematic. Since only strawman agent uses token. We may just mention in it in the text ? 

We next evaluate how accurately \pname detects threats and summarizes them against provenance ground truth. We visualize \pname's performance for DARPA OpTC and THEIA E5 dataset in ~\autoref{fig:det_performance_e5} and the comprehensive results for all the datasets are shown in \autoref{tab:detection_performance}. \pname achieves higher F1 by balancing both precision (improving by 22\%) and recall (improving by 29\%) comparing all the \sota \pids across datasets. \pname consistently outperforms most baselines across datasets or delivers performance comparable to the best-performing baseline. The contrast is especially pronounced on the DARPA OpTC dataset, where most \sota \pids struggle. 
%MK-Oak: the next sentence is not clear, please rewrite.
% Compared to \sota \pids~\cite{rehman2024flash} which was the only paper to evaluate against used DARPA OpTC dataset performed similar to us.

The Strawman Agent does not provide any benefits because without being able to extract the important attack artifacts (\ie processes, files and IP) it cannot find them in the system database. Noticeably, the Strawman Agent used 2000\% (2M vs 200K) more token compared to \pname without providing any significant result because of its aimless pursuit of trying to find attack artifacts. The runtime of \pname is comparable to that of the best-performing \pids, with most runs completing in under 45 minutes.
%MK-Oak: I think somewhere, maybe for threat detection, it may be important to discuss our performance compared to SOTA with respect to run time. -- above

% DARPA OpTC dataset again emerges as the most challenging dataset: most \sota \pids baselines collapse in precisiona nd recall, with Flash performing the best. Despite this, \pname maintains high F1-scores across DARPA OpTC and other DARPA datasets, highlighting its robustness under complex, multi-host traces. 

While \sota \pids either miss most threats or misidentify on irrelevant artifacts, \pname provides structured summaries that align with ground-truth \iocs. Importantly, these summaries not only flag malicious behaviors but also preserve causal chains, allowing analysts to reason about attack progression. In cases where it could not get the right answer or have inconclusive result as the all step budget is used up, it notifies the user of what artifacts it could not find so that the user can focus their effort into identifying those artifacts or provide better guidance to \pname. Overall, these results confirm that \pname is capable of delivering both faithful and operationally efficient detection.

% \input{tables/ablation_detection.tex}
% \begin{figure}[t]
%   \centering
%   \setlength{\tabcolsep}{0pt}
%   \renewcommand{\arraystretch}{0}
%   \begin{tabular}{c}
%     \resizebox{\columnwidth}{!}{\includegraphics{figs/script/ablation/abl_perf_darpa_optc.pdf}}\\[2pt]
%     \resizebox{0.8\columnwidth}{!}{\includegraphics{figs/script/ablation/abl_resources_darpa_optc.pdf}}\\
%     \resizebox{0.8\columnwidth}{!}{\includegraphics{figs/script/ablation/legend.pdf}}\\
%     \resizebox{\columnwidth}{!}{\includegraphics{figs/script/ablation/abl_perf_theia_e5.pdf}}\\[2pt]
%     \resizebox{0.8\columnwidth}{!}{\includegraphics{figs/script/ablation/abl_resources_theia_e5.pdf}}\\
%     \resizebox{0.8\columnwidth}{!}{\includegraphics{figs/script/ablation/legend.pdf}}
%   \end{tabular}
%   \caption{Ablation study of how removing different components affect \pname's threat detection performance on \darpa (OpTC and E5-Theia) dataset. (complete results in \autoref{tab:abl_det_perf}).}
%   \label{fig:det_det_perf_e5}
% \end{figure}
% \subsection{Ablation Study}\label{sec:ablation}
\begin{figure}[t]
  \centering
  \setlength{\tabcolsep}{0pt}
  \renewcommand{\arraystretch}{0}
  \resizebox{0.95\columnwidth}{!}{%
    \begin{tabular}{c}
      \includegraphics[width=\columnwidth]{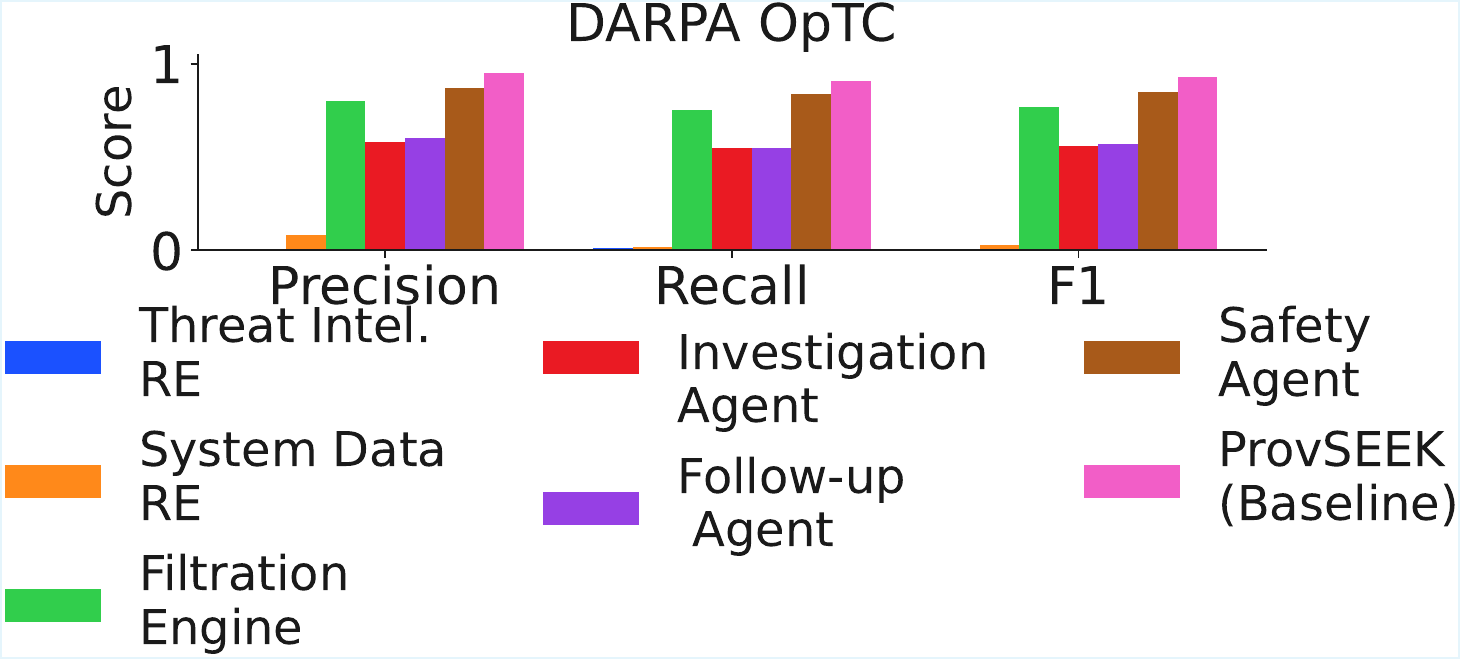}\\[2pt]
      \includegraphics[width=0.8\columnwidth]{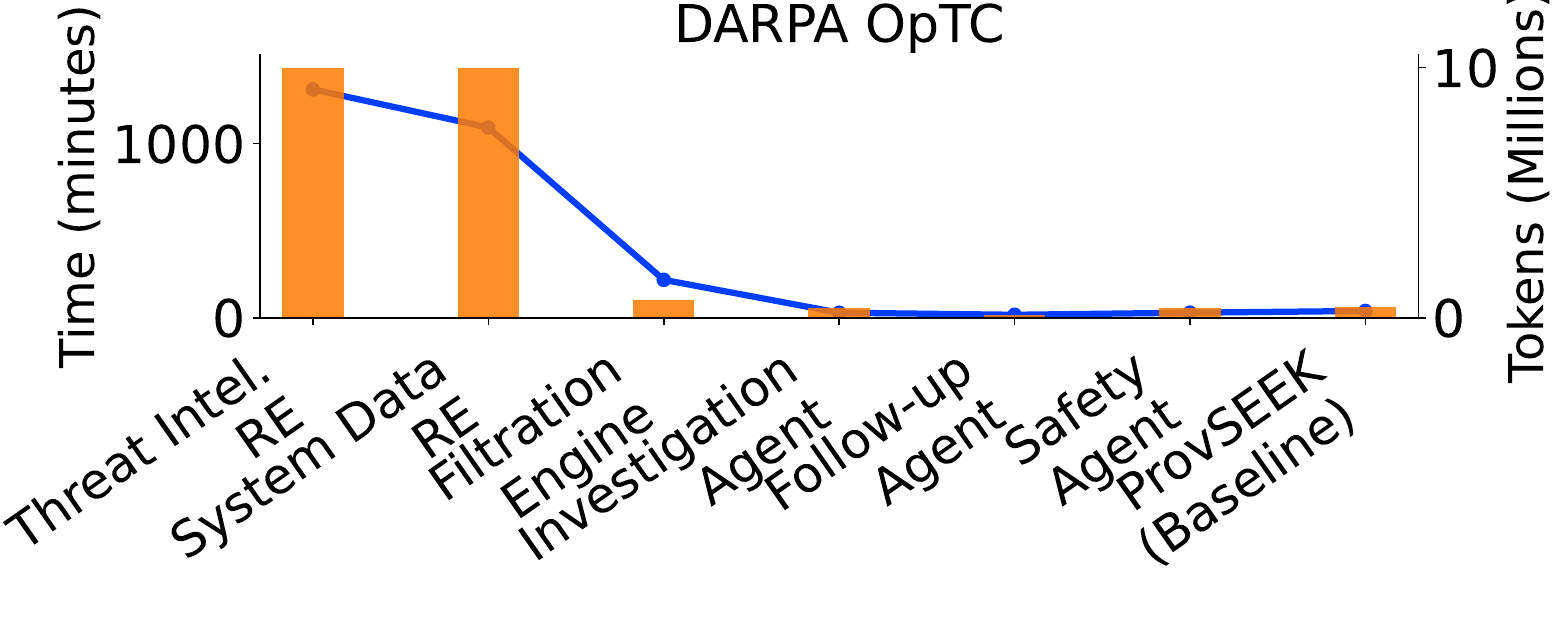}\\
      \includegraphics[width=0.8\columnwidth]{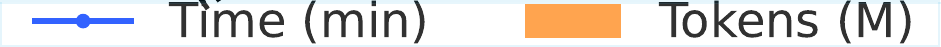}\\
      \includegraphics[width=\columnwidth]{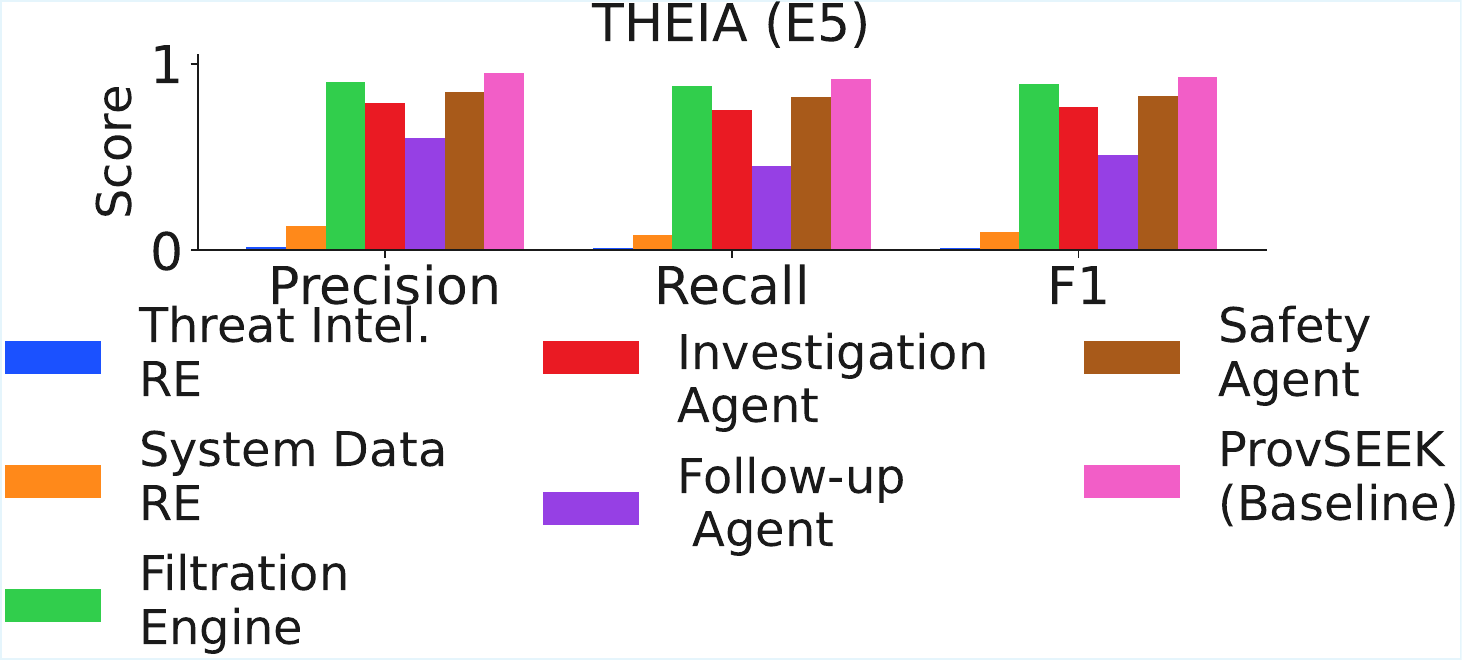}\\[2pt]
      \includegraphics[width=0.8\columnwidth]{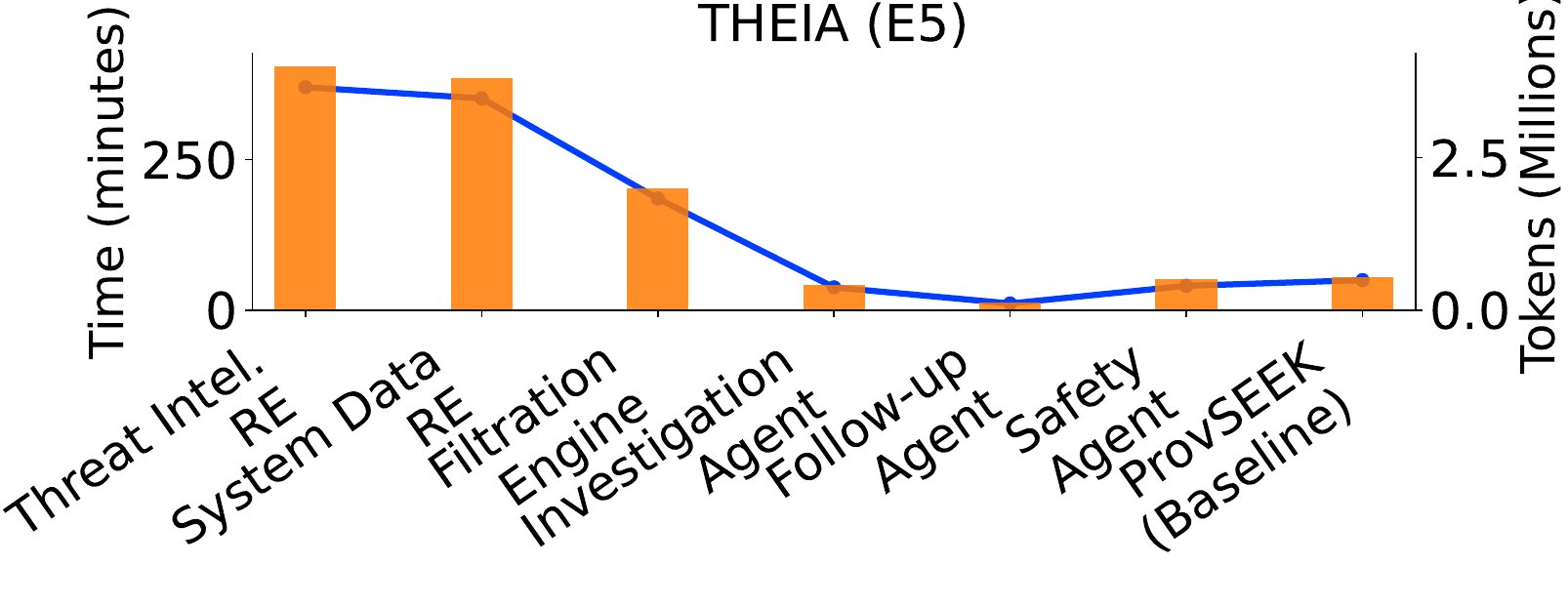}\\
      \includegraphics[width=0.8\columnwidth]{figs/script/ablation/legend.pdf}
    \end{tabular}%
  }
  \caption{Ablation study of how removing different components affect \pname's threat detection performance on \darpa (OpTC and E5-Theia) dataset, (complete results in \autoref{tab:abl_det_perf}).}
  \label{fig:det_det_perf_e5}
\end{figure}

\subsection{Ablation Study}\label{sec:ablation}

We next conduct an ablation study to understand how each component of \pname contributes to threat detection performance and resource usage. We visualize the ablation results for DARPA OpTC and THEIA E5 in \autoref{fig:det_det_perf_e5}, and report the full results across all datasets in \autoref{tab:abl_det_perf}. 

The ablation clearly shows that the \emph{Follow-up Agent} is the most critical component for maintaining high detection performance. Removing the follow-up loop substantially degrades both recall and F1, because the system can no longer decompose complex analyst queries into step-wise evidence gathering and iterative hypothesis refinement. For instance, on CADETS (E3), the F1-score drops from 0.89 with full \pname to 0.55 without the Follow-up Agent (a loss of 0.34), and on THEIA (E5) it falls from 0.93 to 0.51 (a loss of 0.42). Similar drops (e.g., from 0.93 to 0.57 on OpTC) are visible across all datasets in \autoref{tab:abl_det_perf}. At the same time, removing the Follow-up Agent reduces token usage and latency because the system stops issuing additional targeted probes: token usage shrinks by roughly 300K-400K tokens (e.g., 540K to 122K on THEIA (E5)), and runtime falls from tens of minutes to under 10-20 minutes. 

This trade-off highlights that the follow-up loop is where \pname does the heavy lifting of reasoning over artifacts and closing evidence gaps, and that its computational cost is directly tied to more exhaustive and accurate investigations. Ablation study of the Investigation Agent shows a smaller but noticeable effect: removing the Investigation Agent degrades F1 more mildly (e.g., from 0.93 to 0.77 on THEIA (E5)), because the Follow-up Agent can still try to recover evidence, but the system loses the initial structured plan for correlating artifacts and assembling attack chains.

%MK-Oak: the next sentence is not clear. please rewrite.
This is evident in \autoref{fig:det_det_perf_e5}, where removing the Threat Intelligence Retrival Engine (RE) or System Data Retrival Engine (RE) degrades performance, demonstrating that provenance-aware retrieval is essential for both accuracy and scalability. Without Threat Intelligence Retrival Engine (RE), \pname no longer has a principled way to identify which artifacts (files, processes, IPs) are worth querying in the provenance database, and the LLM is forced to ``guess''(\ie hallucinate) from its pre-training which indicators might be relevant. This leads to near-random detection performance while consuming massive resources: on OpTC F1 collapses to 0.00 while token usage explodes to 10M tokens and over 21 hours of compute. A similar pattern appears when we remove the System Data Retrival Engine (RE): detection remains extremely poor (F1 in the 0.03-0.10 range), but token usage and run time remain nearly as high as the ``no-Threat Intelligence Retrival Engine (RE)'' setting (e.g., 10M tokens and 18.18 hours on OpTC). These experiments confirm that naive or unoptimized database querying causes severe context explosion: the agent repeatedly fetches large volumes of low-value events and then spends most of its budget reasoning over irrelevant or noisy artifacts. The combination of CTI-guided RAG and provenance-aware SQL templates in \pname is therefore crucial: Threat Intelligence Retrival Engine (RE) focuses the search on semantically relevant IOCs, while the SQL querying enhancements that enforce type- and schema-aware constraints  keep the retrieved context bounded and interpretable.

The remaining components: the Filtration Engine, Investigation Agent, and Safety Agent, show a finer trade-off between interpretability and overhead. Removing the Filtration Engine step yields only a modest F1 decline for smaller datasets (e.g., 0.93 to 0.88 on THEIA (E5)), but becomes significant for large datasets like CLEARSCOPE (E5). Removing the Filtration Engine, increases token usage and run time because  irrelevant queries are not filtered out and the the system must reason over more raw edges without structured consolidation. 

Investigation Agent ablation study, as discussed above, shows that initial investigation planning matters, but its impact is still smaller than that of the follow-up loop, since the Follow-up Agent continues to search for evidence that can support or refute a hypothesis. In contrast, removing the \emph{Safety Agent} results in only a small performance drop (typically 0.04-0.07 F1), while token usage decreases slightly (e.g., from 423K to 401K on CADETS (E3) and from 570K to 513K on CLEARSCOPE (E5)), and latency changes by only a few minutes. This reflects that the Safety Agent primarily drives extra verification and re-writing passes: it flags over-confident or under-evidenced answers, triggers additional follow-up plans, and enforces evidence-backed reporting. When it is removed, \pname saves the extra queries and rewrites, but at the cost of weaker guarantees that every claim is backed by explicit database evidence. Overall, the ablation results support our design choice: \pname's strongest gains come from the synergy between follow-up reasoning, Threat Intelligence Retrival Engine (RE), and provenance-aware SQL querying, while the other agents refine the balance between accuracy, robustness, and resource usage.

\begin{figure}[t]
  \centering
  \resizebox{0.9\columnwidth}{!}{%
    \begin{tabular}{c}
      \includegraphics{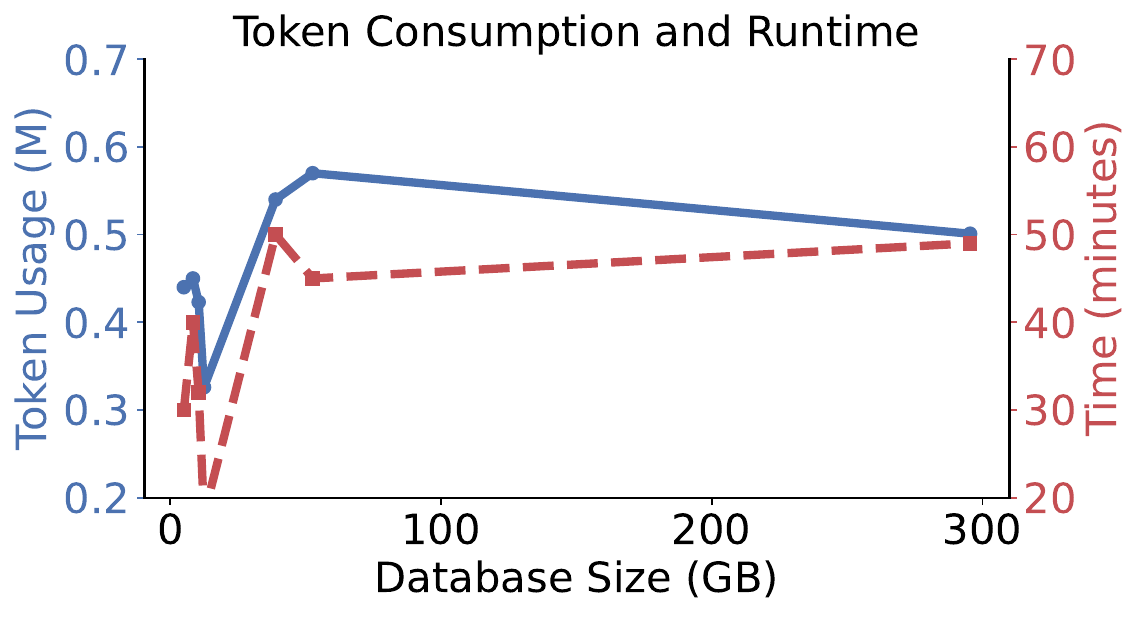}
    \end{tabular}%
  }
  \caption{\pname resource usage with database size.}
  \label{fig:overhead}
\end{figure}

\subsection{Scalability of \pname Study}\label{sec:scale}

We evaluate the scalability of \pname by varying the size of the provenance database and measuring the corresponding runtime and token usage (as seen in \autoref{fig:overhead}). As the database size grows from small \code{CLEARSCOPE} E3 deployments (4.98GB) up to 50x \code{CADET E5} deployment (295.45GB) the token usage increases by 1.42x and time required per investigation increases by 1.63x because from \pname's design: CTI guided RAG narrows the search space to a small set of candidate artifacts, provenance domain-aware SQL enhancements aggressively discard benign edges, and the Filtration Engine further filters out common system events, so the agent reasons over a compact, rare-event based context rather than the full database. 

More interestingly, LLM token usage remains tightly concentrated around $\approx$500K tokens across all datasets (ranging from 326K on \code{THEIA} E3 to 570K on \code{CLEARSCOPE} E5), despite nearly two orders of magnitude variation in database size. As databases get larger, the Follow-up Agent may need to perform additional iterations when RAG extraction is less than optimal. Conversely, in smaller databases, suboptimal RAG behavior can sometimes lead to relatively high token usage because the agent explores a broader fraction of the database when it ``does not know what to look for'' and chases more speculative artifacts. Overall, these results show that \pname scales optimally to the largest provenance database while keeping token usage roughly constant and runtime within tens of minutes, indicating that it is practical to deploy in real-world, large-scale system environments.

\begin{table}[!htb]
\centering
\resizebox{\columnwidth}{!}{
\begin{tabular}{lcccccc}
\toprule
Model & P ($\uparrow$) & R ($\uparrow$) & F1 ($\uparrow$) & \makecell[l]{Token \\Usage ($\downarrow$)} & Cost ($\downarrow$) & \makecell[l]{Time\\ (min.) ($\downarrow$)} \\
\midrule
\multicolumn{7}{c}{\code{CADETS} (E3)} \\
\midrule
GPT-4o         & 0.75 & 0.75 & 0.75 & 276K & \$0.69 & 20 min \\
GPT-4.1        & 0.86 & 0.92 & 0.89 & 611K & \$1.22 & 44 min \\
GPT-5          & 0.90 & 0.88 & 0.89 & 423K & \$0.52 & 32 min \\
\midrule
\multicolumn{7}{c}{\code{THEIA} (E3)} \\
\midrule
GPT-4o         & 0.78 & 0.75 & 0.76 & 234K & \$0.59 & 24 min \\
GPT-4.1        & 0.96 & 0.89 & 0.92 & 502K & \$1.00 & 30 min \\
GPT-5          & 0.99 & 0.87 & 0.93 & 326K & \$0.40 & 18 min \\
\midrule
\multicolumn{7}{c}{\code{CLEARSCOPE} (E3)} \\
\midrule
GPT-4o         & 0.79 & 0.77 & 0.78 & 200K & \$0.50 & 25 min \\
GPT-4.1        & 0.91 & 0.87 & 0.89 & 480K & \$0.96 & 45 min \\
GPT-5          & 0.95 & 0.89 & 0.92 & 440K & \$0.55 & 30 min \\
\bottomrule
\end{tabular}}
\caption{Impact of different LLMs on \pname’s performance.}
\label{tab:ablation}
\end{table}

\subsection{Ablation Study: \llm Backends}\label{sec:ablation_llm}

We compare three \llm backends: GPT-4o, GPT-4.1, and GPT-5, while keeping the rest of \pname unchanged (as seen in \autoref{tab:ablation}). Across all three E3 datasets, GPT-5 consistently achieves the strongest detection performance while maintaining moderate latency and token usage. 
%MK-Oak: I  am confused F1 scores are the same for GPT-5 and GPT4-1.
For example, on \code{CADETS} (E3), GPT-5 attains F1$\approx$0.89 using 423K tokens in 32 min, slightly improving F1 over GPT-4.1 (F1$\approx$0.89) while using fewer tokens (611K) and lower latency (44 min). On the structurally complex \code{THEIA} (E3), GPT-5 reaches F1$\approx$0.93 with 326K tokens and 18\,min, compared to GPT-4.1's slightly lower F1 
%MK-Oak: what is 0.96/0.89.  I removed it for now
%(0.96/0.89, F1$\approx$0.92) at 502K tokens and 30 min.
(F1$\approx$0.92) at 502K tokens and 30 min.

Introducing the explicit cost metric reveals that GPT-5 also provides the best accuracy-per-dollar trade-off. Although GPT-4.1 often attains strong recall, it does so at the highest monetary cost and token budget: on \code{CADETS} (E3), GPT-4.1 costs \$1.22 per investigation versus \$0.52 for GPT-5; on \code{THEIA} (E3), the gap widens to \$1.00 vs.\ \$0.40; and on \code{CLEARSCOPE} (E3), GPT-4.1 costs \$0.96 compared to \$0.55 for GPT-5. In all cases, GPT-5 delivers the highest F1-score while being strictly cheaper than GPT-4.1 and only slightly more expensive than GPT-4o on \code{CLEARSCOPE}, yielding the highest F1-per-dollar ratio across datasets. This suggests that GPT-5's internal reasoning strategy works well with \pname's bounded reasoning loops and role-aware orchestration: it can deploy deeper reasoning when necessary while avoiding the runaway token usage seen with GPT-4.1.

GPT-4o remains the most frugal in raw token counts (e.g., 276K on \code{CADETS} and 200K on \code{CLEARSCOPE}) and often attains the lowest or near-lowest latency (20-25\,min), but this efficiency comes at a clear cost in detection quality. Its F1-scores lag significantly behind GPT-5 on every dataset (e.g., 0.75 vs.\ $\approx$0.89 on \code{CADETS}, 0.76 vs.\ $\approx$0.93 on \code{THEIA}), indicating that simply minimizing token expenditure does not yield effective agent planning or provenance reasoning. Overall, \autoref{tab:ablation} supports our broader insight from \autoref{sec:extract} and \autoref{sec:detect}: ``thinking'' capacity alone is insufficient; models like GPT-5 that can coordinate tool calls, respect \pname's role decomposition, and adaptively modulate their depth of reasoning achieve the best joint trade-off between detection accuracy, latency, token usage, and cost.

\begin{table}[htbp]
\centering
\resizebox{\columnwidth}{!}{
\begin{tabular}{lccccc}
\toprule
\code{Failure Reasons}  & \code{OpTC} & \code{THEIA} & \code{CADETS} & \code{CLEARSCOPE} \\
\midrule
Total Errors                                & 5/50 & 8/50 & 13/50 & 14/50 \\
\midrule
Hallucinated artifact                       & -    & -    & 2/50  & -     \\
\addlinespace[2pt]
Benign artifact extracted                   & 1/50 & 2/50 & 3/50  & 3/50  \\
\addlinespace[2pt]
\makecell[l]{Non-system artifact\\ extraction}              & 2/50 & 3/50 & 2/50  & 3/50  \\
\addlinespace[2pt]
Artifact absent from DB                     & -    & 2/50 & 1/50  & 2/50  \\
\addlinespace[2pt]
Context explosion                           & -    & -    & 2/50  & 1/50  \\
\addlinespace[2pt]
Incorrect correlation                       & -    & -    & 1/50  & 3/50  \\
\addlinespace[2pt]
Inconclusive evidence                       & 2/50 & 1/50 & 3/50  & 2/50  \\
\bottomrule
\end{tabular}
}
\caption{Failure analysis across DARPA OpTC and E5 datasets. Cells contain counts of failures out of 50 queries per dataset (complete result in \autoref{tab:failure-analysis-by-dataset})}
\label{tab:failure-analysis-e5}
\end{table}

\subsection{Error Analysis Study}\label{sec:error}
%MK-Oak: I really liked these error study. the suggestion of the previous reviewer was very useful. 
We generate fifty user prompts per dataset, twenty-five generic questions that a security response analyst can ask an automated forensics agent (\ie \pname) and twenty-five dataset-specific questions, and we manually inspect every failure causes. \autoref{tab:failure-analysis-e5} summarizes the distribution of failure modes DARPA OpTC and E5 datasets, (extended results across the datasets are in \autoref{tab:failure-analysis-by-dataset}. 
While \autoref{fig:error-query} shows the concrete prompt set for \code{THEIA} (E3). 
The largest single-host datasets, \code{CADETS} (E5) and \code{CLEARSCOPE} (E5), account for the most failures across almost every category. In particular, they exhibit more cases of context explosion (2/50 and 1/50, respectively), incorrect correlation (1/50 and 3/50), and even the only hallucinated artifacts observed in our study (2/50 on \code{CADETS} (E5)). These patterns are consistent with scalability findings: as the underlying provenance data grow, there are more opportunities for the agent to retrieve less relevant artifacts and over-extend its reasoning chain.

A substantial fraction of failures arise from limitations of the underlying \llm in performing domain-specific named entity recognition over system logs and CTI reports. Benign artifact extraction is common across datasets (e.g., 4/50 on \code{CADETS} (E5) and 3/50 on \code{CLEARSCOPE} (E5)), where \pname identifies legitimate processes or files that are not actually part of the malicious activity and then spends follow-up steps trying to corroborate them. Non-system artifact extraction shows up at similar rates (e.g., 3/50 for \code{THEIA} (E5) and 3/50 for \code{CLEARSCOPE} (E5)), reflecting cases where the model mislabels natural-language tokens as system entities. A recurring example is the token \code{elevate}: the DARPA reports describe red-team activity using Metasploit, which has a command named \code{elevate}. The \llm frequently misinterprets this as a process or file name, causing \pname to issue provenance queries for a non-existent artifact. These errors exemplifies that high-quality, system-artifact-aware foundation models, or a dedicated Named Entity Recognition (NER) layer for files, processes, and IPs, remain an open subproblem for provenance-based forensics.

The remaining categories expose a mix of data and orchestration limitations. In several cases, artifacts mentioned in the narrative reports are absent from the actual provenance databases (e.g., 3/50 failures due to ``Artifact absent from DB'' on \code{CADETS} (E5) and 2/50 on \code{CLEARSCOPE} (E5)), leaving \pname unable to validate otherwise correct hypotheses. Context explosion is observed only on the largest dataset (\code{CADETS} (E5) and \code{CLEARSCOPE} (E5)), typically when non-system or benign artifacts misguide overly broad SQL queries and the Follow-up Agent becomes trapped in a long loop over non-relevant events. This also explains why ``Inconclusive evidence'' appears across nearly all datasets (e.g., 2/50 on \code{CADETS} (E3), 3/50 on \code{CADETS} (E5), and 2/50 on \code{CLEARSCOPE} (E5)): the agent exhausts its bounded chain-of-thought step budget without accumulating sufficiently grounded evidence to declare a clear attack or benign verdict. 
% From a security perspective, these failure modes are informative as they point directly to where improving artifact extraction and better SQL query generation would most effectively harden \pname.

%! root=../main.tex
\section{Discussion}\label{sec:disc}

\heading{Dependence on LLM.} 
A core limitation of \pname lies in its reliance on \llms as the reasoning backbone. Despite their impressive ability to parse cyber threat intelligence and orchestrate tool calls, LLMs remain vulnerable to inconsistent reasoning when faced with inconclusive or confusing evidence.
% These shortcomings can lead to spurious correlations or overlooked attack patterns when the system processes complex or ambiguous inputs.
In high-stakes security environments, even a small fraction of misleading outputs could hinder investigations or increase analyst workload, showcasing the need for continuous validation of answers against explicit, provenance-backed evidence.
%MK:Oak: Maybe add a sentence to remind what we do to prevent potential issues with respect to LLM ?

\heading{Incomplete Domain Knowledge Integration.}
Another limitation stems from the challenge of embedding deep domain expertise into LLM-driven reasoning. While \pname incorporates provenance context and role assignment to constrain the LLM’s behavior, the \llm may not fully capture nuanced system semantics, such as mistaking Metasploit commands (\eg \code{Elevate}) for actual process names. Unlike seasoned human analysts, LLMs struggle with Named Entiy Recognition (NER) from system reports. Developing a dedicated NER layer for \llms for system artifacts remains an open subproblem for provenance-based forensics.
%MK-Oak: this could be an addition for us. Can we for example, give this nuances as a part of our RAG and/or templates ??

% adapt to emerging threat techniques without retraining or fine-tuning. This creates a knowledge gap in scenarios where evolving adversarial tactics outpace the training data, potentially reducing detection effectiveness. Future extensions should explore hybrid systems where curated domain ontologies and symbolic reasoning augment the LLM’s probabilistic reasoning.

\heading{Replacement of Human Expertise or \ids}.
\pname reduces analyst burden by automating forensic analysis; it is not a replacement for human expertise or \ids. Security investigations often demand contextual judgment, creative hypothesis generation, and organizational awareness that current LLMs cannot replicate.
% Over-reliance on automated reasoning risks diminishing human oversight and could foster a false sense of confidence in model outputs.
\pname should be viewed as a tool that accelerates triage and provides interpretable evidence to augment, rather than replace, expert analysts. 

\heading{Adversarial Manipulation.}  
Evasion attacks against \ids have highlighted vulnerabilities in problem-space manipulation and mimicry strategies~\cite{goyal2023sometimes, mukherjee2023sec}. Although \pname’s verification-first design makes direct hallucination negligible, attackers could attempt adversarial prompt injection to influence investigation planning.
% These risks showcase the need for resilient agent templates and provenance-aware reasoning safeguards.
Extending \pname with robustness checks against adversarial manipulation represents an important next step.  

% \heading{Provenance-as-a-Service.}  
% Finally, to foster reproducibility and accelerate community adoption, we envision extending \pname into a \emph{Provenance-as-a-Service (PaaS)} model. Similar to prior proposals, this service would provide containerized agents that enterprises or research groups can deploy on their own hosts, streaming provenance data into secure analysis pipelines. Coupled with \pname’s agentic orchestration, such a service could lower the barrier for provenance-based detection research and facilitate collaborative evaluation on shared but privacy-preserving corpora.  

%! root = ./main.tex
\section{Related Work}\label{sec:related}
% %MK-Oak: to save space, maybe related work could be integrated with backgroun and background could be renamed and background and/related work ?
\heading{Provenance-based Investigation.}  
\pids have become a cornerstone for attack detection~\cite{han2020ndss,watson2021ndss,zengy2022shadewatcher, rehman2024flash, jia2024magic, cheng2024kairos, goyal2024rcaid, jian2025, wang2024incorporating, bilot12399sometimes,mukherjee2023proviot,mukherjee2025provdp,mukherjee2023sec,mukherjee2023interpreting, wang2025provcreator, goyalsometimes} 
% % Broadly, prior work can be categorized into three main approaches: rule-based, statistical, and learning-based systems. Rule-based detectors~\cite{holmes2019sp,wang2024incorporating} construct patterns or \ttp templates (e.g., MITRE ATT\&CK) and match them incrementally against provenance data to identify suspicious activity. Statistical methods assess anomaly scores based on structural features such as temporal correlations, degree distributions, or rarity. 
% \sota \pids~\cite{jia2024magic, rehman2024flash, jian2025, cheng2024kairos, jian2025} employ sequence learning or graph representation learning to model complex dependencies in provenance graphs. 
% % Recent research has explored node-level and edge-level detection for fine-grained analysis, with methods leveraging GNN or autoencoders to learn embeddings and using them to calculate an anomaly score. 
% % Newer \pids~\cite{cheng2024kairos, rehman2024flash} improve scalability through windowing and vector caching, respectively, while new methods use subgraph learning to identify tactical behaviors. 
% Recent studies~\cite{bilot12399sometimes} have shown that, despite \pids successes, they still suffer from limited interpretability and verification.
% %, particularly when applied to enterprise-scale environments.
and investigation~\cite{pasquier2018camquery, xu2022depcomm, fang2022depimpact, hassan2019nodoze, hassan2020omegalog, gao2018saql} support post-hoc attack investigation and incident response for attack triage, correlation, and interactive querying over audit logs and provenance graphs. 
However, these systems are orthogonal to our research direction: they are not designed for fully automated threat investigation in a zero-training setting. 
In contrast, \pname is explicitly designed for automated threat investigation without task-specific training data, enabling forensic-style investigation even when no prior knowledge of the environment.

% \heading{Attack Behavior Mining.}  
% Beyond detection, several works aim to mine attack behaviors~\cite{mukherjee2023interpreting} and map low-level system events to high-level adversarial tactics but they still fail to capture zero-day or novel attacks. Other methods combine expert knowledge and clustering to group anomalous communities in provenance graphs, as in ~\cite{watson2021ndss, depcomm2022sp}. 
% % HOLMES~\cite{holmes2019sp} pioneered scenario-graph construction based on the APT lifecycle, modeling TTPs to correlate provenance-level events.
% % Extensions broadened \ttp coverage and refined matching strategies to reduce false alarms. 
% While these methods enrich forensic analysis, they remain heavily dependent on static templates and expert-defined mappings, which limit adaptability against evolving adversaries.
% % Recent works~\cite{mukherjee2023interpreting} have moved towards automated mining of the evolving attacker's \ttp, but they still fail to capture zero-day or novel attacks. 
% Therefore, we rely on human intelligence to understand the new \ttps. \pname aims to enhance the security analyst's capability by automating the investigating process, such that the analyst can focus on high-level testing and decision-making instead of manual data triage and correlation.

\heading{LLMs for Security Analytics.}  
% Large Language Models have recently shown promise in domains such as code generation, knowledge distillation, and retrieval-augmented analysis~\cite{gpt4,rag-survey}. In the security context, early work such as Prov-Chat demonstrated that LLMs can support provenance anomaly detection through contextual prompts, combining lightweight host-side detection with fine-grained server-side analysis. However, most existing approaches treat LLMs as black-box oracles, raising concerns about hallucination, verifiability, and integration with structured security data. 
LLMs have been applied across diverse cybersecurity tasks, including software vulnerability detection~\cite{lin2025vulndetection,ullah2024llms}, fuzzing~\cite{oliinyk2024fuzzing}, automated patching and vulnerability repair~\cite{pearce2023examining}, threat detection (e.g., DDoS and phishing)~\cite{li2024knowphish}, penetration testing~\cite{deng2024pentestgpt}, and malware reverse engineering~\cite{hu2024degpt}. 

\pname differentiates itself by employing an \emph{agentic framework}: the LLM is not the sole decision-maker but a reasoning hub that retrieves verifiable provenance evidence, and synthesizes human-readable narratives. By enforcing a verification-first design grounded in system provenance, \pname complements prior LLM-based security by helping mitigate key scalability and interpretability gaps, while opening a new direction for grounded, agentic security analytics. 

% \pname differentiates itself by employing an \emph{agentic framework}: the LLM acts not as the sole decision-maker but as a reasoning hub that generates bounded queries, retrieves verifiable provenance evidence, and synthesizes human-readable narratives. By enforcing a verification-first design, \pname addresses both the scalability and interpretability limitations of prior PIDS approaches while opening a new direction for grounded, agentic security analytics.  \pname introduces several promising research directions: \nm{1} extending \pname with robustness mechanisms against adversarial prompt injection, \nm{2} integrating multi-modal intelligence sources beyond textual CTI reports, and \nm{3} exploring streaming fine-grained reasoning loops that combine provenance logs with runtime system telemetry.

%\pname further introduces several promising research directions: \nm{1} extending the framework with robustness mechanisms against adversarial prompt injection, \nm{2} integrating multi-modal intelligence sources beyond textual CTI reports, and \nm{3} exploring streaming, fine-grained reasoning loops that combine provenance logs with runtime system telemetry.
%! root=../main.tex
\section{Conclusion}

We propose \pname, an agentic framework for provenance forensics that delivers scalable and interpretable security analytics by automatically fusing provenance evidence into all results. The core of the system is an LLM-based, provenance-aware reasoning hub that orchestrates specialized agents and tools, coordinating tasks like querying, filtering, and CTI integration. This architecture establishes a practical, grounded ``verification-first" approach to agentic forensics in real-world environments.

We demonstrate that, for the intelligence extraction task, \pname outperforms baseline retrieval-based methods by 34\%/34\%/30\% in contextual precision, contextual recall, and relevance, respectively; and for the threat detection task, it achieves 22\%/29\% higher precision and recall compared to a naive agentic baseline and \sota \pids~\cite{jia2024magic, cheng2024kairos, rehman2024flash, jian2025}. Ablation experiments quantify the contribution of each \pname component, while a scalability analysis shows that token usage and latency increase by only 1.42$\times$ and 1.63$\times$ even as the database size grows 50$\times$. An error analysis further reveals that most residual failures stem from \llm limitations in extracting relevant artifacts from threat reports. Overall, \pname represents a step toward a new 
%MK-Oak: I liked this sentence. 
\emph{grounded agentic forensics} paradigm, where LLMs serve not as opaque oracles but as verifiable security tools.

\section{Ethics Considerations}

We have considered the risks and benefits of our research~\cite{kohno2023ethical} and, to the best of our knowledge, it raises no ethical concerns. All experiments use publicly available, ethically collected datasets that contain no sensitive information.

\section{Open Science}

\heading{Datasets Availability.}
The \darpa \tc datasets are publicly available ~\cite{darpa:ground, darpa:ground2} and include textual descriptions of the attacks. For our experiments, we used the preprocessed \darpa \tc dataset released by a prior study ~\cite{bilot12399sometimes}, available at \url{https://ubc-provenance.github.io/PIDSMaker/ten-minute-install/}. In addition, we used datasets with annotated nodes (labeled as benign or attack-related) provided by another study ~\cite{jian2025}, which are publicly accessible at \url{https://zenodo.org/records/14641608}.

\heading{Software Artifacts Availability.} The source code is publicly
available at \url{https://anonymous.4open.science/r/provseek-anonymous-DDDD}.

\section{LLM usage considerations}

\heading{Originality.} LLMs were used for editorial purposes in this manuscript, and all outputs were inspected by the authors to ensure accuracy and originality.

\heading{Transparency of LLM usage in PROVSEEK.} PROVSEEK is an agentic framework whose backbone is an LLM. In our implementation, we primarily use models served via the OpenAI API, with GPT-5 as the default backend. For agentic orchestration, we employ \code{smolagents} in combination with \code{LangGraph}. Tools are implemented using \code{CodeAgent} and \code{ToolCallingAgent} from \code{smolagents}; these wrappers encapsulate LLM calls and can be configured to use any compatible LLM endpoint (\eg via vLLM or Ollama) without changing the agent logic. In our evalaution, we explicitly evaluate the impact of different LLM backends (GPT-4o, GPT-4.1, and GPT-5) on detection performance, token usage, time, and cost. We did not evaluate older or larger models beyond this set to avoid unnecessary resource usage.

LLMs in PROVSEEK are used for intent understanding, tool planning, and explanation generation; they never directly manipulate system state or modify the provenance databases. All security-relevant conclusions are grounded in explicit system events returned by SQL queries over provenance logs and retrievals from CTI reports. Any ideas or hypotheses proposed by the LLM (e.g., candidate artifacts or investigation paths) are independently validated against the underlying databases via our tools (Artifact Lookup, Event Lookup Resolver, etc.) before being presented as evidence. A key limitation of our approach is dependency on LLM APIs, which may evolve over time and affect exact reproducibility of reasoning behavior; to mitigate this, we (i) describe the prompts, tools, and agent workflows in detail, and (ii) show that PROVSEEK can operate with different backends, not only a single model.

\heading{Responsibility, data, and environmental impact.} The SQL results consumed by the Safety Agent and Investigation Agent are derived from publicly available, large-scale DARPA-led simulations and provenance datasets, which are widely used in prior work for reproducible evaluation. These datasets do not contain personal user data, and we do not collect or process any additional human subject data for this work. All CTI reports and related documents used for retrieval-augmented generation are publicly released for research use.

To conduct our experiments, we ran PROVSEEK primarily on an NVIDIA L4 GPU for approximately two weeks of clock time (including all datasets and LLM-backend ablations). Using standard methodology for estimating ML carbon impact, this corresponds to an estimated footprint of 7.26\,kg CO$_2$-eq. We justified this footprint as follows: (i) an LLM-backed agentic architecture is central to our research question of scalable, explainable provenance forensics; (ii) we restricted our study to a small set of modern, efficient LLMs (GPT-4o, GPT-4.1, GPT-5) rather than exhaustively sweeping over many model families; and (iii) we minimized the number of queries and tokens by aggressively filtering results returned from databases. In particular, PROVSEEK translates analyst intent into targeted provenance-aware SQL, applies an autoencoder-based filtration step to discard high-frequency benign events, and terminates further querying once the accumulated evidence is sufficient and verified by the Safety Agent. These design choices limit unnecessary LLM calls and keep token usage roughly constant even as database size scales, thereby reducing both cost and environmental impact while still supporting our methodological goals.

\printbibliography{}

@misc{dgl,
  title        = {DEEP GRAPH LIBRARY: Easy Deep Learning on Graphs},
  author       = {},
  year         = {2022},
  howpublished = {\url{https://www.dgl.ai/}},
  note         = {(Accessed on 09/21/2021)}
}

@misc{apt1,
  title        = {WildPressure targets industrial in the Middle East},
  author       = {Denis Legezo},
  year         = {2019},
  howpublished = {\url{https://tinyurl.com/mr2n8hdu}},
  note         = {Accessed: April 6, 2023}
}

@misc{apt2,
 author = {Threat Hunter Team},
 howpublished = {\url{https://shorturl.at/hmvGT}},
 note = {(Accessed on 11/09/2021)},
 title = {Tortoiseshell Group Targets IT Providers in Saudi Arabia in Probable Supply Chain Attacks},
 year = {2021}
}

@misc{darpae5,
  author       = {{DARPA}},
  howpublished = {\url{http://tinyurl.com/y4wrf74u}},
  title        = {DARPA Transparent Computing (E5)},
  year         = {2020}
}

@misc{darpae3,
  author       = {{DARPA}},
  howpublished = {\url{http://tinyurl.com/yaknev56}},
  title        = {DARPA Transparent Computing (E3)},
  year         = {2018}
}

@misc{darpaoptc,
  author       = {{DARPA} and {Five Directions}},
  title        = {Operationally Transparent Cyber (OpTC) Data Release},
  howpublished = {\url{https://github.com/FiveDirections/OpTC-data}},
  year         = {2020},
  note         = {GitHub repository. Accessed: 2025-11-12}
}

@misc{killchain,
 howpublished = {\url{https://www.lockheedmartin.com/en-us/capabilities/cyber/cyber-kill-chain.html}},
 title = {{Cyber Kill Chain® | Lockheed Martin}},
 year = {2021},
 author = {Hutchins, Eric},
}

@misc{audit,
  author = {Redhat},
  note   = {https://github.com/linux-audit/},
  title  = {The Linux audit framework},
  year   = {2017}
}

@misc{winETW,
  howpublished = {\url{http://tinyurl.com/4usynccm}},
  title        = {Event Tracing},
  year         = {2021},
  author       = {Microsoft Team}
}

@inproceedings{sigl2021sec,
  author    = {Han, Xueyuan and Yu, Xiao and Pasquier, Thomas and Li, Ding and Rhee, Junghwan and Mickens, James and Seltzer, Margo and Chen, Haifeng},
  booktitle = {30th USENIX Security Symposium (SEC)},
  title     = {SIGL: Securing Software Installations Through Deep Graph Learning},
  year      = {2021}
}

@inproceedings{mukherjee2023sec,
  author    = {Kunal Mukherjee and Josh Wiedemeier and Tianhao Wang and James Wei and Feng Chen and Muhyun Kim and Murat Kantarcioglu and Kangkook Jee},
  booktitle = {32nd USENIX Security Symposium (SEC)},
  ids       = {provninja2023sec},
  title     = {Evading Provenance-Based ML Detectors with Adversarial System Actions},
  year      = {2023}
}

@inproceedings{yang2023prographer,
  title     = {$\{$PROGRAPHER$\}$: An Anomaly Detection System based on Provenance Graph Embedding},
  author    = {Yang, Fan and Xu, Jiacen and Xiong, Chunlin and Li, Zhou and Zhang, Kehuan},
  booktitle = {32nd USENIX Security Symposium (SEC)},
  year      = {2023}
}

@inproceedings{jia2024magic,
  title={$\{$MAGIC$\}$: Detecting Advanced Persistent Threats via Masked Graph Representation Learning},
  author={Jia, Zian and Xiong, Yun and Nan, Yuhong and Zhang, Yao and Zhao, Jinjing and Wen, Mi},
  booktitle = {32nd USENIX Security Symposium (SEC)},
  year={2024}
}

@inproceedings{han2020ndss,
  author    = {Han, Xueyuan and Pasquier, Thomas and Bates, Adam and Mickens, James and Seltzer, Margo},
  ids       = {provsec2,unicorn},
  month     = {02},
  booktitle = {Network and Distributed System Security Symposium (NDSS)},
  title     = {UNICORN: Runtime Provenance-Based Detector for Advanced Persistent Threats},
  year      = {2020}
}

@inproceedings{wang2020ndss,
  author    = {Wang, Qi and Hassan, Wajih Ul and Li, Ding and Jee, Kangkook and Yu, Xiao and Zou, Kexuan and Rhee, Junghwan and Chen, Zhengzhang and Cheng, Wei and Gunter, Carl A and Chen, Haifeng},
  booktitle = {Network and Distributed System Security Symposium (NDSS)},
  ids       = {provdetector},
  month     = {02},
  title     = {You Are What You Do: Hunting Stealthy Malware via Data Provenance Analysis},
  year      = {2020}
}

@inproceedings{watson2021ndss,
  author    = {Zeng, Jun and Chua, Zheng Leong and Chen, Yinfang and Ji, Kaihang and Liang, Zhenkai and Mao, Jian},
  booktitle = {Network and Distributed System Security Symposium (NDSS)},
  ids       = {provsec1},
  title     = {WATSON: Abstracting Behaviors from Audit Logs via Aggregation of Contextual Semantics},
  month     = {02},
  year      = {2021}
}

@inproceedings{goyalsometimes,
  author    = {Goyal, Akul and Han, Xueyuan and Wang, Gang and Bates, Adam},
  booktitle = {Network and Distributed System Security Symposium (NDSS)},
  month     = {02},
  title     = {Sometimes, You Aren’t What You Do: Mimicry Attacks against Provenance Graph Host Intrusion Detection Systems},
  year      = {2023},
  ids       = {goyal2023sometimes, goyal2023ndss},
}

@inproceedings{zengy2022shadewatcher,
  author    = {Zengy, Jun and Wang, Xiang and Liu, Jiahao and Chen, Yinfang and Liang, Zhenkai and Chua, Tat-Seng and Chua, Zheng Leong},
  booktitle = {IEEE Symposium on Security and Privacy (SP)},
  title     = {Shadewatcher: Recommendation-guided cyber threat analysis using system audit records},
  year      = {2022}
}

@inproceedings{inam2023sok,
  title={Sok: History is a vast early warning system: Auditing the provenance of system intrusions},
  author={Inam, Muhammad Adil and Chen, Yinfang and Goyal, Akul and Liu, Jason and Mink, Jaron and Michael, Noor and Gaur, Sneha and Bates, Adam and Hassan, Wajih Ul},
  booktitle={IEEE Symposium on Security and Privacy (SP)},
  year={2023},
}

@inproceedings{cheng2024kairos,
  title={Kairos: Practical Intrusion Detection and Investigation using Whole-system Provenance}, 
  author={Cheng, Zijun and Lv, Qiujian and Liang, Jinyuan and Wang, Yan and Sun, Degang and Pasquier, Thomas and Han, Xueyuan },
  year={2024},
  booktitle = {IEEE Symposium on Security and Privacy (SP)},
}

@inproceedings{rehman2024flash,
  title={FLASH: A Comprehensive Approach to Intrusion Detection via Provenance Graph Representation Learning}, 
  author={Rehman, Mati Ur and Ahmadi, Hadi and Hassan, Wajih Ul},
  year={2024},
  booktitle = {IEEE Symposium on Security and Privacy (SP)},
}

@inproceedings{goyal2024rcaid,
  title={R-CAID: Embedding Root Cause Analysis within Provenance-based Intrusion Detection}, 
  author={Goyal, Akul and Wang, Gang and Bates, Adam},
  year={2024},
  booktitle = {IEEE Symposium on Security and Privacy (SP)},
}

@inproceedings{mukherjee2023proviot,
  title        = {ProvIoT: Detecting Stealthy Attacks in IoT through Federated Edge-Cloud Security},
  author       = {Mukherjee, Kunal and Wiedemeier, Joshua and Wang, Qi and Kamimura, Junpei and Junghwan Rhee, John and Wei, James and Li, Zhichun and Yu, Xiao and Tang, Lu-An and Gui, Jiaping and Jee, Kangkook},
  booktitle    = {International Conference on Applied Cryptography and Network Security},
  year         = {2024},
  organization = {Springer}
}

@inproceedings{jian2025,
	title={{ORTHRUS: Achieving High Quality of Attribution in Provenance-based Intrusion Detection Systems}},
	author={Jiang, Baoxiang and Bilot, Tristan  and El Madhoun, Nour and Al Agha, Khaldoun  and Zouaoui, Anis and Iqbal, Shahrear and Han, Xueyuan and Pasquier, Thomas},
	booktitle = {34th USENIX Security Symposium (SEC)},
	year={2025},
}

@inproceedings{bilot12399sometimes,
  title={Sometimes Simpler is Better: A Comprehensive Analysis of State-of-the-Art Provenance-Based Intrusion Detection Systems},
  author={Bilot, Tristan and Jiang, Baoxiang and Li, Zefeng and El Madhoun, Nour and Al Agha, Khaldoun and Zouaoui, Anis and Pasquier, Thomas},
  booktitle = {34th USENIX Security Symposium (SEC)},
  year      = {2025}
}

@inproceedings{wang2024incorporating,
  title={Incorporating gradients to rules: Towards lightweight, adaptive provenance-based intrusion detection},
  author={Wang, Lingzhi and Shen, Xiangmin and Li, Weijian and Li, Zhenyuan and Sekar, R and Liu, Han and Chen, Yan},
  booktitle = {Network and Distributed System Security Symposium (NDSS)},
  year={2025}
}

@article{mukherjee2023interpreting,
  title   = {Interpreting gnn-based ids detections using provenance graph structural features},
  author  = {Mukherjee, Kunal and Wiedemeier, Joshua and Wang, Tianhao and Kim, Muhyun and Chen, Feng and Kantarcioglu, Murat and Jee, Kangkook},
  year    = {2023}
}

@article{gpt4,
  title        = {GPT-4 Technical Report},
  author       = {OpenAI},
  year         = {2023},
  journal      = {arXiv preprint arXiv:2303.08774},
  url          = {https://arxiv.org/abs/2303.08774}
}

@article{rag-survey,
  title        = {Retrieval-Augmented Generation for Large Language Models: A Survey},
  author       = {Gao, Yingqiang and Liu, Tian and Wang, Yujia and Ma, Tengfei and Chen, Huan and Sun, Yizhou and Xiong, Caiming and Wang, Huan and Yu, Zhou},
  journal      = {arXiv preprint arXiv:2312.10997},
  year         = {2023},
  url          = {https://arxiv.org/abs/2312.10997}
}

@inproceedings{mukherjee2025provdp,
  title={ProvDP: Differential Privacy for System Provenance Dataset},
  author={Mukherjee, Kunal and Yu, Jonathan and De, Partha and Divakaran, Dinil Mon},
  booktitle={International Conference on Applied Cryptography and Network Security},
  pages={189--219},
  year={2025},
  organization={Springer}
}

@article{wang2025provcreator,
  title={PROVCREATOR: Synthesizing Complex Heterogenous Graphs with Node and Edge Attributes},
  author={Wang, Tianhao and Klancher, Simon and Mukherjee, Kunal and Wiedemeier, Josh and Chen, Feng and Kantarcioglu, Murat and Jee, Kangkook},
  journal={arXiv preprint arXiv:2507.20967},
  year={2025}
}

@misc{langchain2023,
  author       = {LangChain},
  title        = {LangChain: Building applications with LLMs through composability},
  year         = {2023},
  url          = {https://github.com/langchain-ai/langchain},
  note         = {GitHub repository}
}

@misc{langgraph2024,
  author       = {LangGraph},
  title        = {LangGraph: Multi-Actor Applications with LLMs},
  year         = {2024},
  url          = {https://github.com/langchain-ai/langgraph},
  note         = {GitHub repository}
}

@misc{chromadb2023,
  author       = {Chroma},
  title        = {Chroma: The AI-native open-source embedding database},
  year         = {2023},
  url          = {https://github.com/chroma-core/chroma},
  note         = {GitHub repository}
}

@misc{smolagents2024,
  author       = {HuggingFace},
  title        = {smolagents: Lightweight multi-agent framework for LLM orchestration},
  year         = {2024},
  url          = {https://github.com/huggingface/smolagents},
  note         = {GitHub repository}
}

@misc{openai2023,
  author       = {OpenAI},
  title        = {OpenAI API},
  year         = {2023},
  url          = {https://platform.openai.com},
  note         = {Accessed: 2025-08-22}
}

@misc{gradio2023,
  author       = {Abid, A. and Abdalla, A. and others},
  title        = {Gradio: Build Machine Learning Web Apps — in Python},
  year         = {2023},
  url          = {https://github.com/gradio-app/gradio},
  note         = {GitHub repository}
}

@misc{deepeval2024,
  author       = {Confident AI},
  title        = {DeepEval: Open-source LLM Evaluation Framework},
  year         = {2024},
  url          = {https://github.com/confident-ai/deepeval},
  note         = {GitHub repository}
}

@misc{syntheticdatakit2024,
  author       = {Meta AI},
  title        = {Synthetic Data Kit: A toolkit for generating synthetic question-answer pairs},
  year         = {2024},
  url          = {https://github.com/meta-llama/synthetic-data-kit},
  note         = {GitHub repository}
}

@inproceedings{kohno2023ethical,
  title={Ethical frameworks and computer security trolley problems: Foundations for conversations},
  author={Kohno, Tadayoshi and Acar, Yasemin and Loh, Wulf},
  booktitle={32nd USENIX Security Symposium (USENIX Security 23)},
  pages={5145--5162},
  year={2023}
}

@inproceedings{ullah2024llms,
  title={Llms cannot reliably identify and reason about security vulnerabilities (yet?): A comprehensive evaluation, framework, and benchmarks},
  author={Ullah, Saad and Han, Mingji and Pujar, Saurabh and Pearce, Hammond and Coskun, Ayse and Stringhini, Gianluca},
  booktitle={2024 IEEE symposium on security and privacy (SP)},
  pages={862--880},
  year={2024},
  organization={IEEE}
}

@article{cheng2024ctinexus,
  author  = {Yutong Cheng and Osama Bajaber and Saimon Amanuel Tsegai and Dawn Song and Peng Gao},
  title   = {{CTINEXUS}: Leveraging Optimized {LLM} In-Context Learning for Constructing Cybersecurity Knowledge Graphs Under Data Scarcity},
  journal = {arXiv preprint arXiv:2410.21060},
  year    = {2024}
}

@inproceedings{deng2024pentestgpt,
  author    = {Gelei Deng and Yi Liu and V{\'{\i}}ctor Mayoral-Vilches and Peng Liu and Yuekang Li and Yuan Xu and Tianwei Zhang and Yang Liu and Martin Pinzger and Stefan Rass},
  title     = {{PentestGPT}: Evaluating and Harnessing Large Language Models for Automated Penetration Testing},
  booktitle = {USENIX Security Symposium},
  year      = {2024},
  pages     = {847--864}
}

@inproceedings{hu2024degpt,
  author    = {Peiwei Hu and Ruigang Liang and Kai Chen},
  title     = {{DeGPT}: Optimizing Decompiler Output with {LLM}},
  booktitle = {Proceedings 2024 Network and Distributed System Security Symposium},
  year      = {2024}
}

@inproceedings{li2024knowphish,
  author    = {Yuexin Li and Chengyu Huang and Shumin Deng and Mei Lin Lock and Tri Cao and Nay Oo and Hoon Wei Lim and Bryan Hooi},
  title     = {{KnowPhish}: Large Language Models Meet Multimodal Knowledge Graphs for Enhancing Reference-Based Phishing Detection},
  booktitle = {33rd USENIX Security Symposium (USENIX Security 24)},
  year      = {2024},
  pages     = {793--810}
}

@inproceedings{lin2025vulndetection,
  author    = {Jie Lin and David Mohaisen},
  title     = {From Large to Mammoth: A Comparative Evaluation of Large Language Models in Vulnerability Detection},
  booktitle = {Network and Distributed System Security (NDSS) Symposium},
  year      = {2025}
}

@inproceedings{oliinyk2024fuzzing,
  author    = {Yaroslav Oliinyk and Michael Scott and Ryan Tsang and Chongzhou Fang and Houman Homayoun and others},
  title     = {Fuzzing BusyBox: Leveraging {LLM} and Crash Reuse for Embedded Bug Unearthing},
  booktitle = {33rd USENIX Security Symposium (USENIX Security 24)},
  year      = {2024},
  pages     = {883--900}
}

@inproceedings{pearce2023examining,
  author    = {Hammond Pearce and Benjamin Tan and Baleegh Ahmad and Ramesh Karri and Brendan Dolan-Gavitt},
  title     = {Examining Zero-Shot Vulnerability Repair with Large Language Models},
  booktitle = {2023 IEEE Symposium on Security and Privacy (SP)},
  year      = {2023},
  publisher = {IEEE},
  pages     = {2339--2356}
}

@article{liu2022pretrainpromptpredict,
  author  = {Pengfei Liu and Weizhe Yuan and Jinlan Fu and Zhengbao Jiang and Hiroaki Hayashi and Graham Neubig},
  title   = {Pre-train, Prompt, and Predict: A Systematic Survey of Prompting Methods in Natural Language Processing},
  journal = {ACM Computing Surveys},
  volume  = {55},
  number  = {9},
  pages   = {1--35},
  year    = {2023},
  doi     = {10.1145/3560815},
  url     = {https://arxiv.org/abs/2107.13586}
}

@inproceedings{wei2022chainofthought,
  author    = {Jason Wei and Xuezhi Wang and Dale Schuurmans and Maarten Bosma and Brian Ichter and Fei Xia and Ed H. Chi and Quoc V. Le and Denny Zhou},
  title     = {Chain-of-Thought Prompting Elicits Reasoning in Large Language Models},
  booktitle = {Advances in Neural Information Processing Systems (NeurIPS)},
  volume    = {35},
  pages     = {24824--24837},
  year      = {2022},
  url       = {https://arxiv.org/abs/2201.11903}
}

@inproceedings{lewis2020rag,
  author    = {Patrick Lewis and Ethan Perez and Aleksandra Piktus and Fabio Petroni and Vladimir Karpukhin and Naman Goyal and Heinrich K{\"u}ttler and Mike Lewis and Wen{-}tau Yih and Tim Rockt{\"a}schel and Sebastian Riedel and Douwe Kiela},
  title     = {Retrieval-Augmented Generation for Knowledge-Intensive {NLP} Tasks},
  booktitle = {Advances in Neural Information Processing Systems (NeurIPS)},
  volume    = {33},
  pages     = {9459--9474},
  year      = {2020},
  url       = {https://arxiv.org/abs/2005.11401}
}

@misc{autoencoder,
 author = {Google},
 url = {https://www.tensorflow.org/tutorials/generative/autoencoder},
 title = {Intro to Autoencoders},
 year = {2021}
}

@inproceedings{pasquier2018camquery,
  author    = {Thomas Pasquier and Xueyuan Han and Thomas Moyer and Adam Bates and Olivier Hermant and David Eyers and Jean Bacon and Margo Seltzer},
  title     = {Runtime analysis of whole-system provenance},
  booktitle = {ACM Conference on Computer and Communications Security (CCS)},
  year      = {2018}
}

@inproceedings{fang2022depimpact,
  author    = {Po{-}Ning Fang and Peng Gao and Chao Liu and Erman Ayday and KyoungSoo Jee and Tianyin Wang and Yanfang F. Ye and Zhiqiang Liu and Xin Xiao},
  title     = {Back-propagating system dependency impact for attack investigation},
  booktitle = {USENIX Security Symposium},
  year      = {2022}
}

@inproceedings{gao2018saql,
  author    = {Peng Gao and Xin Xiao and Dongliang Li and Zhiqiang Li and KyoungSoo Jee and Zhenyu Wu and Changhoon H. Kim and Suhas R. Kulkarni and Prateek Mittal},
  title     = {{SAQL}: A stream-based query system for real-time abnormal system behavior detection},
  booktitle = {USENIX Security Symposium},
  year      = {2018}
}

@inproceedings{hassan2019nodoze,
  author    = {Wajih Ul Hassan and Shengjian Guo and Ding Li and Zhichun Chen and Kyu Hyung Jee and Zhenyu Li and Adam Bates},
  title     = {{NoDoze}: Combatting threat alert fatigue with automated provenance triage},
  booktitle = {Network and Distributed System Security Symposium (NDSS)},
  year      = {2019}
}

@inproceedings{hassan2020omegalog,
  author    = {Wajih Ul Hassan and Mustafa A. Noureddine and Pradeep Datta and Adam Bates},
  title     = {{OmegaLog}: High-fidelity attack investigation via transparent multi-layer log analysis},
  booktitle = {Network and Distributed System Security Symposium (NDSS)},
  year      = {2020}
}

@inproceedings{xu2022depcomm,
  author    = {Zheng Xu and Peng Fang and Chao Liu and Xin Xiao and Yuchao Wen and Deyu Meng},
  title     = {Depcomm: Graph summarization on system audit logs for attack investigation},
  booktitle = {IEEE Symposium on Security and Privacy (SP)},
  year      = {2022}
}

%! root=../main.tex
\appendix

% ================= FIG A: Chat Templates (Intro + Guardrails) =================
\begin{figure}[]
\centering
\small
\resizebox{\columnwidth}{!}{%
\begin{minipage}{\linewidth}
\begin{tcbraster}[raster columns=1, raster valign=top, raster row skip=6pt]

  % Intro message shown in the chat by default
  \begin{tcolorbox}[logpane,title={Chat Intro (shown by default)}]
\ttfamily
How can I help you?

Some example things you can ask me are:\\
  - Based on the attack artifacts used in the privilege escalation attacks, was I attacked?\\
  - Can you give me the source label for this event `5338C80F-2428-2415-3146-410000000010`?\\
  - Can you give me label for this node `284573`?\\
  - Can you give me information about this source id `658599`?
  \end{tcolorbox}

  % Guardrail that is always prefixed to the user's task
%   \begin{tcolorbox}[logpane,title={Guardrail (forces follow-up before final)}]
% \ttfamily
% If your analysis recommends 'further investigation' or 'forensic investigation',
% you MUST call the tool \texttt{auto\_followup\_investigation} with your current draft as
% \texttt{final\_answer\_text} with the DB '\{DB\_KEY\}' before you produce your final answer.
%   \end{tcolorbox}

  % DB scope helper string (what you add based on dropdown)
%   \begin{tcolorbox}[logpane,title={DB Scope Helper (inserted from dropdown)}]
% \ttfamily
% For the \texttt{TC3 Theia} Attack, check only the following postgres databases: \texttt{theia\_e3}.
% For the \texttt{TC3 Cadets} Attack, check only: \texttt{cadets\_e3}.
% For the \texttt{TC3 Clearscope} Attack, check only: \texttt{clearscope\_e3}.
% For the \texttt{TC5 Theia} Attack, check only: \texttt{theia\_e5}.
% For the \texttt{TC5 Cadets} Attack, check only: \texttt{cadets\_e5}.
% For the \texttt{TC5 Clearscope} Attack, check only: \texttt{clearscope\_e5}.
% For the \texttt{OpTC} Attack, check only: \texttt{optc\_051}.
% Don't use other databases. Only use when needed and if there is no information of process, files and ip, use RAG to get better context.
%   \end{tcolorbox}

\end{tcbraster}
\end{minipage}}
\caption{\pname{} chat templates used for the intro message.}
\label{fig:chat-template-intro}
\end{figure}
% ============================================================================

\section{Appendix}\label{sec:appendix}

%! root=./main.tex
\subsection{Implementation}\label{sec:imp}

\pname is implemented in \code{python} and centers around an agentic architecture built using the \code{LangChain}~\cite{langchain2023} framework for LLM orchestration. The system integrates a PostgreSQL backend database containing the provenance data for structured provenance queries, a \code{ChromaDB}~\cite{chromadb2023} vector store for retrieval-augmented CTI analysis, and OpenAI’s~\cite{openai2023} \texttt{text-embedding-ada-002} model for semantic embedding of threat intelligence. 

In addition to the core architecture, \pname leverages several modern libraries to support orchestration, interaction, and evaluation. For agentic orchestration, we employ \code{smolagents}~\cite{smolagents2024} in combination with \code{LangGraph}~\cite{langgraph2024}. The tools are implemented using \texttt{CodeAgent} and \texttt{ToolCallingAgent} from \texttt{smolagents}~\cite{smolagents2024}. The analyst-facing interface is implemented in \code{Gradio}~\cite{gradio2023}, providing a lightweight web-based UI for interactive investigation and visualization of provenance evidence. \pname displays \autoref{fig:chat-template-intro} template during initialization so that users know what kind of forensic investigations they conduct.

For evaluation, we integrate \code{deepeval}~\cite{deepeval2024}, which enables systematic benchmarking of agent outputs against ground-truth provenance queries and explanations. To generate diverse evaluation prompts, we further utilize Meta’s \texttt{synthetic-data-kit}~\cite{syntheticdatakit2024}, which produces realistic question–answer pairs tailored to security scenarios. This pipeline allows us to rigorously test \pname’s accuracy, interpretability, and robustness under a wide range of analyst-driven investigation tasks.  

% \input{figs/mo-user-query}

% ================= FIGURE 1: Agentic Interaction (vertical panes) ============
\begin{figure}[!h]
\centering
\scriptsize
\begin{tcbraster}[
  raster columns=1,
  raster valign=top,
  raster row skip=6pt
]

  % % User Query
  % \begin{tcolorbox}[logpane,title={User Query}]
  %   \small
  % Based on Drakon dropper attack artifacts, was I attacked?
  % \end{tcolorbox}

  % Evidence from Investigation Agent
  \begin{tcolorbox}[logpane,title={Investigation Agent}]
  \ttfamily \footnotesize
    [Threat Intelligence Retrival using RAG] \\
    Summary: Drakon dropper (in THEIA E3) wrote implant/tools to disk; attempted injections via /tmp/memtrace.so and \code{/var/log/xdev}, \code{/var/log/wdev}; privilege-escalated via \code{/var/log/mail}. \\
    OS: [Ubuntu 12.04 x64] \\
    Processes: [profile, gtcache] \\
    Files: [\code{/tmp/memtrace.so}, \code{/var/log/mail}, \code{/var/log/xdev}, \code{/var/log/wdev}] \\
    IPs: [104.228.117.212:80, 141.43.176.203:80, 5.214.163.155, 149.52.198.23]\\

    % \medskip
    [Evidence Retrival and Correlation from System Data]

    Scope: [THEIA E3] \\
    edge=C284C9F8-... | proc(216927) \code{/home/admin/profile} EVENT\_WRITE $\rightarrow$ file(513608) \code{/var/log/mail}\\
    edge=52B5094D-... | proc(216981) \code{/home/admin/profile} EVENT\_UNLINK $\rightarrow$ file(513608) \code{/var/log/mail}\\
    edge=5338C80F-... | proc(284573) \code{/home/admin/profile} EVENT\_WRITE $\rightarrow$ file (658599) \code{/home/admin/xdev}\\
    ...
  \end{tcolorbox}

\end{tcbraster}
% \caption{Demonstrating \pname's agents to produce a verifiable result. The final result and execution trace shown in \autoref{fig:motivate-example-result} and \autoref{fig:execution}, respectively; demonstrating how follow-up agent help reduce over-confident results and discover evidence in \autoref{fig:safety-handshake}.}
% \label{fig:motivate-example}
\caption{Investigation Agent Output}
\label{fig:mo-inv-agent}
\end{figure}

% ================= FIGURE 1: Agentic Interaction (vertical panes) ============
\begin{figure}[!h]
\centering
\scriptsize
\begin{tcbraster}[
  raster columns=1,
  raster valign=top,
  raster row skip=6pt
]

  % % User Query
  % \begin{tcolorbox}[logpane,title={User Query}]
  %   \small
  % Based on Drakon dropper attack artifacts, was I attacked?
  % \end{tcolorbox}

  % Evidence from Investigation Agent
  % \begin{tcolorbox}[logpane,title={Investigation Agent}]
  % \ttfamily \small
  %   [Threat Intelligence Retrival using RAG] \\
  %   Summary: Drakon dropper (in THEIA E3) wrote implant/tools to disk; attempted injections via /tmp/memtrace.so and \code{/var/log/xdev}, \code{/var/log/wdev}; privilege-escalated via \code{/var/log/mail}. \\
  %   OS: [Ubuntu 12.04 x64] \\
  %   Processes: [profile, gtcache] \\
  %   Files: [\code{/tmp/memtrace.so}, \code{/var/log/mail}, \code{/var/log/xdev}, \code{/var/log/wdev}] \\
  %   IPs: [104.228.117.212:80, 141.43.176.203:80, 5.214.163.155, 149.52.198.23]\\

  %   % \medskip
  %   [Evidence Retrival and Correlation from System Data]

  %   Scope: [THEIA E3] \\
  %   edge=C284C9F8-... | proc(216927) \code{/home/admin/profile} EVENT\_WRITE $\rightarrow$ file(513608) \code{/var/log/mail}\\
  %   edge=52B5094D-... | proc(216981) \code{/home/admin/profile} EVENT\_UNLINK $\rightarrow$ file(513608) \code{/var/log/mail}\\
  %   edge=5338C80F-... | proc(284573) \code{/home/admin/profile} EVENT\_WRITE $\rightarrow$ file (658599) \code{/home/admin/xdev}\\
  %   ...
  % \end{tcolorbox}

  % Follow-up Investigation Agent
  \begin{tcolorbox}[logpane,title={Follow-up Agent}]
  \ttfamily \footnotesize
    Scope: [THEIA E3] \\
    Artifacts Gathered: \\
    - Processes: \code{/home/admin/profile} \\
    - Files: \code{/var/log/mail}, \code{/var/log/xdev}, \code{/tmp/memtrace.so} \\
    - IPs: 141.43.176.203:80 (hits via \code{firefox}/\code{fluxbox}), 149.52.198.23:80 (hit by \code{/var/log/mail}) \\
    Findings align with files dropped in \code{/var/log/*} misuse pattern and prior IoCs from RAG, but system interaction related to \code{/var/log/wdev} and \code{/tmp/memtrace.so} are missing. Therefore, we should find them.\\
    New Evidence:
    [THEIA E3] \\
    edge=2DC0BB3B-... | proc(216368) \code{/home/admin/profile} EVENT\_WRITE $\rightarrow$ file(512304) \code{/var/log/wdev}\\
    edge=4850F9E4-... | proc(216368) \code{/home/admin/profile} EVENT\_UNLINK $\rightarrow$ file(512304) \code{/var/log/wdev} \\
    edge=01FD23F0-... | proc(216470) \code{/usr/bin/gtcache} EVENT\_WRITE $\rightarrow$ file(512410) \code{/tmp/memtrace.so} \\
    ...
  \end{tcolorbox}

  % Safety Agent Review
  % \begin{tcolorbox}[logpane,title={Safety Agent}]
  %  \small
  %  \textbf{status: \notflag}\\
  % \textbf{Does all claim are backed by edge IDs: \yesflag} \\
  % \textbf{Context}:
  %  Evidence on THEIA E3 DB of defense evasion via log tampering and masquerading, plus suspicious HTTP egress likely for C2/exfiltration. Firefox wrote \code{/home/admin/profile}. A process labeled \code{/home/admin/profile} then wrote and unlinked multiple log-like files (\code{/var/log/mail}, \code{/var/log/wdev}, \code{/var/log/xdev}). \code{Firefox} and \code{fluxbox} initiated anomalous HTTP connections to 141.43.176.203:80. Critically, a process named \code{/var/log/mail} initiated HTTP to 149.52.198.23:80, strongly suggesting execution of a masqueraded payload from a log path. 
  % \end{tcolorbox}

\end{tcbraster}
% \caption{Demonstrating \pname's agents to produce a verifiable result. The final result and execution trace shown in \autoref{fig:motivate-example-result} and \autoref{fig:execution}, respectively; demonstrating how follow-up agent help reduce over-confident results and discover evidence in \autoref{fig:safety-handshake}.}
% \label{fig:motivate-example}
\caption{Follow-up Agent Output}
\label{fig:mo-fol-agent}
\end{figure}
% ========== FIG B: Auto Follow-up Investigation Plan Template ================
\begin{figure*}[]
\centering
\small
\resizebox{\linewidth}{!}{%
\begin{minipage}{\linewidth}
\begin{tcolorbox}[logpane,title={Follow-up Investigator Plan (auto-generated)}]
\ttfamily
You are a Digital Forensics and Incident Response follow-up investigator.
Execute the following plan using the available tools. Be concise and avoid redundant queries. After your plan execution Finally,
call \texttt{safety\_check} and \texttt{create\_attack\_summary} with all evidence gathered. \\

Plan: \\
DB: \textit{\{db\}} \\
Step budget: \textit{\{followup\_steps\}} \\
- investigate process: run \texttt{check\_artifacts\_in\_postgres(\{db\}:\{process\})}\, (for each extracted process)\\
- investigate file:    run \texttt{check\_artifacts\_in\_postgres(\{db\}:\{file\})}\, (for each extracted file)\\
- investigate ip:      run \texttt{check\_artifacts\_in\_postgres(\{db\}:\{port\})}\, (for each extracted IP)\\
- get edge type $\rightarrow$ \code{get\_event\_type\_from\_source\_id} \& \code{get\_event\_type\_from\_destination\_id}\\\
- aggregate edge ids $\rightarrow$ \code{get\_source\_labels\_from\_edge\_ids} \& \code{get\_destination\_labels\_from\_edge\_ids}\\
- aggregate node labels $\rightarrow$ \code{get\_process\_labels\_from\_ids} \& \code{get\_file\_labels\_from\_ids} \& \code{get\_ip\_labels\_from\_ids} \\
- check is the answer is safe by triggering safety\_check with the compiled context \\
- produce \texttt{create\_attack\_summary} with the compiled context
\end{tcolorbox}
\end{minipage}}
\caption{Prompt template for the follow-up investigator to create a tool-grounded plan for evidence aggregation and safety review.}
\label{fig:followup-plan-prompt}
\end{figure*}
% ============================================================================

% ================= FIGURE 1: Agentic Interaction (vertical panes) ============
\begin{figure*}[]
\centering
\small
\begin{tcbraster}[
  raster columns=1,
  raster valign=top,
  raster row skip=6pt
]

  \begin{tcolorbox}[logpane,title={Safety Agent}]
   \footnotesize
   \textbf{status: \notflag}\\
  \textbf{Does all claim are backed by edge IDs: \yesflag} \\
  \textbf{Context}:
   Evidence on THEIA E3 DB of defense evasion via log tampering and masquerading, plus suspicious HTTP egress likely for C2/exfiltration. Firefox wrote \code{/home/admin/profile}. A process labeled \code{/home/admin/profile} then wrote and unlinked multiple log-like files (\code{/var/log/mail}, \code{/var/log/wdev}, \code{/var/log/xdev}). \code{Firefox} and \code{fluxbox} initiated anomalous HTTP connections to 141.43.176.203:80. Critically, a process named \code{/var/log/mail} initiated HTTP to 149.52.198.23:80, strongly suggesting execution of a masqueraded payload from a log path. 
  \end{tcolorbox}

\end{tcbraster}
% \caption{Demonstrating \pname's agents to produce a verifiable result. The final result and execution trace shown in \autoref{fig:motivate-example-result} and \autoref{fig:execution}, respectively; demonstrating how follow-up agent help reduce over-confident results and discover evidence in \autoref{fig:safety-handshake}.}
% \label{fig:motivate-example}
\caption{Safety Agent Output}
\label{fig:mo-saf-agent}
\end{figure*}

\begin{figure}[]
  \centering
  \footnotesize

  \begin{tcolorbox}[pane,title={Safety Agent (Step 8)}]
  \textbf{status: \flagged}\\
  \textbf{Does all claim are backed by edge IDs: \noflag} \\
  \textbf{Concerns} (summarized):
  \begin{itemize}[leftmargin=*]
    \item Use full network containment (EDR quarantine/VLAN), not outbound-only.
    \item Acquire forensic images before recovery/undelete; operate on copies.
    \item ``Likely via browser'' is a hypothesis need to corroborate with browser artifacts. Treat initial vector \code{Firefox} $\rightarrow$ \code{/home/admin/profile} as hypothesis pending browser history/download provenance.
    \item Phrase \code{/var/log/mail} as consistent with staging/tampering; but require corroboration for \code{/var/log/wdev}.
    \item Monitor-first for listed IPs and explore these IP for C2 connections.
  \end{itemize}
  \textbf{Update Evidence Gap} (excerpt): Treat current evidence as hypothesis and further investigation  using Follow-up agent to scope impact; recover logs from database.
  \end{tcolorbox}

  \begin{tcolorbox}[logpane,title={Follow-Up Agent (Step 9)}]
    \textbf{Required Evidence (from draft)}: \\
      process: [\code{/home/admin/profile}] \\
      files: [\code{/var/log/wdev}] \\
      IPs: [\code{141.43.176.203:80}, \code{149.52.198.23}, \code{104.228.117.212}, \code{5.214.163.155}]\\

      \textbf{Evidence Gathered (Relevant Events or Edges)}:
    \begin{itemize}[leftmargin=*]
        \item \code{AB9231DC-...} : \code{/usr/bin/firefox} 
        $\rightarrow$ WRITE \code{/home/admin/profile}
        % \item \code{C284C9F8-...} : \code{/home/admin/profile} $\rightarrow$ WRITE \code{/var/log/mail}
        \item \code{4850F9E4-...} : \code{/home/admin/profile} $\rightarrow$ WRITE \code{/var/log/wdev} 
        \item \code{52B5094D-...} : \code{/home/admin/profile} $\rightarrow$ UNLINK \code{/var/log/mail}
        \item \code{/usr/bin/fluxbox} $\rightarrow$ \code{141.43.176.203:80} (multiple SENDTO)
        % \item label \code{/var/log/mail} $\rightarrow$ \code{149.52.198.23:80} (edge \code{424E0D67-...})
    \end{itemize}
    \textbf{Negatives (scoped):} \code{./deploy/archive/*}, \code{104.228.117.212:80}, \code{5.214.163.155} - no matches in this dataset.
  \end{tcolorbox}

  \begin{tcolorbox}[pane,title={Safety Agent (Step 10)}]
  \textbf{status: \notflag}\\
  \textbf{Does all claim are backed by edge IDs: \yesflag} \\
  \textbf{Context} (excerpt): Evidence on THEIA E3 DB of defense evasion ...
  \end{tcolorbox}

  \caption{Safety agent flags over-confident answers and answers that do not contain verifiable evidence. It also generates Follow-up investigation plan to bridge the gap.}
  \label{fig:mo-fol-safe-agent}
\end{figure}
\begin{figure}[]
\centering
\small
\resizebox{\columnwidth}{!}{%
\begin{minipage}{\columnwidth}
\begin{tcbraster}[raster columns=1, raster valign=top, raster row skip=6pt]

    %   % Attack summary skeleton
    \begin{tcolorbox}[logpane,title={Attack Summary Skeleton (\texttt{create\_attack\_summary})}]
        \ttfamily
        You are a cyber security and system security expert with experience in data forensic. You have a good knowledge of MITRE Attack TTPs used by Advanced Persistent Threat attackers. If you have knowledge of event's UUID/ID or node's UUID/ID or edge's UUID/ID please include them as evidence. If you have event's description or node's description or edge's description please include them as evidence and to make sense of context.
        The user has compiled results from multiple Postgres-checking tools.\\
        Create a short custom summary of the potential attack:

        Summary: ...
        OS: []
        Processes: []
        Files: []
        IP Address: []

        Context: \{\code{answer}\}
    \end{tcolorbox}

\end{tcbraster}
\end{minipage}}
\caption{Prompt template for evidence-backed summarization.}
\label{fig:mo-summ}
\end{figure}
\begin{figure}[!h]
  \centering
  \footnotesize
  \begin{tcolorbox}[pane,title={Execution Trace (Agent Tool Calls)}]
  \code{User Query:} Based on Drakon dropper attack artifacts, was I attacked? \\
  \code{Step 0} \, \code{Plan Generator} \hfill \\
  \code{Step 1} \, \code{CTI Retriever} \hfill \\
  \code{Step 2-6} \, \code{Artifact Lookup}, \code{Prov. SQL Explorer}, \code{Evt. Lookup Resolver}, \code{Type-Aware Correlator}, \code{Summarizer}\\
  % \code{Step 3} \,  \hfill \\
  % \code{Step 4} \,  \hfill \\
  % \code{Step 5} \,  \hfill \\
  % \code{Step 6} \, \code{Summarizer} \hfill \\
  \code{Step 7} \, \code{Safety Checker} (\flaged) \\
  \code{Step 8} \, \code{Confidence Validator} \hfill \\
  \code{Step 9} \, \code{Follow-up Planner} \hfill \\
  \code{Step 10} \, \code{Plan Generator} \hfill \\
  \code{Step 11} \, \code{Plan Validator} \hfill \\
  \code{Step 12-16} \, \code{Artifact Lookup}...\code{Summarizer} \hfill \\
  % \code{Step 16} \, \code{Summarizer} \hfill \\
  \code{Step 17} \, \code{Safety Checker} (\notflag)\\ Return \textbf{final assessment} with context and actions.
  \end{tcolorbox} 
  \caption{\pname's execution trace for resolving query.}
  \label{fig:mo-trace}
\end{figure}
\begin{figure*}[]
\centering
\small
\resizebox{1\linewidth}{!}{%
\begin{minipage}{1.\linewidth}
\begin{tcbraster}[raster columns=1, raster valign=top, raster row skip=6pt]

  % Role / Prompt
  \begin{tcolorbox}[logpane,title={Safety Agent (Prompt)}]
    \ttfamily
    You are a safety auditor who is needed to generate incident-response reports and triage plans. Your job is to detect overconfidence, unverifiable claims, evidence gap, unsafe instructions, and missing caveats in the current answer. Only approve drafts whose claims are backed by explicit edge IDs and/or node IDs and/or artifacts. Otherwise, mark as flagged and suggest fixes. If flagged, trigger the follow-up agent investigator to close gaps.

    \begin{itemize}[leftmargin=*]
    \item \textbf{Evidence binding.} Each claim about processes/files/IPs should references concrete edge or node IDs.
    \item \textbf{No speculation.} Avoid attribution/C2 claims without artifacts; use hedging + next steps if unsure.
    \item \textbf{Procedural safety.} No risky actions (e.g., tampering live evidence) before imaging/preservation.
    \item \textbf{Scope hygiene.} DB scope is correct; no querying unrelated databases.
    \item \textbf{Data sensitivity.} No leakage of secrets/keys/PII; redact if needed.
    \item \textbf{Mitigation language.} “Monitor-first” vs “block now” is justified with evidence and risk.
    \end{itemize}

    If your analysis recommends ``further investigation'' or ``forensic investigation'', you MUST call the tool \code{auto\_followup\_investigation} with your current draft as \code{final\_answer\_text} before you produce your final answer.\\[2pt]

    Output the answer as:
    \{
    "status": "\notflag" | "\flaged",
    "concerns": [".."],          % empty if status == "safe"
    "suggested\_rewrite": ".."    % optional: concise, evidence-backed rewrite only
    \}
    \\[2pt]
    Current Answer: \{\code{answer}\}
  \end{tcolorbox}

%   % JSON schema (machine-checkable gate)
%   \begin{tcolorbox}[logpane,title={Output Schema (machine-checkable)}]
%     \ttfamily
%     \{
%     "status": "safe" | "flagged",
%     "concerns": ["..."],          % empty if status == "safe"
%     "suggested\_rewrite": "..."    % optional: concise, evidence-backed rewrite only
%     \}
%     \end{tcolorbox}

%   % Handshake (safety -> follow-up -> safety)
  \begin{tcolorbox}[logpane,title={Safety Investigation Loop (control loop)}]
    \ttfamily
    \textbf{Step S1 (Audit)}: Inspect draft. If missing evidence, set \texttt{status="flagged"} and
    populate \texttt{concerns}.\\
    \textbf{Step S2 (Plan)}: Emit follow-up asks (process/file/IP) needed to clear concerns.\\
    \textbf{Step S3 (Investigate)}: Follow-up agent runs: 
    \texttt{check\_artifacts\_in\_postgres} $\rightarrow$ collect edge\_ids $\rightarrow$
    \texttt{get\_source\_labels\_from\_edge\_ids}, \texttt{get\_destination\_labels\_from\_edge\_ids}
    $\rightarrow$ \texttt{create\_attack\_summary}.\\
    \textbf{Step S4 (Re-audit)}: Re-run Safety Agent on revised draft.\\
    \textbf{Step S5 (Summarize/Plan)}: If the answer is  \notflag, then run summarizer tool else plan and investigate.\\[2pt]

    \textbf{Typical states:}\\
    \notflag\; \texttt{status="safe"} \,\,\,$\Rightarrow$ return final assessment. \\ \quad
    \flaged\; \texttt{status="flagged"} \,\,\,$\Rightarrow$ trigger follow-up plan.
  \end{tcolorbox}

\end{tcbraster}
\end{minipage}}
\caption{Prompt template for Safety Agent and an example closed-loop handshake with the Follow-up Investigator to turn flagged answers into evidence-backed, verifiable answers.}
\label{fig:safety-agent-prompt}
\end{figure*}

\begin{figure*}[]
\centering
\small
\resizebox{\linewidth}{!}{%
\begin{minipage}{\linewidth}
\begin{tcbraster}[raster columns=1, raster valign=top, raster row skip=6pt]

  % RAG prompt template
  \begin{tcolorbox}[logpane,title={RAG Prompt Template (\texttt{rag\_with\_reasoner})}]
    \ttfamily
    You are a cyber security and system security expert with experience in data forensic. You have a good knowledge of MITRE Attack TTPs used by steamlthy actors such as Advanced Persistent Threat attackers, Ransomeware and Malware writers. You have read threat reports and are trying to answer questions.
    Be concise and specific about the process names, ip addresses, and file paths when producing a result. If there isn't sufficient information, provide a better query to improve the RAG results.

    Give the answer in the below format:

    Answer:
    Summary: []
    OS: []
    Processes: []
    Files: []
    IP Address: []\\

    Example 1:\\
    Question: For the TC5 Theia Attack, can you summarize ``8.6 THEIA - Firefox Drakon APT BinFmt-Elevate Inject''?\\
    Summary: The attacker gains C2 connections to the target by exploiting Firefox by browsing a website. The attacker then gains root privileges using the installed elevate driver. Further the attacker injected a shellcode     (using process injection technique) into the sshd process which wrote a file sshdlog into the disk.\\
    OS: [Ubuntu 12.04]\\
    Processes: [sshd, scp, insmod]\\
    Files: [/var/log/sshdlog, read\_scan.ko, load\_helper.ko, /e5/dist/sshd-linux-x64]\\
    IP Address: [128.55.12.167:55430, 128.55.12.167:8108, 128.55.12.110, 189.141.204.211, 208.203.20.42]\\

    Context: \{\code{context from vectorDB}\}
    
    Question: \{\code{user query}\}

  \end{tcolorbox}

\end{tcbraster}
\end{minipage}}
\caption{Prompt template for the tool using retrieval (RAG).}
\label{fig:rag-prompts}
\end{figure*}

\begin{table*}[]
\centering
\resizebox{0.8\linewidth}{!}{
\begin{tabular}{lcccc|cccc}
\toprule
\cmidrule(lr){1-5} \cmidrule(lr){6-9}
Method & \makecell{Contextual\\Precision($\uparrow$)} & \makecell{Contextual\\Recall($\uparrow$)} & Relevance($\uparrow$) & Faithfulness($\uparrow$) & \makecell{Contextual\\Precision($\uparrow$)} & \makecell{Contextual\\Recall($\uparrow$)} & Relevance($\uparrow$) & Faithfulness($\uparrow$) \\
\midrule
\multicolumn{5}{c|}{\code{CADETS} (E3)} & \multicolumn{4}{c}{\code{CADETS} (E5)} \\
\midrule
Vanilla RAG & 0.53 (\emph{-0.37}) & 0.52 (\emph{-0.39}) & 0.59 (\emph{-0.32}) & 0.58 (\emph{-0.31}) & 0.58 (\emph{-0.35}) & 0.69 (\emph{-0.25}) & 0.59 (\emph{-0.32}) & 0.52 (\emph{-0.38}) \\
\addlinespace[2pt] 
\makecell{RAG + \\Type filtering} & 0.70 (\emph{-0.20}) & 0.69 (\emph{-0.22}) & 0.80 (\emph{-0.11}) & 0.82 (\emph{-0.07}) & 0.63 (\emph{-0.30}) & 0.82 (\emph{-0.12}) & 0.77 (\emph{-0.14}) & 0.63 (\emph{-0.27}) \\
\midrule
\textbf{\pname} & \textbf{0.90} & \textbf{0.91} & \textbf{0.91} & \textbf{0.89} & \textbf{0.93} & \textbf{0.94} & \textbf{0.91} & \textbf{0.90} \\
\midrule
\multicolumn{5}{c|}{\code{THEIA} (E3)} & \multicolumn{4}{c}{\code{THEIA} (E5)} \\
\midrule
Vanilla RAG & 0.51 (\emph{-0.41}) & 0.58 (\emph{-0.31}) & 0.66 (\emph{-0.24}) & 0.55 (\emph{-0.33}) & 0.60 (\emph{-0.33}) & 0.55 (\emph{-0.32}) & 0.52 (\emph{-0.38}) & 0.69 (\emph{-0.25}) \\
\addlinespace[2pt]
\makecell{RAG + \\Type filtering} & 0.71 (\emph{-0.21}) & 0.84 (\emph{-0.05}) & 0.65 (\emph{-0.25}) & 0.63 (\emph{-0.25}) & 0.62 (\emph{-0.31}) & 0.63 (\emph{-0.24}) & 0.62 (\emph{-0.28}) & 0.69 (\emph{-0.25}) \\
\midrule
\textbf{\pname} & \textbf{0.92} & \textbf{0.89} & \textbf{0.90} & \textbf{0.88} & \textbf{0.93} & \textbf{0.87} & \textbf{0.90} & \textbf{0.94} \\
\midrule
\multicolumn{5}{c|}{\code{CLEARSCOPE} (E3)} & \multicolumn{4}{c}{\code{CLEARSCOPE} (E5)} \\
\midrule
Vanilla RAG & 0.65 (\emph{-0.26}) & 0.53 (\emph{-0.41}) & 0.54 (\emph{-0.34}) & 0.50 (\emph{-0.40}) & 0.59 (\emph{-0.31}) & 0.60 (\emph{-0.34}) & 0.68 (\emph{-0.21}) & 0.65 (\emph{-0.26}) \\
\addlinespace[2pt] 
\makecell{RAG + \\Type filtering} & 0.76 (\emph{-0.15}) & 0.77 (\emph{-0.17}) & 0.80 (\emph{-0.08}) & 0.82 (\emph{-0.08}) & 0.72 (\emph{-0.18}) & 0.82 (\emph{-0.12}) & 0.84 (\emph{-0.05}) & 0.81 (\emph{-0.10}) \\
\midrule
\textbf{\pname} & \textbf{0.91} & \textbf{0.94} & \textbf{0.88} & \textbf{0.90} & \textbf{0.90} & \textbf{0.94} & \textbf{0.89} & \textbf{0.91} \\
\midrule
\multicolumn{5}{c|}{\code{DARPA OpTC}} & \multicolumn{4}{c}{} \\
\midrule
Vanilla RAG & 0.32 (\emph{-0.60}) & 0.25 (\emph{-0.65}) & 0.27 (\emph{-0.66}) & 0.19 (\emph{-0.71}) & - & - & - & - \\
\addlinespace[2pt]
\makecell{RAG + \\Type filtering} & 0.44 (\emph{-0.48}) & 0.41 (\emph{-0.49}) & 0.30 (\emph{-0.63}) & 0.35 (\emph{-0.55}) & - & - & - & - \\
\midrule
\textbf{\pname} & \textbf{0.92} & \textbf{0.90} & \textbf{0.93} & \textbf{0.90} & - & - & - & - \\
\bottomrule
\end{tabular}}
\caption{Intelligence extraction performance on \darpa dataset.}
\label{tab:intel_performance}
\end{table*}

\subsection{Extended Artifact Extraction Performance}\label{sec:appendix_intel_perf}

\autoref{tab:intel_performance} reports detailed artifact extraction results for \pname and different RAG pipelines across all publically available provenance datasets. These numbers complement the high-level trends discussed in \autoref{sec:extract}: \pname reliably recovers key IOCs (files, processes, IPs) needed for downstream provenance queries, with only modest variation. The table also highlights that stronger RAG pipelines tend to improve contextual recall (fewer missed indicators) without degrading precision, which is critical in a security setting where missed IOCs can hide entire attack paths. Overall, \autoref{tab:intel_performance} shows that the CTI extraction stage is robust enough to lead the subsequent provenance analysis with high-quality artifacts.

% % eval detection table

\begin{table*}[]
\centering
\resizebox{1.\textwidth}{!}{
\begin{tabular}{lccccc|ccccc}
\toprule
\cmidrule(lr){1-6} \cmidrule(lr){7-11}
Method & Precision($\uparrow$) & Recall($\uparrow$) & F1+score($\uparrow$) & Token Usage($\downarrow$) & Time($\downarrow$) & Precision($\uparrow$) & Recall($\uparrow$) & F1+score($\uparrow$) & Token Usage($\downarrow$) & Time($\downarrow$) \\
\midrule
\multicolumn{6}{c|}{\code{CADETS} (E3)} & \multicolumn{5}{c}{\code{CADETS} (E5)} \\
\midrule
\kairos~\cite{cheng2024kairos} & 0.01 (\emph{-0.94}) & 0.00 (\emph{-0.91}) & 0.00 (\emph{-0.93}) & - & 8.00 hr (\emph{+7.50 hr}) & 0.00 (\emph{-0.85}) & 0.01 (\emph{-0.84}) & 0.00 (\emph{-0.85}) & - & 14.00 hr (\emph{+13.18 hr}) \\
\magic~\cite{jia2024magic}     & 0.01 (\emph{-0.94}) & 0.95 (\emph{0.04}) & 0.02 (\emph{-0.91}) & - & 38.00 hr (\emph{+37.50 hr}) & 0.00 (\emph{-0.85}) & 1.00 (\emph{+0.15}) & 0.00 (\emph{-0.85}) & - & 87.00 hr (\emph{+86.18 hr}) \\
\flash~\cite{rehman2024flash}  & 0.92 (\emph{-0.03}) & 0.93 (\emph{0.02}) & 0.92 (\emph{-0.00}) & - & 58.00 hr (\emph{+57.50 hr}) & 0.36 (\emph{-0.49}) & 0.49 (\emph{-0.36}) & 0.00 (\emph{-0.85}) & - & 207.00 hr (\emph{+206.18 hr}) \\
\orthrus~\cite{jian2025}       & 0.25 (\emph{-0.70}) & 0.05 (\emph{-0.86}) & 0.08 (\emph{-0.85}) & - & 4.00 hr  (\emph{+3.50 hr})  & 0.17 (\emph{-0.68}) & 0.02 (\emph{-0.83}) & 0.04 (\emph{-0.81}) & - & 7.00 hr (\emph{+6.18 hr}) \\
Strawman Agent             & 0.01 (\emph{-0.89}) & 0.01 (\emph{-0.87}) & 0.01 (\emph{-0.88}) & 2M (\emph{+1.60M}) & 2.00 hr (\emph{+1.50 hr}) & 0.02 (\emph{-0.83}) & 0.03 (\emph{-0.82}) & 0.02 (\emph{-0.83}) & 6M (\emph{+5.50M}) & 5.50 hr (\emph{+4.68 hr}) \\
\midrule
\pname              & 0.90 & 0.88 & 0.89 & 423.0K & 32.0 min & 0.85 & 0.85 & 0.85 & 501.0K & 49.0 min \\
\midrule
\multicolumn{6}{c|}{\code{THEIA} (E3)} & \multicolumn{5}{c}{\code{THEIA} (E5)} \\
\midrule
\kairos~\cite{cheng2024kairos}   & 1.00 (\emph{+0.01}) & 0.03 (\emph{-0.84}) & 0.06 (\emph{-0.87}) & - & 1.50 hr (\emph{+1.20 hr}) & 0.00 (\emph{-0.95}) & 0.01 (\emph{-0.91}) & 0.00 (\emph{-0.93}) & - & 5.00 hr (\emph{+4.17 hr}) \\
\magic~\cite{jia2024magic}       & 0.00 (\emph{-0.99}) & 0.97 (\emph{+0.10}) & 0.00 (\emph{-0.00}) & - & 23.00 hr (\emph{+22.70 hr}) & 0.00 (\emph{-0.95}) & 0.01 (\emph{-0.91}) & 0.00 (\emph{-0.93}) & - & 24.00 hr (\emph{+23.17 hr}) \\
\flash~\cite{rehman2024flash}    & 0.00 (\emph{-0.99}) & 0.18 (\emph{-0.69}) & 0.00 (\emph{-0.00}) & - & 12.00 hr (\emph{+11.70 hr}) & 0.00 (\emph{-0.95}) & 0.62 (\emph{-0.3}) & 0.00 (\emph{-0.93}) & - & 99.00 hr (\emph{+98.17 hr}) \\
\orthrus~\cite{jian2025}         & 0.52 (\emph{-0.47}) & 0.40 (\emph{-0.47}) & 0.45 (\emph{+0.45}) & - & 45.00 min (\emph{+27.00 min}) & 0.87 (\emph{-0.08}) & 1.00 (\emph{+0.08}) & 0.93 (\emph{-0.00}) & - & 6.50 hr (\emph{+5.67 hr}) \\
Strawman Agent               & 0.05 (\emph{-0.94}) & 0.08 (\emph{-0.79}) & 0.06 (\emph{-0.87}) & 2M (\emph{+1.67M}) & 1.50 hr (\emph{+1.20 hr}) & 0.02 (\emph{-0.93}) & 0.03 (\emph{-0.89}) & 0.02 (\emph{-0.91}) & 4M (\emph{+3.46M}) & 3.00 hr (\emph{+2.17 hr}) \\
\midrule
\pname                          & 0.99 & 0.87 & 0.93 & 326.0K & 18.0 min & 0.95 & 0.92 & 0.93 & 540.0K & 50.0 min \\
\midrule
\multicolumn{6}{c|}{\code{CLEARSCOPE} (E3)} & \multicolumn{5}{c}{\code{CLEARSCOPE} (E5)} \\
\midrule
\kairos~\cite{cheng2024kairos}   & 0.00 (\emph{-0.95}) & 0.01 (\emph{-0.88}) & 0.00 (\emph{-0.92}) & - & 33.0 min (\emph{+3 min}) & 0.25 (\emph{-0.65}) & 0.00 (\emph{-0.85}) & 0.00 (\emph{-0.87}) & - & 4.00 hr (\emph{3.25 hr}) \\
\magic~\cite{jia2024magic}       & 0.00 (\emph{-0.95}) & 0.97 (\emph{0.08}) & 0.00 (\emph{-0.92}) & - & 2.50 hr (\emph{+2 hr}) & 0.00 (\emph{-0.9}) & 1.00 (\emph{0.15}) & 0.00 (\emph{-0.87}) & - & 15.00 hr (\emph{14.25 hr}) \\
\flash~\cite{rehman2024flash}    & 0.00 (\emph{-0.95}) & 0.01 (\emph{-0.88}) & 0.00 (\emph{-0.92}) & - & 37.00 hr (\emph{+36.5 hr}) & 0.00 (\emph{-0.9}) & 0.29 (\emph{-0.56}) & 0.00 (\emph{-0.87}) & - & 49.00 hr (\emph{48.25 hr}) \\
\orthrus~\cite{jian2025}         & 0.25 (\emph{-0.7}) & 0.05 (\emph{-0.84}) & 0.08 (\emph{-0.84}) & - & 10.0 min (\emph{+20 min}) & 0.33 (\emph{-0.57}) & 0.07 (\emph{-0.78}) & 0.12 (\emph{-0.76}) & - & 4.00 hr (\emph{3.25 hr}) \\
Strawman Agent               & 0.03 (\emph{-0.92}) & 0.07 (\emph{-0.82}) & 0.04 (\emph{-0.88}) & 6M (\emph{+5.56M}) & 5.50 hr (\emph{+5 hr}) & 0.07 (\emph{-0.83}) & 0.01 (\emph{-0.84}) & 0.02 (\emph{-0.86}) & 7M (\emph{6.43M}) & 6.00 hr (\emph{5.25 hr}) \\
\midrule
\pname                           & 0.95 & 0.89 & 0.92 & 440.0K & 30.0 min & 0.90 & 0.85 & 0.87 & 570.0K & 45.0 min \\
\midrule
\multicolumn{6}{c|}{\code{DARPA OpTC}} & \multicolumn{5}{c}{} \\
\midrule
\kairos~\cite{cheng2024kairos}        & 0.01 (\emph{-0.94}) & 0.00 (\emph{-0.91}) & 0.00 & - & 8.00 hr (\emph{+7.20 hr }) & - & - & - & - & - \\
\magic~\cite{jia2024magic}            & 0.01 (\emph{-0.94}) & 0.95 (\emph{0.04}) & 0.02 & - & 38.00 hr (\emph{+37.20 hr}) & - & - & - & - & - \\
\flash~\cite{rehman2024flash}         & 0.92 (\emph{-0.03}) & 0.93 (\emph{0.02}) & 0.92 & - & 58.00 hr (\emph{+57.20 hr}) & - & - & - & - & - \\
\orthrus~\cite{jian2025}              & 0.25 (\emph{-0.7}) & 0.05 (\emph{-0.86}) & 0.08 & - & 4.00 hr (\emph{+3.20 hr}) & - & - & - & - & - \\
Strawman Agent                    & 0.00 (\emph{-0.95}) & 0.01 (\emph{-0.9}) & 0.00 (\emph{-0.93}) & 10M (\emph{+9.5M}) & 8.00 hr (\emph{+7.20 hr}) & - & - & - & - & - \\
\midrule
\pname              & 0.95 & 0.91 & 0.93 &  450.0K  &  40 min  & - & - & - & - & - \\
\bottomrule
\end{tabular}}
\caption{Comparing threat detection performance  of \pname against \sota \pids and Strawman Agent using ground truth IoCs.}
\label{tab:detection_performance}
\end{table*}

\subsection{Extended Threat Detection Performance}\label{sec:appendix_detection_perf}

\autoref{tab:detection_performance} provides the full breakdown of \pname's threat detection performance, including per-dataset precision, recall, and F1-scores for each \sota \pids and Strawman Agent. These results extend \autoref{sec:detect} by showing that the trends we highlight there hold consistently across all DARPA TC datasets: \pname maintains high precision (low false-alarm rates) while preserving strong recall even on structurally complex datasets such as \code{THEIA} and \code{OpTC}. The table also makes clear that detectors which better coordinate tool calls and reason over multi-hop evidence yield more reliable security decisions.

% % eval ablation detection table

\begin{table*}[]
\centering
\resizebox{1.\textwidth}{!}{
\begin{tabular}{lccccc|ccccc}
\toprule
\cmidrule(lr){1-6} \cmidrule(lr){7-11}
Removing & Precision($\uparrow$) & Recall($\uparrow$) & F1+score($\uparrow$) & Token Usage($\downarrow$) & Time($\downarrow$) & Precision($\uparrow$) & Recall($\uparrow$) & F1+score($\uparrow$) & Token Usage($\downarrow$) & Time($\downarrow$) \\
\midrule

\multicolumn{6}{c|}{\code{CADETS} (E3)} & \multicolumn{5}{c}{\code{CADETS} (E5)} \\
\midrule
 \makecell[l]{Threat Intelligence RE}                 & 0.01 (\emph{-0.89}) & 0.01 (\emph{-0.87}) & 0.01 (\emph{-0.88}) & 2M (\emph{+1.60M}) & 2.52 hr (\emph{+1.90 hr}) & 0.02 (\emph{-0.83}) & 0.04 (\emph{-0.81}) & 0.03 (\emph{-0.82}) & 6M (\emph{+5.5M}) & 9.78 hr (\emph{+8.96 hr}) \\
 \makecell[l]{System Data RE}   & 0.09 (\emph{-0.81}) & 0.02 (\emph{-0.86}) & 0.03 (\emph{-0.86}) & 2M (\emph{+1.60M}) & 2.52 hr (\emph{+1.90 hr}) & 0.10 (\emph{-0.75}) & 0.08 (\emph{-0.77}) & 0.09 (\emph{-0.76}) & 5.50M (\emph{+5M}) & 8.97 hr (\emph{+8.15 hr}) \\
 Filtration Engine                  & 0.87 (\emph{-0.03}) & 0.82 (\emph{-0.06}) & 0.84 (\emph{-0.05}) & 1M (\emph{+577K}) & 1.26 hr (\emph{+43.60 min}) & 0.82 (\emph{-0.03}) & 0.78 (\emph{-0.07}) & 0.80 (\emph{-0.05}) & 1M (\emph{+499K}) & 1.63 hr (\emph{+48.80 min}) \\
 Investigation Agent & 0.85 (\emph{-0.05}) & 0.85 (\emph{-0.03}) & 0.85 (\emph{-0.04}) & 376K (\emph{-47K}) & 28.40 min (\emph{-3.60 min}) & 0.78 (\emph{-0.07}) & 0.77 (\emph{-0.08}) & 0.77 (\emph{-0.08}) & 437K (\emph{-64K}) & 42.70 min (\emph{+6.30 min}) \\
 Follow-up Agent     & 0.60 (\emph{-0.30}) & 0.50 (\emph{-0.38}) & 0.55 (\emph{-0.34}) & 113K (\emph{-310K}) & 8.60 min (\emph{-23.40 min}) & 0.59 (\emph{-0.26}) & 0.44 (\emph{-0.41}) & 0.50 (\emph{-0.35}) & 126K (\emph{-375K}) & 12.30 min (\emph{+36.70 min}) \\
 Safety Agent        & 0.87 (\emph{-0.03}) & 0.80 (\emph{-0.08}) & 0.83 (\emph{-0.06}) & 401K (\emph{-22K}) & 30.30 min (\emph{-1.70 min}) & 0.80 (\emph{-0.05}) & 0.78 (\emph{-0.07}) & 0.79 (\emph{-0.06}) & 481K (\emph{-20K}) & 47.00 min (\emph{+2.00 min}) \\
\midrule
 \pname              & 0.90 & 0.88 & 0.89 & 423.0K & 32.0 min & 0.85 & 0.85 & 0.85 & 501.0K & 49.0 min \\
\midrule
\multicolumn{6}{c|}{\code{THEIA} (E3)} & \multicolumn{5}{c}{\code{THEIA} (E5)} \\
\midrule
\makecell[l]{Threat Intelligence RE}                              & 0.03 (\emph{-0.96}) & 0.05 (\emph{-0.82}) & 0.04 (\emph{-0.89}) & 1.10M (\emph{+774K}) & 1.01 hr (\emph{+42.70 min}) & 0.02 (\emph{-0.93}) & 0.01 (\emph{-0.91}) & 0.01 (\emph{-0.92}) & 4M (\emph{+3.46M}) & 6.17 hr (\emph{+5.34 hr}) \\
 \makecell[l]{System Data RE}               & 0.10 (\emph{-0.89}) & 0.08 (\emph{-0.79}) & 0.09 (\emph{-0.84}) & 2M (\emph{+1.67M}) & 1.84 hr (\emph{+1.54 hr}) & 0.13 (\emph{-0.82}) & 0.08 (\emph{-0.84}) & 0.10 (\emph{-0.84}) & 3.80M (\emph{+3.26M}) & 5.86 hr (\emph{+5.03 hr}) \\
 Filtration Engine                              & 0.95 (\emph{-0.04}) & 0.82 (\emph{-0.05}) & 0.88 (\emph{-0.05}) & 380K (\emph{+54K}) & 21.00 min (\emph{+3.00 min}) & 0.90 (\emph{-0.05}) & 0.88 (\emph{-0.04}) & 0.89 (\emph{-0.04}) & 2M (\emph{+1.46M}) & 3.09 hr (\emph{+2.25 hr}) \\
 Investigation Agent             & 0.87 (\emph{-0.12}) & 0.81 (\emph{-0.06}) & 0.84 (\emph{-0.09}) & 280K (\emph{-46K}) & 15.50 min (\emph{+2.50 min}) & 0.79 (\emph{-0.16}) & 0.75 (\emph{-0.17}) & 0.77 (\emph{-0.17}) & 410K (\emph{-130K}) & 38.00 min (\emph{+12.00 min}) \\
 Follow-up Agent                 & 0.66 (\emph{-0.33}) & 0.56 (\emph{-0.31}) & 0.61 (\emph{-0.32}) & 150K (\emph{-176K}) & 8.30 min (\emph{+9.70 min}) & 0.60 (\emph{-0.35}) & 0.45 (\emph{-0.47}) & 0.51 (\emph{-0.42}) & 122K (\emph{-418K}) & 11.30 min (\emph{+34.70 min}) \\
 Safety Agent                    & 0.90 (\emph{-0.05}) & 0.84 (\emph{-0.05}) & 0.87 (\emph{-0.05}) & 400K (\emph{+40K}) & 27.30 min (\emph{+2.70 min}) & 0.85 (\emph{-0.05}) & 0.82 (\emph{-0.03}) & 0.83 (\emph{-0.04}) & 513K (\emph{-57K}) & 40.50 min (\emph{+4.50 min}) \\
\midrule
 \pname                          & 0.99 & 0.87 & 0.93 & 326.0K & 18.0 min & 0.95 & 0.92 & 0.93 & 540.0K & 50.0 min \\
\midrule
\multicolumn{6}{c|}{\code{CLEARSCOPE} (E3)} & \multicolumn{5}{c}{\code{CLEARSCOPE} (E5)} \\
\midrule
\makecell[l]{Threat Intelligence RE}                              & 0.01 (\emph{-0.94}) & 0.00 (\emph{-0.89}) & 0.00 (\emph{-0.92}) & 6M (\emph{+5.56M}) & 6.82 hr (\emph{+6.32 hr}) & 0.02 (\emph{-0.88}) & 0.03 (\emph{-0.82}) & 0.02 (\emph{-0.85}) & 7M (\emph{+6.43M}) & 9.21 hr (\emph{+8.46 hr}) \\
\makecell[l]{System Data RE}                & 0.07 (\emph{-0.88}) & 0.04 (\emph{-0.85}) & 0.05 (\emph{-0.87}) & 6M (\emph{+5.56M}) & 6.82 hr (\emph{+6.32 hr}) & 0.08 (\emph{-0.82}) & 0.05 (\emph{-0.8}) & 0.06 (\emph{-0.81}) & 6M (\emph{+5.43M}) & 7.89 hr (\emph{+7.14 hr}) \\
Filtration Engine                               & 0.87 (\emph{-0.08}) & 0.82 (\emph{-0.07}) & 0.84 (\emph{-0.07}) & 1.2M (\emph{+760K}) & 1.36 hr (\emph{+51.80 min}) & 0.84 (\emph{-0.06}) & 0.80 (\emph{-0.05}) & 0.82 (\emph{-0.05}) & 910K (\emph{+340K}) & 1.20 hr (\emph{+26.8 min}) \\
Investigation Agent              & 0.85 (\emph{-0.1}) & 0.70 (\emph{-0.19}) & 0.77 (\emph{-0.15}) & 378K (\emph{-62K}) & 25.80 min (\emph{+4.20 min}) & 0.77 (\emph{-0.13}) & 0.73 (\emph{-0.12}) & 0.75 (\emph{-0.12}) & 510K (\emph{-60K}) & 40.3 min (\emph{+4.7 min}) \\
Follow-up Agent                  & 0.60 (\emph{-0.35}) & 0.50 (\emph{-0.39}) & 0.55 (\emph{-0.37}) & 113K (\emph{-327K}) & 7.70 min (\emph{+22.30 min}) & 0.65 (\emph{-0.25}) & 0.44 (\emph{-0.41}) & 0.52 (\emph{-0.35}) & 130K (\emph{-440K}) & 10.3 min (\emph{+34.7 min}) \\
Safety Agent                     & 0.90 (\emph{-0.05}) & 0.84 (\emph{-0.05}) & 0.87 (\emph{-0.05}) & 400K (\emph{-40K}) & 27.30 min (\emph{+2.70 min}) & 0.85 (\emph{-0.05}) & 0.82 (\emph{-0.03}) & 0.83 (\emph{-0.04}) & 513K (\emph{-57K}) & 40.5 min (\emph{+4.5 min}) \\
\midrule
\pname                           & 0.95 & 0.89 & 0.92 & 440.0K & 30.0 min & 0.90 & 0.85 & 0.87 & 570.0K & 45.0 min \\
\midrule
\multicolumn{6}{c|}{\code{DARPA OpTC}} & \multicolumn{5}{c}{} \\
\midrule
\makecell[l]{Threat Intelligence RE}                 & 0.00 (\emph{-0.95}) & 0.01 (\emph{-0.90}) & 0.00 (\emph{-0.93}) & 10M (\emph{+9.5M}) & 21.82 hr (\emph{+19.82 hr}) & - & - & - & - & - \\
\makecell[l]{System Data RE}   & 0.08 (\emph{-0.87}) & 0.02 (\emph{-0.89}) & 0.03 (\emph{-0.90}) & 10M (\emph{+9.5M})  & 18.18 hr (\emph{+16.18 hr}) & - & - & - & - & - \\
Filtration Engine                  & 0.80 (\emph{-0.15}) & 0.75 (\emph{-0.16}) & 0.77 (\emph{-0.16}) & 700K (\emph{+350K})  & 3.64 hr (\emph{+1.64 hr})   & - & - & - & - & - \\
Investigation Agent & 0.58 (\emph{-0.37}) & 0.55 (\emph{-0.36}) & 0.56 (\emph{-0.36}) & 400K (\emph{-50K})  & 30 min (\emph{-10.90 min})  & - & - & - & - & - \\
Follow-up Agent     & 0.60 (\emph{-0.35}) & 0.55 (\emph{-0.36}) & 0.57 (\emph{-0.36}) & 120K (\emph{-330K})  & 18.10 min (\emph{-22 min}) & - & - & - & - & - \\
Safety Agent        & 0.87 (\emph{-0.08}) & 0.84 (\emph{-0.07}) & 0.85 (\emph{-0.07}) & 390K (\emph{-60K})  & 30 min (\emph{-8.00 min}) & - & - & - & - & - \\
\midrule
\pname              & 0.95 & 0.91 & 0.93 &  450.0K  &  40 min  & - & - & - & - & - \\
\bottomrule
\end{tabular}}
\caption{Ablation study evaluating how removing each component of \pname impacts threat detection performance.}
\label{tab:abl_det_perf}
\end{table*}

\subsection{Extended Ablation Details}\label{sec:appendix_ablation_det}

\autoref{tab:abl_det_perf} expands the ablation results summarized in \autoref{sec:ablation} by providing per-component precision, recall, F1-score, token usage, and runtime across all datasets. The table quantifies the qualitative trends visible in \autoref{fig:det_det_perf_e5}: removing RAG or the SQL enhancements severely harms detection accuracy while inflating both token usage and latency, reflecting context explosion over irrelevant events; dropping the Follow-up Agent produces the largest F1 degradation among the orchestration components, underscoring its central role in iterative evidence gathering; and disabling the Safety Agent yields only modest accuracy drops but slightly reduced resource use, confirming that it primarily contributes additional verification and re-writing passes. Taken together, \autoref{tab:abl_det_perf} shows that \pname's design is not redundant, each module contributes a distinct performance gain of that matters in operational deployments.

% % 50 error analysis user query
\begin{figure*}[t]
\centering
\scriptsize
\begin{tcbraster}[
  raster columns=1,
  raster valign=top,
  raster row skip=6pt
]
  % User Query
  \begin{tcolorbox}[logpane,title={Generic Queries}]
    \small
      1. Based on the system artifacts you see, can you determine whether this host was successfully compromised? \\
    2. For this intrusion scenario, list every file artifact that appears to be linked to the attack. \\
    3. Which file paths in the environment are associated with the suspected dropper and backdoor activity? \\
    4. Enumerate all files that the suspected malicious activity touched (read, write, or delete) on this system. \\
    5. What file nodes (UUIDs and paths) are associated with the suspected incident in the provenance data?\\
    6. Based on the system artifacts related to a potential privilege-escalation attempt, can you tell me if a successful attack occurred? \\
    7. Which log and configuration files were accessed during the suspected compromise window? \\
    8. In this scenario, which files does the suspected malicious component interact with across the system?\\
    9. Retrieve the complete list of file artifacts that appear to belong to this incident. \\
    10. Which files under sensitive configuration and user home directories are associated with the suspected attack? \\
    11. What files tied to the suspected dropper and any follow-on backdoor activity can you find in the ground-truth or labeled data? \\
    12. Identify all file paths that show attacker-like activity (e.g., transient shared objects, log tampering, profile modifications). \\
    13. Which file nodes in the provenance data capture the dropper’s actions and the subsequent backdoor behavior? \\
    14. Which files were written, read, or deleted in this case according to the database-backed provenance logs? \\
    15. Provide all file UUIDs, paths, and node sources that appear related to this incident. \\
    16. What file-based artifacts best characterize this compromise from a forensic standpoint? \\
    17. Which browser-related configuration or native-messaging files and related artifacts are implicated in this scenario? \\
    18. List all file artifacts (including temporary directories, log directories, and user profile paths) associated with this attack set. \\
    19. In the labeled ground-truth for this case, which file objects are explicitly marked as part of the attack? \\
    20. Summarize the file-level evidence for this scenario based on the available provenance or database records. \\
    21. In this incident, how did the attacker’s processes behave and what process identifiers (PIDs or node IDs) are involved? \\
    22. Which attacker-controlled or suspicious processes (with subject strings) are present in the provenance for this compromise? \\
    23. List all malicious or suspicious process subjects and node IDs linked to the attack in this dataset. \\
    24. Retrieve the browser and helper processes used by the attacker in this scenario, along with their identifiers. \\
    25. How do the dropper and backdoor processes appear in the data (process subjects plus node IDs), and how are they related? \\
  \end{tcolorbox}

  \begin{tcolorbox}[logpane,title={Database Specific Queries}]
    \small
    26. Which process nodes represent the execution of gtcache and pass\_mgr in the THEIA E3 Drakon case? \\
    27. Show all process subjects under \code{admin/Downloads/firefox} that are associated with the THEIA E3 Drakon attack. \\
    28. For THEIA E3, what attacker process behavior is recorded for the Drakon/Firefox incident set? \\
    29. Identify all process UUIDs and subjects that correspond to attacker activity in THEIA E3 Drakon/Firefox. \\
    30. Which processes launched from the malicious Firefox bundle participate in the THEIA E3 Drakon/Firefox compromise? \\
    31. Summarize the attacker’s process tree (subjects and node IDs) for the THEIA E3 Drakon/Firefox attacks. \\
    32. What native-messaging host processes are implicated in the THEIA E3 Drakon/Firefox scenario? \\
    33. Which Firefox-related processes (downloads, backdoor, in-memory execution) show up in THEIA E3? \\
    34. In THEIA E3, which processes correspond to the Drakon browser extension executing its payload? \\
    35. Provide all process-level artifacts (subjects and node indices) for the Drakon/Firefox attack in THEIA E3. \\
    36. How does the sequence of Firefox and helper processes look in the THEIA E3 Drakon/Firefox ground-truth data? \\
    37. Which attacker processes are linked specifically to the in-memory Drakon backdoor in THEIA E3? \\
    38. From the THEIA E3 database, what evidence do we have about the lifecycle of attacker processes in Drakon/Firefox? \\
    39. List every process subject string tied to the Drakon dropper and Firefox backdoor in the THEIA E3 attack set. \\
    40. In the THEIA E3 Drakon/Firefox attack chain, which file artifacts appear in the provenance graph? \\
    41. For the DARPA TC3 THEIA-E3 Drakon/Firefox scenario, list all files that were created, modified, or deleted by the attacker. \\
    42. Using the THEIA E3 Drakon/Firefox dataset, which file paths are associated with the malicious browser extension and backdoor activity? \\
    43. From the THEIA E3 Drakon dropper + Firefox backdoor events, enumerate the key log, profile, and shared-object files involved in the compromise. \\
    44. Looking at the THEIA E3 Drakon/Firefox traces, what files in /etc, /var/log, /tmp, and /home/admin are tied to the attack? \\
    45. In the THEIA E3 Drakon/Firefox incident, which attacker-controlled processes (by subject string and internal ID) are visible in the provenance data? \\
    46. For the THEIA E3 Drakon dropper scenario, summarize the process subjects and their node IDs that correspond to the malicious Firefox and native-messaging hosts activity. \\
    47. Within the THEIA E3 Drakon/Firefox attack set, which process nodes show the transition from the downloaded Firefox binary to /usr/bin/firefox? \\
    48. For the THEIA E3 Drakon/Firefox backdoor, identify all attacker processes tied to the gtcache native messaging host and provide their IDs. \\
    49. Using the THEIA E3 Drakon/Firefox provenance, which process subjects and file artifacts together best characterize the attacker’s behavior path? \\
    50. Enumerate the key log, profile, and shared-object files involved in the compromisein Drakon dropper + Firefox backdoor events.
  \end{tcolorbox}

\end{tcbraster}
\caption{Fifty user prompts to invoke different failure case in DARPA Theia E3. Twenty-five user prompts are generic questions that a security response analyst will ask an automated forensics agent (\ie \pname) and twenty-five are dataset specific questions.}
\label{fig:error-query}
\end{figure*}

\begin{table*}[]
\centering
\resizebox{\linewidth}{!}{
\begin{tabular}{lccccccc}
\toprule
\code{Failure Reasons}  & \code{OpTC} & \code{THEIA (E3)} & \code{THEIA (E5)} & \code{CADETS (E3)} & \code{CADETS (E5)} & \code{CLEARSCOPE (E3)} & \code{CLEARSCOPE (E5)} \\
\midrule
Total Errors                                & 5/50 & 5/50 & 8/50 & 6/50 & 13/50 & 3/50 & 14/50 \\
\midrule
Hallucinated artifact                       & - & - & - & - & 2/50 & - & - \\
\addlinespace[2pt]
Benign artifact extracted                   & 1/50 & 2/50 & 2/50 & - & 3/50 & 1/50 & 3/50 \\
\addlinespace[2pt]
Non-system artifact extraction              & 2/50 & 1/50 & 3/50 & 1/50 & 2/50 & 1/50 & 3/50 \\
\addlinespace[2pt]
Artifact absent from DB                     & - & 1/50 & 2/50 & 2/50 & 1/50 & - & 2/50 \\
\addlinespace[2pt]
Context explosion                           & - & - & - & - & 2/50 & - & 1/50 \\
\addlinespace[2pt]
Incorrect correlation                       & - & - & - & 1/50 & 1/50 & - & 3/50 \\
\addlinespace[2pt]
Inconclusive evidence                       & 2/50 & 1/50 & 1/50 & 2/50 & 3/50 & 1/50 & 2/50 \\
\bottomrule
\end{tabular}
}
\caption{Failure analysis across datasets. Cells contain counts of failures out of 50 queries per dataset.}
\label{tab:failure-analysis-by-dataset}
\end{table*}

\subsection{Extended Error Analysis}\label{sec:appendix_error_queries}

\autoref{tab:failure-analysis-by-dataset} reports detailed detail error analysis study across all publically available provenance datasets. 
\autoref{fig:error-query} visualizes the prompts used for error analysis over 50 representative user queries. 
We observe that the prompts that cause the most errors cluster around underspecified or highly abstract questions (for example, queries that omit concrete artifacts such as general questions), as well as prompts that implicitly require cross-scenario reasoning beyond the available provenance window.

% TODO: S&P submission edits 
% \subsection{Implementation}

% \pname is implemented in \code{python} using an agentic architecture using \code{LangChain} for LLM orchestration and integrates PostgreSQL, \code{ChromaDB}, and OpenAI's \code{text-embedding-ada-002} model for provenance-aware CTI retrieval.
% For multi-agent orchestration, we combine \code{smolagents} and \code{LangGraph}, with tools implemented via \code{CodeAgent} and \code{ToolCallingAgent}. An analyst-facing \code{Gradio} UI supports interactive investigations through a chat-style template and provenance visualizations. For evaluation, \pname integrates \code{deepeval} and Meta's \code{synthetic-data-kit} to benchmark agent outputs and generate realistic security-focused test prompts.

\balance{}
\clearpage
\end{document}